\documentclass[%
 preprint,
 superscriptaddress,
nofootinbib,
 amsmath,amssymb,
 aip,
 jap,
 graphicx
]{revtex4-2}

\usepackage{graphicx}% Include figure files
\usepackage{dcolumn}% Align table columns on decimal point
\usepackage{tabularx}
\usepackage{booktabs}
\usepackage{bm} % bold math
\usepackage[version=4]{mhchem}
\usepackage[section]{placeins} % keep tables/figures within their section
\usepackage{array, makecell} % to split long table cells in two lines
\usepackage{siunitx}
\usepackage{pdftexcmds}
\usepackage[journal=rsc]{chemstyle}

\usepackage{hyperref}% add hypertext capabilities
\usepackage[capitalise]{cleveref} % For easy cross references

\usepackage[myheadings]{fullpage} % avoid overlap of footnote and page number
\usepackage{caption}
\newcommand{\sub}[1]{$_{\text{#1}}$}
\makeatletter
\newcommand\kv[2]{%
  \ifnum\pdf@strcmp{\unexpanded{#1}}{V}=0 %
     \expandafter\@firstoftwo
  \else
    \expandafter\@secondoftwo
  \fi
    {\textit{#1}\!\sub{#2}}
    {#1\sub{#2}}%
}
\makeatother
\makeatletter
\newcommand\kvc[3]{%
  \ifnum\pdf@strcmp{\unexpanded{#1}}{V}=0 %
     \expandafter\@firstoftwo
  \else
    \expandafter\@secondoftwo
  \fi
    {\textit{#1}\!\sub{#2}$^{#3}$}
    {#1\sub{#2}$^{#3}$}
}
\makeatother

\DeclareSIUnit\angstrom{\protect \text {Å}}

\hypersetup{
    colorlinks=true,                
    breaklinks=true,                
    urlcolor= black,                
    linkcolor= blue,                
    bookmarksopen=false,
    filecolor=black,
    citecolor=magenta,
}

\usepackage{subcaption} % for SI
\newcommand{\angstrom}{\mbox{\normalfont\AA}} % avoid issue with \AA in math mode
\DeclareSIUnit\angstrom{\text {Å}}
\usepackage{booktabs} 

\begin{document}

\title{Machine-learning structural reconstructions for accelerated point defect calculations}

\author{Irea Mosquera-Lois}
\affiliation{Thomas Young Centre \& Department of Materials, Imperial College London, London SW7 2AZ, UK}

\author{Seán R. Kavanagh}
\affiliation{Thomas Young Centre \& Department of Materials, Imperial College London, London SW7 2AZ, UK}
%\affiliation{Thomas Young Centre \& Department of Chemistry, University College London, 20 Gordon Street, London WC1H 0AJ, UK}

\author{Alex M. Ganose}
\affiliation{Thomas Young Centre \& Department of Chemistry, Imperial College London, London W12 0BZ, UK}

\author{Aron Walsh}
\email{a.walsh@imperial.ac.uk}
\affiliation{Thomas Young Centre \& Department of Materials, Imperial College London, London SW7 2AZ, UK}
\affiliation{Department of Physics, Ewha Womans University, Seoul 03760, Korea}

\date{\today}% It is always \today, today,

% For npj format:
% \author[1]{\fnm{Irea} \sur{Mosquera-Lois}}
% \author[1]{\fnm{Seán R.} \sur{Kavanagh}}
% \author[2]{\fnm{Alex M.} \sur{Ganose}}
% \author[1,3]{\fnm{Aron} \sur{Walsh}}

\begin{abstract} 
Defects dictate the properties of many functional materials. To understand the behaviour of defects and their impact on physical properties, it is necessary to identify the most stable defect geometries. However, global structure searching is computationally challenging for high-throughput defect studies or materials with complex defect landscapes, like alloys or disordered solids. Here, we tackle this limitation by harnessing a machine-learning surrogate model to qualitatively explore the defect structural landscape. By learning defect motifs in a family of related metal chalcogenide and mixed anion crystals, the model successfully predicts favourable reconstructions for unseen defects in unseen compositions for 90\% of cases, thereby reducing the number of first-principles calculations by 73\%. Using CdSe$_x$Te$_{1-x}$ alloys as an exemplar, we train a model on the end member compositions and apply it to find the stable geometries of all inequivalent vacancies for a range of mixing concentrations, thus enabling more accurate and faster defect studies for configurationally complex systems.
\end{abstract}

\maketitle

\section{Introduction}

Defects control the properties of many functional materials and devices\cite{Sambur_2023}, like solar cells\cite{shockleyStatisticsRecombinationsHoles1952,kimUpperLimitPhotovoltaic2020}, batteries\cite{maierThermodynamicsElectrochemicalLithium2013,squiresLowElectronicConductivity2022}, catalysts\cite{liDefectEngineeringFuel2020,pastorElectronicDefectsMetal2022,Kehoe2011}, and quantum computers\cite{Ivdy2018,Weber2010,Thomas2023,Dreyer_2018}. 
To discover better materials for these applications it is thus necessary to predict how their defects behave. 
However, defect calculations are computationally demanding. 
The large supercells and high level of theory required to obtain robust predictions typically limit point defect analysis to in-depth studies of specific materials. 
In a move towards data-driven defect workflows\cite{Yan_2024}, defect databases\cite{Davidsson2023,Sluydts2016,Bertoldo2022,huang_unveiling_2023,medasani_predicting_2016,Rahman_23,ivanov2023database} and surrogate models have been developed to predict defect properties, like the dominant defect type\cite{medasani_predicting_2016}, formation\cite{Kumagai_2021,Deml2015,Broberg2023,MannodiKanakkithodi2022,Varley2017,Wan2021,Wexler2021,Frey2020,Sharma_20,Baldassarri_23,Park_23_exploring,kazeev_sparse_2023,Choudhary_2023_can,zhao_machine_2023,Manzoor2021,Rahman_23} and migration\cite{Manzoor2021} energies, and charge transition levels\cite{Polak2022,Varley2017,Rahman_23}. 
By learning the relationship between defect structure and properties, these models enable high-throughput studies that quickly evaluate and screen a group of materials based on their defect behaviour.
\cite{Witman2023,Wexler2021,Frey2020,Baldassarri_23} 

Despite progress in accelerating defect predictions, most high-throughput studies are limited in scope. 
Typically, their training datasets are generated assuming the ideal defect structure inherited from the crystal host, which often lies within a local minimum, thereby trapping a gradient-based optimisation algorithm in a metastable arrangement\cite{Arrigoni2021,Kavanagh_rapid,mosquera-lois_search_2021,Mosquera-Lois2023}. By yielding incorrect geometries, the predicted defect properties, such as equilibrium concentrations\cite{Mosquera-Lois2023,wang2023fourelectron,Kavanagh_rapid}, charge transition levels\cite{Mosquera-Lois2023,wang2023fourelectron,Kavanagh_rapid} and recombination rates\cite{Kavanagh_rapid}, are rendered inaccurate.\cite{Morris2009,Mulroue2011,al-mushadani_free-energy_2003,Kehoe2011,kononov_2023}. 
%Finding the most stable defect geometry requires exploring the configurational landscape with methods such as random sampling or targeted distortions. 
However, defect structure searching is often too expensive for high-throughput studies that target thousands of defects\cite{Baldassarri_23} or materials with complex (defect) energy landscapes, like alloys, disordered solids, and low-symmetry crystals. 

\begin{figure}[ht]
    %\centering
    \caption{Schematic of a machine-learning surrogate model used to accelerate defect structure searching. The computationally efficient model learns the plausible defect reconstructions (local minima in the potential energy surface) and thus reduces the number of candidate structures relaxed with expensive first-principles density functional theory (DFT) calculations.}\label{fig:surrogate_pes}
    \includegraphics[width=0.55\textwidth]{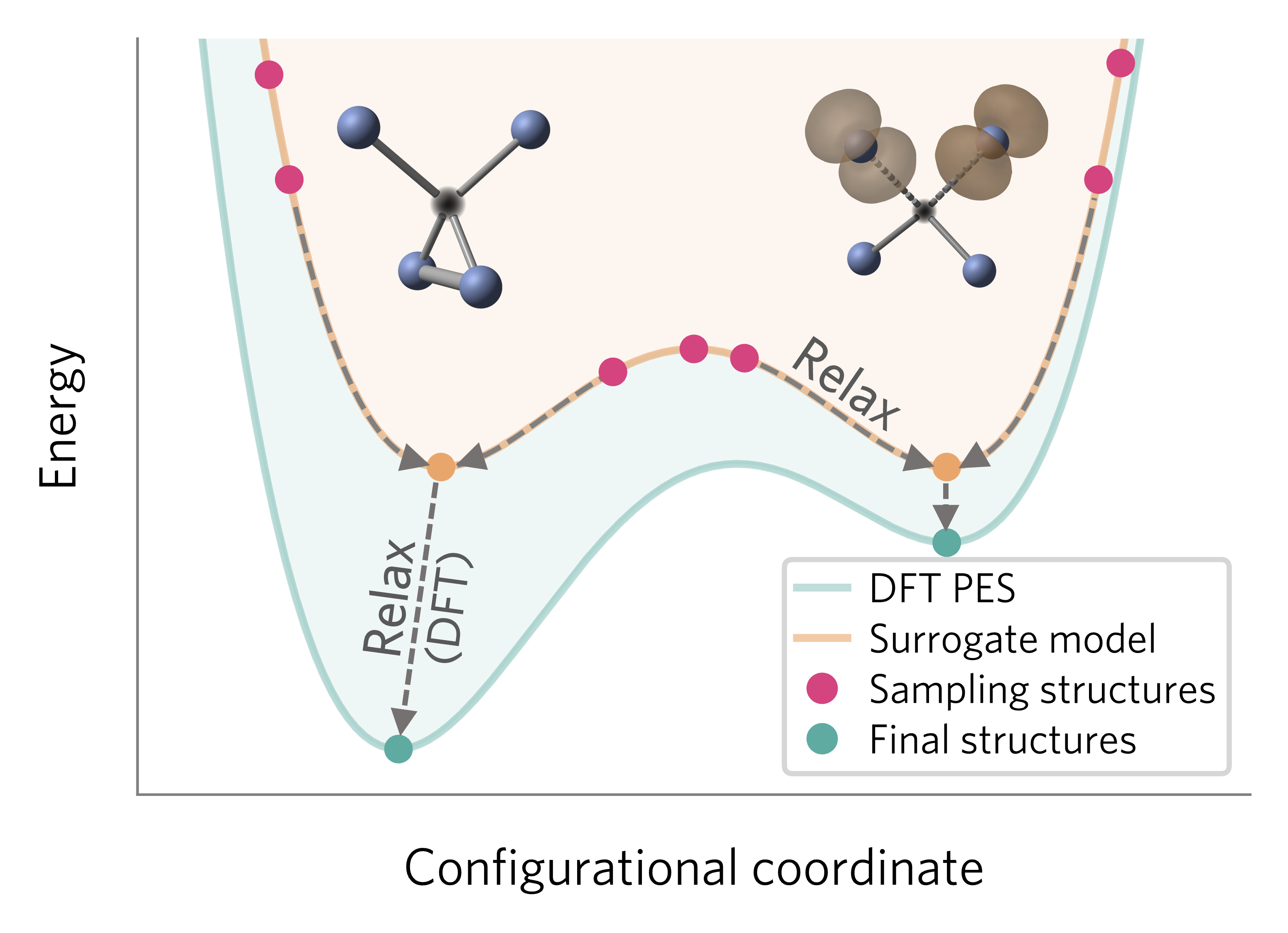}       
\end{figure}

In this study, we aim to reduce the computational burden of defect structure searching by introducing a machine-learning surrogate model. 
%We use a set of first-principles point defect structures, energies, forces and stresses to fine-tune a universal machine-learning force field (MLFF) and qualitatively explore the defect landscape. 
We build a dataset containing a set of point defect structures, energies, forces and stresses from first-principles, and use it to fine-tune a universal machine-learning force field (MLFF) and qualitatively explore the energy landscape across 132 defects. 
Defect reconstructions often follow common motifs\cite{Mosquera-Lois2023}, especially when comparing similar defects in families of related compounds.   
By learning the plausible reconstructions undergone by defects in similar hosts, a surrogate model can be used to optimise the initial sampling structures and thus identify the promising, low-energy configurations (\ref{fig:surrogate_pes}), as previously shown for surface adsorbates\cite{Schaarschmidt_22,Lan2023}.

\section{Results}
To assess the ability of a surrogate model to learn defect reconstructions, we will focus on one of the most common --- and often strongest in terms of energy-lowering --- reconstruction motifs: dimerisation\cite{Mosquera-Lois2023,lany_metal-dimer_2004,Kang2017,wilson_defect_2008,zhao_correlations_2016,Agoston_2009,Han2017,meggiolaro_tin_2020,erhart_first-principles_2005,sokol_oxygen_2010,evarestov_oxygen_1996,kotomin_radiation-induced_1998,burbano_sources_2011,scanlon_possibility_2012,godinho_energetic_2009,scanlon_nature_2011,keating_analysis_2012,wilson_defect_2008,walsh_interplay_2009,whalley_h-center_2017,agiorgousis_strong_2014,whalley_giant_2021,motti_defect_2019,xiao_defect_2016,Na-Phattalung_2006,Han2017}
\footnote{Dimers/trimers have been previously reported for numerous vacancies and interstitials, including \kvc{V}{Se}{0} in ZnSe, \ce{CuInSe2} and \ce{CuGaSe2}\cite{lany_metal-dimer_2004}, \kvc{V}{S}{0} in ZnS\cite{lany_metal-dimer_2004}, 
\kvc{V}{Cd}{0} in \ce{CdTe}\cite{Kavanagh_rapid}, \kvc{V}{Sb}{0,+1,+2} in $\rm Sb_2S/Se_3$\cite{mosquera-lois_search_2021,wang2023fourelectron}, 
\kvc{V}{Ti}{0,-1} and \kvc{V}{Zr}{0} in \ce{CaZrTi2O7} \cite{Mulroue2011}, \kvc{V}{Sb}{0} in \ce{Sb2O5}\cite{Li_2023}, \kvc{O}{i}{0} in \ce{In2O3}\cite{Agoston_2009}, \ce{ZnO}\cite{erhart_first-principles_2005}, \ce{Al2O3}\cite{sokol_oxygen_2010}, \ce{MgO}\cite{evarestov_oxygen_1996,kotomin_radiation-induced_1998}, \ce{CdO}\cite{burbano_sources_2011}, \ce{SnO2}\cite{scanlon_possibility_2012,godinho_energetic_2009}, \ce{PbO_2}\cite{scanlon_nature_2011}, \ce{CeO2}\cite{keating_analysis_2012}, \ce{BaSnO3}\cite{Scanlon_2013_basno3},
\ce{In2ZnO4}\cite{walsh_interplay_2009} and \ce{LiNi_{0.5}Mn_{1.5}O4}\cite{Cen2023}, $\rm Ag_{i}^{0}$ in AgCl and AgBr\cite{wilson_defect_2008},~\kvc{V}{I}{-},~\kvc{I}{i}{0},~\kvc{Pb}{i}{0}, $\rm Pb_{CH_3NH_3}^{0}$ and $\rm I_{CH_3NH_3}^{0}$ in \ce{CH3NH3PbI3} \cite{agiorgousis_strong_2014,whalley_giant_2021,whalley_h-center_2017,motti_defect_2019}, $\rm Pb_i$ in \ce{CsPbBr3}\cite{Kang2017}, \ce{(CH3NH3)3Pb2I7}\cite{zhao_correlations_2016}, 
\ce{(CH3NH3)2Pb(SCN)2I2}\cite{xiao_defect_2016} and \kv{Sn}{i} in \ce{CH3NH3SnI3}\cite{meggiolaro_tin_2020}.
}.
While cation dimerisation has been reported in several hosts (AgCl/Br, \ce{CuInSe2}, \ce{CuGaSe2}, ZnS/Se, CdTe, $\rm Sb_2S/Se_3$, \ce{CH3NH3PbI3}, \ce{CsPbBr3}, \ce{(CH3NH3)3Pb2I7}, \ce{CH3NH3SnI3})\cite{Mosquera-Lois2023,lany_metal-dimer_2004,Kang2017,wilson_defect_2008,zhao_correlations_2016}, anion dimers are more common and will be the focus of our study. 

\begin{figure}[h!]
    %\centering
    \includegraphics[width=0.92\linewidth]{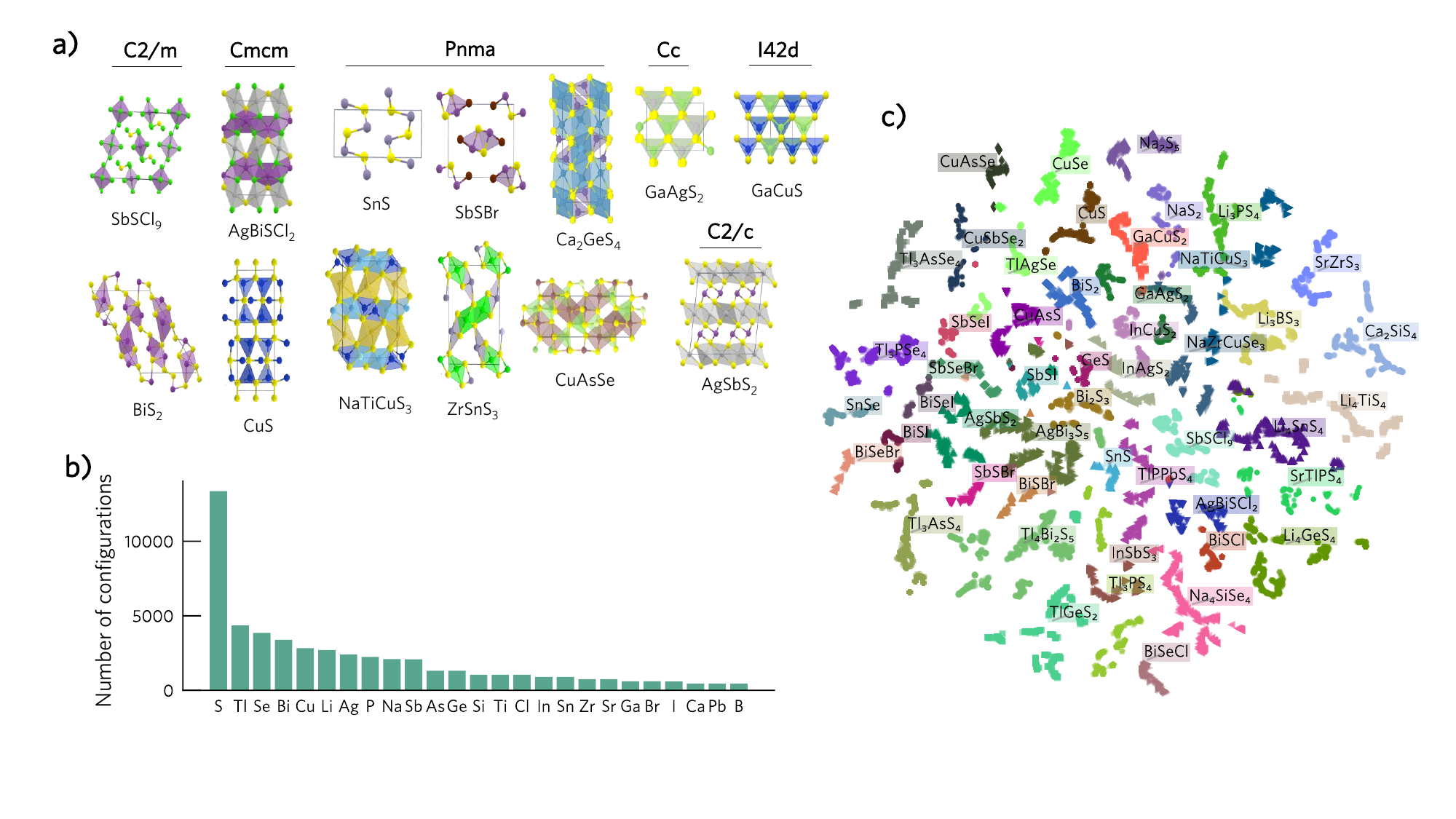}
    \caption{Distribution of the defect dataset generated with first-principles calculations. 
    a) Example host structures and their respective space groups. 
    b) Number of configurations containing each element. 
    c) Two-dimensional projection of structural similarity for defect configurations.  Each configuration is represented with the feature vector generated by the M3GNet model\cite{Chen2022,Qi_2023} (trained on the bulk formation energies of the Materials Project database) and the vector dimensions are reduced using t-distributed stochastic neighbour embedding (t-SNE)\cite{vanDerMaaten2008,hinton2002stochastic}. The defect configurations are coloured by their host composition (with similar colours indicating compositions with similar MEGNet\cite{Chen2019} feature vectors), showing that related chemical systems cluster near each other. For clarity, in (b) and (c) 10 evenly spaced steps are selected from each relaxation trajectory.}
    \label{fig:dataset}
\end{figure}

To target dimerisation, we consider cation vacancies in low-symmetry metal sulfides/selenides, where their covalent character and soft structures favour dimer formation\cite{Mosquera-Lois2023,wang2023fourelectron,Han2017}. Our first-principles dataset spans 50 hosts (exemplified in \ref{fig:dataset}a) and 132 neutral cation vacancies, covering 25 elements (\ref{fig:number_dimers}b) and 6 space groups. The configurational landscape of each vacancy was explored with the ShakeNBreak method\cite{shakenbreak2022,Mosquera-Lois2023} by applying 15 chemically-guided distortions to the unperturbed defect structure, followed by geometry optimisation with DFT (\ref{sec:methods}) -- resulting in a diverse set of trajectories for each defect and the dataset shown in \ref{fig:dataset}c. 

%

%\FloatBarrier
\subsection{Defect reconstructions}    
By analysing our first-principles dataset, we find that 29.9\% of the neutral defects undergo symmetry-breaking reconstructions missed by both the standard modelling approach but \emph{also} when applying a rattle distortion (with energy differences between the identified ground state and the relaxed ideal configuration greater than \SI{0.5}{eV}; \ref{stab:energy_lowerings}, \ref{sfig:delta_E_vs_holes}). 
Rattle distortions (i.e. randomised displacements) have been used in recent studies\cite{Witman2023} as the prevalence of defect reconstructions have become more recognised. While rattling helps to break the symmetry of the initial defect configuration and escape PES saddle points, it often fails to identify reconstructions with significant energy barriers (i.e. bond formation), highlighting the need for structure searching.  

\begin{figure}[ht]
\includegraphics[width=0.9\textwidth]{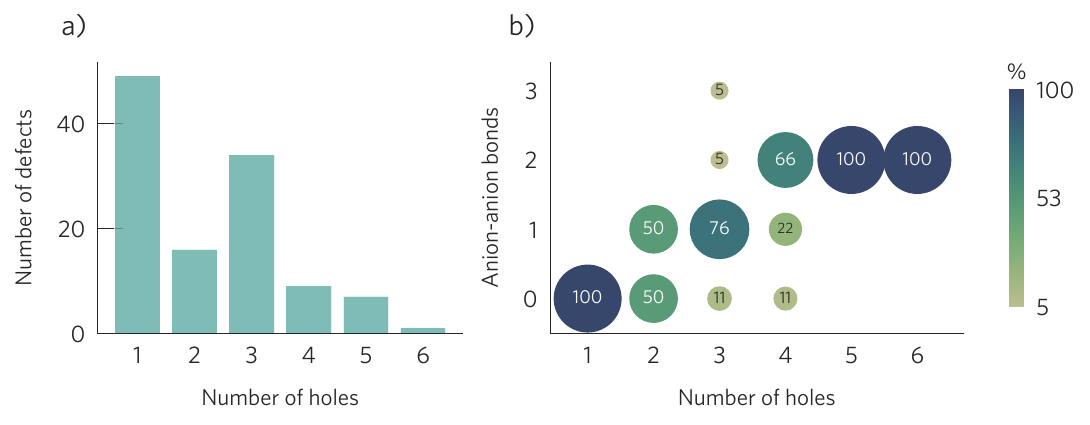}
    \caption{Analysis of the point defect dataset. a) Distribution of number of holes produced upon defect formation. b) Correlation between the number of anion--anion bonds formed and the number of holes created per defect. The label, colour, and size of the circles indicate the percentage of defects with that number of anion--anion bonds for defects for a given number of holes created.
    }
    \label{fig:number_dimers}
\end{figure}

The identified reconstructions are driven by anion--anion bond formation, with the number of new bonds determined by the number of valence electrons lost upon defect formation (\ref{sfig:reconstructions}). 
In general, energy-lowering structural reconstructions at defects tend to be driven by the localisation of excess charge introduced by the defect formation, through various bonding (re-)arrangements. Here, excess charge refers to the change in valence electrons available for bonding, and in fact is the chemical guiding principle used in ShakeNBreak to target likely distortion pathways.\footnote{For instance, upon forming a neutral antimony vacancy (\kvc{V}{Sb}{0}) in \ce{Sb2(S/Se)3} (where Sb is in the +3 oxidation state), we have removed three bonding electrons and so we have three excess holes. Further changes in the defect charge state will then alter this excess charge (e.g. 2 excess holes in the -1 charge state, or zero excess charge in the `fully-ionised' -3 charge state).}
Defects resulting in one hole (e.g. \kv{V}{Li} in \ce{Li4SnS4}) can easily accommodate the missing charge without strong reconstructions, while defects with two or more holes (e.g. \kv{V}{Bi} in \ce{BiSI}) tend to form anion dimers or trimers, as shown in \ref{fig:number_dimers}b. As a result, anion--anion bonds are more favourable for more positive defect charge states, and can stabilise unexpected defect oxidation states, as observed previously for \kvc{V}{Sb}{+1} in $\rm Sb_2(S/Se)_3$\cite{Mosquera-Lois2023,wang2023fourelectron} and \kvc{O}{i}{+1,+2} in several metal oxides\cite{Mosquera-Lois2023,erhart_first-principles_2005,Agoston_2009}. 

There are some exceptions to this trend, where systems are able to accommodate three or more holes without undergoing strong reconstructions. 
One example is hosts with \textit{d/f} metals that adopt multiple stable oxidation states (e.g. Fe, Co, Cu), which can accommodate a hole by adopting a higher oxidation state\cite{Kehoe2011}. To verify this trend, we compared two isostructural $\rm A^{III}B^{I}S_2$ systems which only differ in the identity of the B cation: \kvc{V}{Ga}{0} in $\rm Ga{\bf Cu}S_2$ and $\rm Ga{\bf Ag}S_2$; and \kvc{V}{In}{0} in $\rm In{\bf Cu}S_2$ and $\rm In{\bf Ag}S_2$ (\ref{sfig:V_Ga_GaAgS2}). In $\rm (Ga/In){\bf Ag}S_2$, two of the holes localise in a S--S bond formed by the vacancy nearest neighbours (NN), while the third hole is split between the remaining two NNs. In contrast, in $\rm (Ga/In){\bf Cu}S_2$, no dimer forms since three holes are localised in three of the vacancy NNs and five of the Cu ions closer to the vacancy --- with these Cu ions showing shorter Cu--S bonds. The different behaviour of Cu and Ag can be rationalised by considering their second ionisation energies (I2(Cu): 20.3 eV, I2(Ag): 21.5 eV)\cite{nist}, where the low I2 (and thus higher \textit{d} states) of Cu(I) favours cation oxidation, while the higher I2 of Ag(I) results in a sulfur dimer accommodating two of the holes. %(see \ref{sfig:V_Ga_GaAgS2_bandgaps}).  

In addition to systems with \textit{d/f} elements, defects with nearby anion--anion bonds can localise the positive charge in these bonds and thus avoid forming new ones. This behaviour is exemplified by \ce{RhSe2}, where the two symmetry-inequivalent Rh vacancies show different reconstructions. The first vacancy site is surrounded by four Se--Se bonds (\ref{sfig:RhSe2_reconstructions}b), and thus can accommodate the four holes by depopulating the anti-bonding orbitals of these bonds. In contrast, the second site has only one Se--Se bond neighbouring the vacancy (\ref{sfig:RhSe2_reconstructions}c), and thus has to form an additional Se dimer to accommodate the positive charge. 

Beyond chalcogenide dimers, other rearrangements to accommodate positive charge involve chalco-halide (e.g. S--Cl formed by \kvc{V}{Bi}{0} in \ce{AgBiSCl2}) and halide-halide bond formation (e.g. Cl dimers formed by \kvc{V}{Sb}{0} in \ce{SbSCl9}) (\ref{sfig:SbSCl9_reconstructions}). Here we note that the zero-dimensional character of \ce{SbSCl9} enables this defect to undergo strong distortions forming two Cl dimers (\ref{sfig:SbSCl9_reconstructions}). Overall, we highlight the common reconstruction motifs exhibited by different defects in various host structures (\ref{sfig:reconstructions}), facilitating the requisite diversity for a model to learn the plausible reconstructions for a group of related defects. 

\subsection{Model training}
To develop a model that can be applied for defect structure searching in \emph{unseen} compositions, we first split our dataset by composition into training, validation and test sets (\ref{sfig:split_2d_proj}), amounting to 68\%, 5\% and 27\% of defects, respectively.\footnote{
For the composition-wise splitting, we were aiming for a balanced distribution of the constituent elements and similar host systems (e.g. \ce{Li4SnS4}, \ce{Li4GeS4}, \ce{Li4TiS4}) into the train-validation-test splits. 
} 
The validation set is then augmented with 5\% of the configurations selected for the systems in the training set, to ensure that the diversity of the training set is also included for validation.\footnote{
The validation set was built so that it includes both some unseen compositions but also unseen configurations from a large diversity of defects and compositions. As a result, it will validate the extrapolation to unseen systems but also the transferability of the model to a large diversity of defects and compositions. 
} 
This results in training, validation and test sets of 11,955 (63\%), 2,100 (11\%), and 4,830 (26\%) configurations, respectively, where configuration denotes a point defect structure with its associated energy, forces and stresses. 

To sample the training data, we compared two approaches: i) a manual method where we sample 10 evenly spaced frames from each relaxation (MS) and ii) the Dimensionality-Reduced Encoded Clusters approach (DIRECT)\cite{Qi_2023}, which aims to select a robust training set from a complex configurational space. 
Surprisingly, we find that, when using datasets of similar sizes, the MS approach performs better --- with the DIRECT approach only outperforming MS when the final DIRECT dataset is larger than the MS one (\ref{stab:sampling_methods}). This is because the DIRECT approach mainly samples structures from the initial ionic steps (\ref{sfig:direct_ionic_step}), which correspond to high distortions and thus lead to larger errors for the low energy structures (\ref{sfig:distribution_mae_sampling}).

As a surrogate model, we aim for a method that takes an initial defect structure and outputs the energy and structure of the locally relaxed configuration. Machine-learning force fields are ideal for this task since they can map regions of the potential energy surface (PES) by learning the energies, forces, and, optionally stresses of a set of training structures. Specifically, we focus on universal graph-based MLFFs, which are trained on relaxation data from diverse databases of bulk crystals\cite{Chen2022,Deng2023,Merchant2023,Batatia_2024}, and thus already incorporate general chemical behaviour. Accordingly, we use a universal MLFF as a base model and fine-tune it with a training set of defect configurations.  
We have compared different model architectures (M3GNet\cite{Chen2022}, CHGNet\cite{Deng2023} and MACE\cite{Batatia_2022}), elemental reference energies, structure featurisation parameters (graph cutoffs, readout layers) and fine-tuning strategies, which are discussed in detail in the Supporting Information (SI) (\cref{ssec:training}). In addition, we compared a model trained on just defect structures, and both defect and bulk structures, with the second case improving performance (\ref{stab:errors_defect_with_bulk} and \ref{sfig:bulk_data_error_distribution}). 
From these benchmarks, the optimal model architecture and parameters were selected: a M3GNet model\cite{Chen2022} with radial and 3-body cutoffs of \SI{5}{\angstrom} and \SI{4}{\angstrom}, respectively, and the weighted atom readout function\cite{Chen2022,m3gnet_repo} (further details in \ref{sec:methods}). 

Overall, we note that the mean absolute errors for the \emph{absolute} energies in the validation and test sets are significant ($\mathrm{MAE_{E,test}}=31.2~\mathrm{meV/atom}$, \ref{tab:errors_mae}), but comparable to those obtained in MLFFs used for bulk structure searching of carbon ($\mathrm{MAE_{E,test}}=64.8~\mathrm{meV/atom}$)\cite{Salzbrenner_2023}. 
However, a more meaningful metric for our purpose is the error for the \emph{relative} energies of each defect configuration relative to its ground state structure ($\mathrm{MAE_{E,rel,test}} = 11.3~\mathrm{meV/atom}$). Further, we mostly care about the low-energy region of the potential energy surface, which can be measured by calculating the relative energy errors for configurations less than $\approx 5~\mathrm{eV}$ above the global minimum, resulting in MAEs of $3.6~\mathrm{meV/atom}\approx 0.29~\mathrm{eV}$ for an 80 atom supercell.

Beyond these metrics, we calculate the Spearman correlation coefficient ($\rho$) to measure how well the MLFF and DFT energies are monotonically related (i.e. if greater DFT energies correspond to greater MLFF energies\cite{Pickard2022}\footnote{Note that the Spearman coefficient is calculated for each defect \emph{independently}, and then averaged across defects.}). 
While the value of $\rho$ for the test set is significantly lower than those obtained with MLFFs developed for \emph{bulk} structure searching for a \emph{single composition} (0.72 versus 0.98--0.999\cite{Pickard2022}), this was expected considering that our dataset spans a diverse range of compositions and a wide range of energies. 
While the errors are high, we note that this does not prevent the model from being used as a \emph{qualitative} surrogate of the DFT PES for structure searching (i.e. identification of local minima), as previously observed for surface adsorbates\cite{Musielewicz2022,Jung2023}. 

\begin{table}[ht]
\caption{Mean absolute errors (MAE) 
%and root mean square errors (RMSE) 
of the training, validation and test sets for energies (E), forces (F) and stresses (S). Distributions of the absolute errors are shown in \ref{sfig:bulk_data_error_distribution}.}\label{tab:errors_mae}
\vspace{10pt}
\begin{tabular}{ccccc}
\hline
Split & 
\thead{$\rm MAE_{E}$ \\(meV/atom)}  & 
$\rho$ &
\makecell{$\rm MAE_{F}$ \\(meV/\AA)} & 
\thead{$\rm MAE_{S}$ \\(GPa)} \\
\hline
Train & 18.8 &  -  & 56.5  &  0.10\\
Val   & 27.0 & 0.86 & 93.4  &  0.13\\
Test  & 31.2 & 0.72 &  86.2  &  0.18\\
\hline 
\end{tabular}
\end{table}

\subsection{Model performance}
To evaluate the model performance, we apply the trained model to a robust test set, which includes 13 unseen compositions and 32 defects (accounting for 26\% and 26.5\% of the total number of compositions and defects in our dataset, respectively; \ref{sfig:split_2d_proj}). For each defect, the MLFF is used to relax the 15 distorted structures generated with ShakeNBreak\cite{shakenbreak2022} to sample the defect PES. The MLFF-relaxed structures are then compared to identify the different local minima in the MLFF PES using the SOAP fingerprint\cite{bartok_2013} of the defect site.\footnote{
We note that using the SOAP fingerprint of the defect site was more robust than considering the energies or the root mean squared displacement between the structures. The first case can miss local minima if these have similar energies in the MLFF PES, while the second was more sensitive to structural differences far from the defect site.
}
These local minima are then further relaxed with DFT. 
By comparing the ground state identified from the MLFF+DFT approach and full DFT search, we find the former to correctly identify the DFT ground state for 88\% of test defects, while simultaneously reducing the number of DFT calculations required by 73\% (\ref{tab:performance_test}) and accelerating structure searching by a factor of 13 (\ref{ssubsec:speedup}). In addition, it identifies a more favourable structure than the ones found in the DFT search for \kv{V}{Ge,9} in \ce{TlGeS2}, with an energy lowering of \SI{0.5}{eV} (\ref{sfig:v_Ge_TlGeS2}).

\begin{table}
\fontsize{10.75pt}{10.75pt}\selectfont
\caption{
Performance of the trained model on the test set. The column ``Local min. (DFT)" denotes the number of distinct structures identified in the DFT PES with ShakeNBreak. The column ``Local min. (MLFF)" denotes the number of distinct structures identified in the MLFF PES, which are then further relaxed with DFT. ``GS identified" shows whether the ground state structure located with DFT was also identified with the MLFF+DFT approach. 
}
\label{tab:performance_test}
\setlength{\tabcolsep}{6.pt} % Default value: 6pt
\renewcommand{\arraystretch}{0.7} % Default value: 1
\begin{tabular}{llcccc}
\toprule[1pt]
Hosts & Defect & 
Local min. (DFT) &
Local min. (MLFF) & 
Symmetry broken & 
GS identified \\
\midrule[0.1pt]
\ce{BiSeBr} & \textit{V}$_{\rm Bi_{0}}$ & 4 & 5 & Yes & Yes \\
\ce{BiSeCl} & \textit{V}$_{\rm Bi_{0}}$ & 3 & 4 & Yes & Yes \\
\ce{BiSeI} & \textit{V}$_{\rm Bi_{0}}$ & 6 & 3 & Yes & No \\
\midrule[0.1pt]
\ce{CuAsS} & \textit{V}$_{\rm As_{4}}$ & 2 & 4 & No & Yes \\
\ce{CuAsS} & \textit{V}$_{\rm Cu_{0}}$ & 2 & 1 & No & Yes \\
\midrule[0.1pt]
\ce{CuS} & \textit{V}$_{\rm Cu_{0}}$ & 1 & 3 & No & Yes \\
\ce{CuS} & \textit{V}$_{\rm Cu_{3}}$ & 3 & 2 & No & Yes \\
\ce{CuSe} & \textit{V}$_{\rm Cu_{0}}$ & 1 & 1 & Yes & Yes \\
\ce{CuSe} & \textit{V}$_{\rm Cu_{4}}$ & 2 & 1 & Yes & Yes \\
\midrule[0.1pt]
\ce{Li4SnS4} & \textit{V}$_{\rm Li_{0}}$ & 4 & 3 & No & Yes \\
\ce{Li4SnS4} & \textit{V}$_{\rm Li_{4}}$ & 3 & 3 & No & Yes \\
\ce{Li4SnS4} & \textit{V}$_{\rm Li_{8}}$ & 3 & 3 & No & Yes \\
\ce{Li4SnS4} & \textit{V}$_{\rm Sn_{16}}$ & 5 & 8 & Yes & No \\
\midrule[0.1pt]
\ce{Ca2SnS4} & \textit{V}$_{\rm Ca_{0}}$ & 3 & 3 & Yes & No \\
\ce{Ca2SnS4} & \textit{V}$_{\rm Ca_{4}}$ & 5 & 3 & Yes & Yes \\
\ce{Ca2SnS4} & \textit{V}$_{\rm Sn_{8}}$ & 2 & 3 & Yes & Yes \\
\midrule[0.1pt]
\ce{Na2S5} & \textit{V}$_{\rm Na_{0}}$ & 2 & 2 & No & Yes \\
\ce{Na2S5} & \textit{V}$_{\rm Na_{4}}$ & 2 & 3 & No & Yes \\
\midrule[0.1pt]
\ce{Sb2S3} & \textit{V}$_{\rm Sb_1}$ & 7 & 4 & Yes & Yes \\
\ce{Sb2S3} & \textit{V}$_{\rm Sb_2}$ & 9 & 7 & Yes & Yes \\
\midrule[0.1pt]
\ce{Tl3PS4} & \textit{V}$_{\rm Tl_{0}}$ & 2 & 3 & No & Yes \\
\ce{Tl3PS4} & \textit{V}$_{\rm Tl_{4}}$ & 5 & 3 & No & Yes \\
\ce{Tl3PS4} & \textit{V}$_{\rm P_{12}}$ & 7 & 5 & Yes & Yes \\
\midrule[0.1pt]
\ce{Tl4Bi2S5} & \textit{V}$_{\rm Tl_{0}}$ & 2 & 2 & No & Yes \\
\ce{Tl4Bi2S5} & \textit{V}$_{\rm Tl_{3}}$ & 1 & 1 & No & Yes \\
\ce{Tl4Bi2S5} & \textit{V}$_{\rm Tl_{11}}$ & 1 & 1 & No & Yes \\
\ce{Tl4Bi2S5} & \textit{V}$_{\rm Bi_{16}}$ & 3 & 2 & Yes & Yes \\
\ce{Tl4Bi2S5} & \textit{V}$_{\rm Bi_{20}}$ & 6 & 5 & Yes & Yes \\
\midrule[0.1pt]
\ce{TlGeS2} & \textit{V}$_{\rm Tl_{0}}$ & 5 & 1 & Yes & No \\
\ce{TlGeS2} & \textit{V}$_{\rm Ge_{8}}$ & 4 & 4 & Yes & Yes \\
\ce{TlGeS2} & \textit{V}$_{\rm Ge_{9}}$ & 5 & 4 & Yes & Yes \\
\midrule[0.2pt]
Mean &  & 4 & 3 & 0.51 & 0.88 \\
\bottomrule[1pt]
\end{tabular}
\end{table}

The 12\% of failed cases, where the MLFF ground state structure differed from the DFT one, mostly involve complex hosts. For instance, \kv{V}{Sn} in \ce{Li4SnS4} has a complex DFT energy surface, which traps most of the relaxations in very high energy basins (\ref{sfig:Li4SnS4_snb_plots}). PESs of similar complexity are displayed by the iso-structural systems that were included in the training set (\ce{Li4GeS4} and \ce{Li4TiS4}; \ref{sfig:Li4SnS4_snb_plots}), which biases our training data to the high energy region of the PES for these compositions and thus hinders learning the low-energy region. Accordingly, the training data for these systems can be improved by reducing the magnitude of the distortion used by ShakeNBreak to generate their sampling structures; which would improve model performance. Other defects for which the surrogate model misses the most stable structure are \kv{V}{Tl,0} in \ce{TlGeS2} and \kv{V}{Bi} in \ce{BiSeI} --- yet in both cases the DFT and MLFF+DFT structures are very similar and differ by small energy differences (0.1 and 0.2 eV, respectively) (\ref{sfig:BiSeI_v_Bi} and \ref{sfig:TlGeS2_v_Tl0}). In all failed cases, while the model misses the full DFT ground state, it still correctly predicts a favourable reconstruction, that lowers the energy compared to the relaxed ideal configuration. 

Beyond identifying the correct ground state in the majority of cases, the model has indirectly learned the correlation between the number of holes and the number of formed dimers. For defects with 1 missing electron, the candidate structures generated by the surrogate model rarely contained anion--anion bonds; while for defects with more missing electrons, the model often identifies at least one local minima with a dimer. 

The decreased performance observed for out-of-sample compositions less similar to the training set posed the question of what performance could be achieved if targeting a family of more related systems. To consider more similar host compositions, we select the chalcohalide systems from our dataset and split them composition-wise into training, validation and test sets as described in \cref{ssec:chalcohalides}. After training the model on the training set and applying it to the unseen test defects (details on \cref{ssec:chalcohalides}), we find that the model identifies the correct ground state for all test cases, and achieves lower mean absolute errors compared to the full model.
%By training a model for only the chalcohalides, we observed increased performance with a smaller training set size (see \ref{ssec:chalcohalides}). 
This suggests that higher accuracy can be achieved when targeting more similar host structures, which is likely the case in most high-throughput defect studies. 
%To apply our approach to target a different compositional space or a different defect type, we note that a training set has to be generated first through first-principles calculations to then fine-tune the universal bulk model. 

Our current trained model is limited to neutral cation vacancies in metal sulphides/selenides. However, the approach can be extended to a different compositional space or defect type by first generating a custom training set through first-principles calculations and using it to fine-tune the universal bulk MLFF.  

\subsection{Application to alloys}
Beyond high-throughput studies of many single-phase materials, the surrogate model can also be used to accelerate structure searching in alloys or disordered solids, which is computationally challenging due to the high number of local host compositions and inequivalent defects to consider\cite{hu_first-principles_2023}. 
The distinct local or site environments of a given defect can significantly affect its properties\cite{zhao_machine_2023,Piochaud_2014,Guan_2020,rio_formation_2011,zhang_influence_2015,zhang_atomic-level_2017,arora_effect_2021,zhao_defect_2016,manzoor_influence_2022,zhao_effect_2018,li_first_2019,manzoor_factors_2021,Manzoor2021,Muzyk_2011}, altering formation and migration energies by up to 1.5 eV\cite{zhao_machine_2023,Guan_2020,Muzyk_2011,zhao_effect_2018,li_first_2019,Manzoor2021,manzoor_factors_2021}. Properly sampling various site environments is key to characterise the defect behaviour in such cases. 

\begin{figure}[ht]
\centering\includegraphics[width=0.95\textwidth]{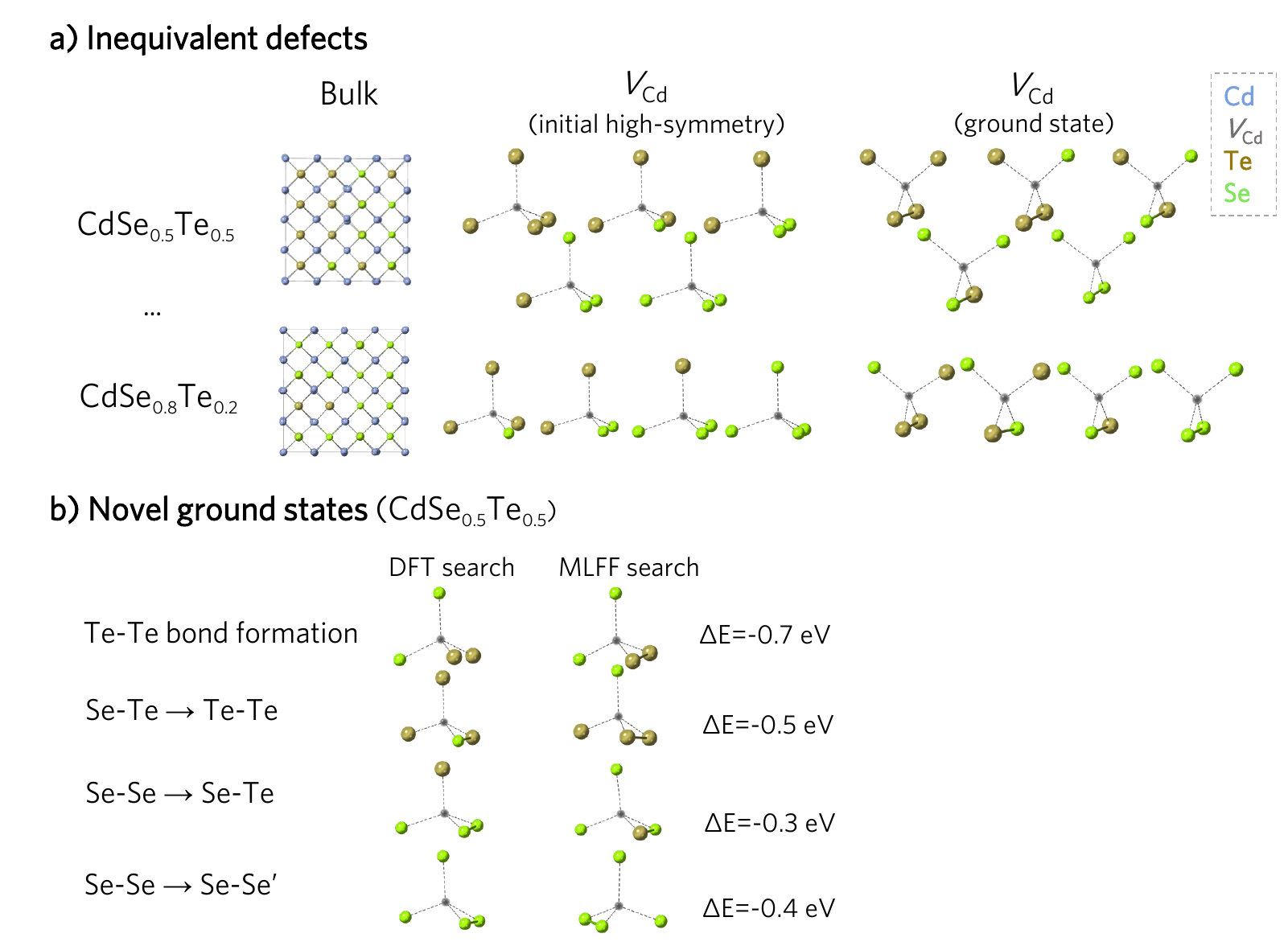}
\caption{a) Inequivalent defect environments for two of the CdSe$_x$Te$_{1-x}$ alloys ($x=0.5,0.8$). b) Examples of the ground state configurations \emph{only} identified through the finer MLFF+DFT search. These reconstructions are driven by either forming a dimer (e.g. Te--Te bond formation), forming a more favourable anion--anion bond (Te--Te instead of Se--Te; Se--Te instead of Se--Se) or forming the same type of anion bond (Se-Se) but breaking weaker anion-cation bonds between the defect nearest neighbours and the defect next nearest neighbours. 
%The energy differences are calculated through DFT.
}
\label{fig:alloy}
\end{figure}

We consider the case of cadmium vacancies in the $\rm CdSe_{x}Te_{(1-x)}$ ($x=0, 0.2, 0.3, 0.5, 0.6, 0.8, 0.9, 1$) pseudo-binary alloy. For each composition, a supercell is generated through random substitution of Te sites, and the Cd sites with a unique nearest neighbour chemical environment are considered\footnote{
As a proof of concept, we only consider the Cd sites with a unique nearest neighbour chemical environment (e.g. Cd surrounded by 4 Te; by 3 Te and 1 Se; by 2 Se and 2 Te etc), which is common in defect studies of alloys.
} (\ref{fig:alloy}a). 
The configurational landscape of each vacancy is explored with the ShakeNBreak method (14 sampling structures), using the relaxations from the pure compositions as the training and validation data while the mixed systems ($0<x<1$) are reserved as the test set.

After fine-tuning the surrogate model (MLFF) on the training configurations (details in \ref{sec:methods}), it is applied to all alloys to perform the structure searching calculations, allowing a more extensive sampling than for the DFT search (31 sampling structures). From the MLFF-relaxed structures, the unique configurations are selected for further relaxation with DFT and compared with the results from the DFT-only search. This comparison shows that the model successfully identifies the ground state for all defects, even in cases where the defects form Te-Se bonds not seen in the training set -- which only included the Te-Te and Se-Se bonds formed by \kv{V}{Cd}(CdTe) and \kv{V}{Cd}(CdSe), respectively. 

More significantly, for 70\% of the defects, the model identifies a \emph{novel}, more favourable ground state missed in the coarser DFT search (with a mean energy lowering of \SI{-0.4}{eV}, \cref{ssec:alloy}). These reconstructions are driven by forming a more favourable anion--anion bond (\ref{fig:alloy}b) and missed in the DFT-only search due to the coarser sampling performed.
This illustrates the advantage of the faster surrogate model to tackle defects with complex configurational landscapes, that require a more exhaustive exploration than a DFT-based search would allow, like alloys, compositionally disordered materials\cite{wang_cation_2022,williford_effects_1999,Quadir_2022,morrow_understanding_2023}, and low-symmetry or multinary systems with many degrees of freedom in their PES.    

\FloatBarrier
\section{Conclusion}
By building a dataset for defect structure searching, we have demonstrated the prevalence of defect reconstructions missed by the standard modelling approach -- and thus the need to perform structure searching in high-throughput defect studies. 
To reduce the associated computational burden, we have developed a surrogate model by fine-tuning a universal machine-learning force field on defect configurations. 
By qualitatively learning the defect configurational landscapes, the trained model successfully predicts low-energy defect structures for \emph{unseen} defects in \emph{unseen} compositions, thereby reducing the number of DFT calculations by 73\%. 
While our current model is limited to neutral cation vacancies in metal chalcogenides, the methodology can be applied to different defect types or compositional spaces. 
Accordingly, the approach could accelerate defect screening studies that target a range of dopants in a family of related host compounds. 
In addition, our openly-available dataset could be used to measure the out-of-distribution performance of universal MLFFs\cite{riebesell2023matbench} by testing the ability to extrapolate from learned bulk motifs to defect environments.

Beyond accelerating structure searching in high-throughput studies, this approach is ideal for systems with a complex defect landscape, like alloys, disordered, or low-symmetry materials where their many inequivalent defects make it intractable to explicitly calculate all of them with accurate DFT methods. 
By using a surrogate model, we can consider a range of alloy compositions and all inequivalent defects, while performing a more exhaustive sampling of the PES --- thereby identifying more favourable reconstructions missed in the (coarser) DFT-based search. 
Beyond (pseudo-)binary alloys, this approach could be extended to model more chemically complex systems, like high-entropy alloys, where the MLFF could be trained on defects of the constituent binary systems and applied to the ternary, quaternary, or high-entropy alloys.

A current limitation of this strategy is the handling of defects in distinct charge states, which have different energy landscapes and structural configurations (e.g. a defect in two different charge states can have a common local structure with different energies). The approach could handle the potential energy landscape for each charge state independently (e.g. training a separate model for defects in the -1 charge state). %Alternatively, instead of grouping defects by charge state, they could be grouped by the number of electrons/holes created upon defect formation (e.g. \kvc{V}{Li}{+1}, \kvc{V}{Ca}{0}, \kvc{V}{Bi}{-1}, \kvc{V}{Si}{-2}). 
To consider different charge states simultaneously, the net charge state can be encoded as a graph global attribute\cite{shimizu_using_2022}. However, a more descriptive encoding could be achieved by using fourth-generation MLFFs that include atomic charges\cite{ko_general-purpose_2021}. Beyond accounting for the defect charge, another improvement could be MLFFs that are fine-tuned on-the-fly during geometry optimisations. 
As shown for surface absorbates\cite{Musielewicz2022,Jung2023}, this strategy would accelerate the defect geometry optimisation by skipping many ionic steps that are performed with the surrogate model. 
% Delta learning (gamma -> std -> ncl)
%Another strategy that could accelerate defect modelling could be delta learning. Typically, defect calculations are performed in steps by gradually increasing the calculation accuracy (e.g. from $\Gamma$-point sampling to denser sampling to spin-orbit coupling), making it an ideal problem for delta-learning models. 
Overall, we note the promise of surrogate models to accelerate and increase the accuracy of defect modelling, whether this is by improving structure searching, accounting for metastable configurations\cite{Kavanagh2022,MosqueraLois2023}, enabling the calculation of defect formation entropies\cite{shimizu_using_2022,MosqueraLois2023}, accelerating defect migration studies\cite{pols_how_2023} or going beyond the dilute limit\cite{morrow_understanding_2023}. 

\section{Methods}\label{sec:methods}
\subsection{High-throughput vacancies in chalcogenide hosts}

\emph{Defect initialisation.} The conventional supercell approach for modelling defects in periodic solids was used\cite{Freysoldt_2014}. To reduce periodic image interactions, supercell dimensions of at least \SI{10}{\angstrom} in each direction\cite{Zunger_2008_assessment} were employed. 
To explore the configurational landscape of each defect, we used the ShakeNBreak code\cite{shakenbreak2022}, with a distortion increment of 0.1 and the default rattle standard deviation (10\% of the nearest neighbour distance in the bulk supercell). 
This strategy results in 14 sampling structures. In addition to these, to ensure that dimerisation was properly sampled, we also generated a sampling structure where two defect neighbours were pushed towards each other with a separation of \SI{2}{\angstrom}, resulting in a total of 15 initial configurations.

\emph{First-principles calculations.} All reference calculations were performed with Density Functional Theory using the exchange-correlation functional HSE06\cite{Heyd2003} and the projector augmented wave method\cite{Kresse_1996}, as implemented in the Vienna Ab initio Simulation Package\cite{Kresse_1993,Kresse_1994}. 
Calculations for the pristine unit cells were performed using a plane wave energy cutoff of \SI{585}{eV} and sampling reciprocal space with a Monkhorst-Pack mesh of density 900 \textit{k}-points/site. The convergence thresholds for the geometry optimisations were set to $10^{-6}$ eV and $10^{-5}$eV/\AA~for energy and forces, respectively.
Defect relaxations were performed with the $\Gamma$-point approximation, which is accurate enough for defect structure searching\cite{Mosquera-Lois2023}, and with a plane wave energy cutoff of \SI{350}{eV}. The energy and force thresholds for defect relaxations were set to $10^{-4}$~eV and $10^{-2}$~eV/\AA, respectively. 
To automate the generation of input files, we designed a workflow using aiida\cite{Pizzi2016,Uhrin2021,Huber2020}, pymatgen\cite{Ong2013,Jain2013,Ong2015}, 
pymatgen-analysis-defects\cite{pymatgen_analysis_defects},
ASE\cite{ase-paper}, doped\cite{doped} and ShakeNBreak\cite{shakenbreak2022}. This code is available from \url{https://github.com/ireaml/defects_workflow.git}. The datasets and trained models are available from the Zenodo repository with DOI 10.5281/zenodo.10514567. 

\emph{Surrogate machine learning model.} To generate the training and test set for the machine learning model, we processed the DFT defect relaxation data by removing unreasonably high-energy configurations (e.g. structures with positive energies), as they decreased model performance. After cleaning the data, 10 evenly-spaced ionic steps were selected from each relaxation. We used the M3GNet model\cite{Chen2022}, as implemented in Ref. \citenum{m3gnet_repo}, with radial and 3 body cutoffs of \SI{5}{\angstrom} and \SI{4}{\angstrom}, respectively, and the weighted atom readout function. The loss function was defined as a combination of the mean squared errors for the energies, forces and stresses, with respective weights of 1, 1 and 0.1\cite{Chen2022,m3gnet_repo}. For fine-tuning, the model was initialised with the weights from the trained bulk crystal model\cite{m3gnet_repo} and then trained on the defect training set (see \ref{ssubsec:params} for further details). 
A batch size of 4 and an exponential learning rate scheduler with an initial rate of $5\times10^{-4}$ were used. The model was trained on a Quadro RTX6000 GPU until the validation errors were converged (30 epochs, 5.3 hours) (\ref{sfig:training_evolution}). 

\emph{Model application.} MLFF geometry optimisation was performed with the FIRE algorithm\cite{Bitzek_2006}, as implemented in the ASE package\cite{ase-paper}, until the mean force was lower than $10^{-5}$~eV/\AA~or the number of ionic steps exceeded 1500, which were found to be reasonable thresholds. After relaxing the sampling structures with the model, we identified the different local minima or configurations by calculating the cosine distance between the SOAP descriptor\cite{bartok_2013} for the defect site of each configuration, which was found to be an effective metric for identifying different defect motifs. The parameters used to generate the SOAP descriptor were: $r=5$~\AA, $n_{max}=10, l_{max}=10, \sigma=1.0$~\AA, for the local cutoff, number of radial basis functions, maximum degree of spherical harmonics, and the standard deviation of the Gaussian functions used to expand the atomic density, respectively.

\subsection{Application to the $\mathrm{CdSe_xTe_{(1-x)}}$ alloy}
\emph{Data generation.} %
To generate the supercells for the mixed compositions in $\rm CdSe_xTe_{(1-x)}$ ($x=0.2, 0.3, 0.5, 0.6, 0.8, 0.9$), we used random substitution of Te sites with Se. 
%
% AW - save this if a reviewer asks...
% the main advantage of SQS is for thermodynamics of mixing (representing the ensemble of environments) 
%
%We note that to generate supercells special quasi-random structures\cite{Zunger_1990} could be used. However, for this proof of concept of defect structure searching on alloys, random substitution is a reasonable approximation, and the results should not be affected by using SQS. 
For each supercell, we consider the Cd sites with a unique nearest neighbour chemical environment as vacancy sites, and generate the vacancy high-symmetry structures with pymatgen\cite{Ong2013,Jain2013,Ong2015}. For the DFT-based exploration of the PES, we apply ShakeNBreak with default parameters, generating 14 sampling structures, which were relaxed with DFT as previously described.  

\emph{Model training.} To generate the dataset, we again processed the defect relaxation data by removing unreasonably high-energy configurations ($>15$~eV above the defect ground state configuration). As the training set, we used a combination of defect and bulk configurations: 45 evenly-spaced frames from the 14 relaxations of \kv{V}{Cd} in CdTe and CdSe, and 20 frames from the relaxations of each pristine system, resulting in a total of 1420 frames. For validation, we selected 5 unseen configurations from the 14 relaxations of \kv{V}{Cd} in CdTe and CdSe (total of 140 frames). The M3GNet surrogate model was trained with similar parameters as previously described and until the errors were converged (80 epochs, see \ref{sfig:training_evolution_alloy}).

\emph{Finer exploration of the PES.} To perform a finer exploration of the PES with the surrogate model, we applied a set of bond distortions generated by ShakeNBreak (-0.6, -0.5, -0.4, -0.3, -0.2) to all unique pair combinations of nearest neighbours (e.g. for a \kv{V}{Cd} surrounded by two Te and two Se anions, Te(1), Te(2), Se(1) and Se(2), we considered the pairs Te(1)-Te(2), Te(1)-Se(1), Te(1)-Se(2), Te(2)-Se(1), Te(2)-Se(2), and Se(1)-Se(2)). By default, for a defect with two missing electrons like \kvc{V}{Cd}{0}, ShakeNBreak only applies the bond distortions to the two atoms closest to the defect. This is typically a reliable approach for most \emph{pure} systems, but can miss reconstructions for alloys with complex defect environments (e.g. \kv{V}{Cd} surrounded by a mix of Te and Se anions). 
The model application and analysis were performed as described in the previous section.

%\backmatter
\section*{Declarations}
% {\bf Supplementary information.}

{\bf Data \& code availability.}
The code used to generate the defect dataset is available from \url{https://github.com/ireaml/defects_workflow.git}.
The datasets and trained models are available from the Zenodo repository with DOI 10.5281/zenodo.10514567. 

{\bf Acknowledgments.}
We thank David O. Scanlon for discussions on defect symmetry breaking.
I.M.L. acknowledges Imperial College London for funding a President’s
PhD scholarship. 
S.R.K. acknowledges the EPSRC Centre for Doctoral Training in the Advanced Characterisation of Materials (CDT-ACM)(EP/S023259/1) for funding a PhD studentship.
A.M.G. is supported by EPSRC Fellowship EP/T033231/1.
A. W. is supported by EPSRC project EP/X037754/1.
We are grateful to the UK Materials
and Molecular Modelling Hub for computational resources, which is partially funded
by EPSRC (EP/P020194/1 and EP/T022213/1). This work used the ARCHER2 UK National
Supercomputing Service (https://www.archer2.ac.uk) via our membership of the UK’s
HEC Materials Chemistry Consortium, which is funded by EPSRC (EP/L000202). We acknowledge the Imperial College London's High Performance Computing services for computational resources. 

{\bf Author contributions.}
Conceptualisation \& Project Administration: All authors. Investigation and methodology: I.M.-L. Supervision: S.R.K., A.M.G., A.W. Writing - original draft: I.M.-L. Writing - review \& editing: All authors. Resources and funding acquisition: A.M.G., A.W. These author contributions are defined according to the CRediT contributor roles taxonomy.

{\bf Competing interests.}
The authors declare no competing interests.

\bibliographystyle{rsc}
\bibliography{biblio}

\providecommand*{\mcitethebibliography}{\thebibliography}
\csname @ifundefined\endcsname{endmcitethebibliography}
{\let\endmcitethebibliography\endthebibliography}{}
\begin{mcitethebibliography}{131}
\providecommand*{\natexlab}[1]{#1}
\providecommand*{\mciteSetBstSublistMode}[1]{}
\providecommand*{\mciteSetBstMaxWidthForm}[2]{}
\providecommand*{\mciteBstWouldAddEndPuncttrue}
  {\def\EndOfBibitem{\unskip.}}
\providecommand*{\mciteBstWouldAddEndPunctfalse}
  {\let\EndOfBibitem\relax}
\providecommand*{\mciteSetBstMidEndSepPunct}[3]{}
\providecommand*{\mciteSetBstSublistLabelBeginEnd}[3]{}
\providecommand*{\EndOfBibitem}{}
\mciteSetBstSublistMode{f}
\mciteSetBstMaxWidthForm{subitem}
{(\emph{\alph{mcitesubitemcount}})}
\mciteSetBstSublistLabelBeginEnd{\mcitemaxwidthsubitemform\space}
{\relax}{\relax}

\bibitem[Sambur and Brgoch(2023)]{Sambur_2023}
J.~Sambur and J.~Brgoch, \emph{Chem. Mater.}, 2023, \textbf{35},
  7351--7354\relax
\mciteBstWouldAddEndPuncttrue
\mciteSetBstMidEndSepPunct{\mcitedefaultmidpunct}
{\mcitedefaultendpunct}{\mcitedefaultseppunct}\relax
\EndOfBibitem
\bibitem[Shockley and Read(1952)]{shockleyStatisticsRecombinationsHoles1952}
W.~Shockley and W.~T. Read, \emph{Phys. Rev.}, 1952, \textbf{87},
  835--842\relax
\mciteBstWouldAddEndPuncttrue
\mciteSetBstMidEndSepPunct{\mcitedefaultmidpunct}
{\mcitedefaultendpunct}{\mcitedefaultseppunct}\relax
\EndOfBibitem
\bibitem[Kim \emph{et~al.}(2020)Kim, Márquez, Unold, and
  Walsh]{kimUpperLimitPhotovoltaic2020}
S.~Kim, J.~A. Márquez, T.~Unold and A.~Walsh, \emph{Energy Environ. Sci.},
  2020, \textbf{13}, 1481--1491\relax
\mciteBstWouldAddEndPuncttrue
\mciteSetBstMidEndSepPunct{\mcitedefaultmidpunct}
{\mcitedefaultendpunct}{\mcitedefaultseppunct}\relax
\EndOfBibitem
\bibitem[Maier(2013)]{maierThermodynamicsElectrochemicalLithium2013}
J.~Maier, \emph{Angew. Chem. Int. Ed.}, 2013, \textbf{52}, 4998--5026\relax
\mciteBstWouldAddEndPuncttrue
\mciteSetBstMidEndSepPunct{\mcitedefaultmidpunct}
{\mcitedefaultendpunct}{\mcitedefaultseppunct}\relax
\EndOfBibitem
\bibitem[Squires \emph{et~al.}(2022)Squires, Davies, Kim, Scanlon, Walsh, and
  Morgan]{squiresLowElectronicConductivity2022}
A.~G. Squires, D.~W. Davies, S.~Kim, D.~O. Scanlon, A.~Walsh and B.~J. Morgan,
  \emph{Phys. Rev. Mater.}, 2022, \textbf{6}, 085401\relax
\mciteBstWouldAddEndPuncttrue
\mciteSetBstMidEndSepPunct{\mcitedefaultmidpunct}
{\mcitedefaultendpunct}{\mcitedefaultseppunct}\relax
\EndOfBibitem
\bibitem[Li \emph{et~al.}(2020)Li, Wang, Zhang, Tao, Wang, Zou, Wang, Chen, and
  Wang]{liDefectEngineeringFuel2020}
W.~Li, D.~Wang, Y.~Zhang, L.~Tao, T.~Wang, Y.~Zou, Y.~Wang, R.~Chen and
  S.~Wang, \emph{Adv. Mater.}, 2020, \textbf{32}, 1907879\relax
\mciteBstWouldAddEndPuncttrue
\mciteSetBstMidEndSepPunct{\mcitedefaultmidpunct}
{\mcitedefaultendpunct}{\mcitedefaultseppunct}\relax
\EndOfBibitem
\bibitem[Pastor \emph{et~al.}(2022)Pastor, Sachs, Selim, Durrant, Bakulin, and
  Walsh]{pastorElectronicDefectsMetal2022}
E.~Pastor, M.~Sachs, S.~Selim, J.~R. Durrant, A.~A. Bakulin and A.~Walsh,
  \emph{Nat. Rev. Mater.}, 2022, \textbf{7}, 503--521\relax
\mciteBstWouldAddEndPuncttrue
\mciteSetBstMidEndSepPunct{\mcitedefaultmidpunct}
{\mcitedefaultendpunct}{\mcitedefaultseppunct}\relax
\EndOfBibitem
\bibitem[Kehoe \emph{et~al.}(2011)Kehoe, Scanlon, and Watson]{Kehoe2011}
A.~B. Kehoe, D.~O. Scanlon and G.~W. Watson, \emph{Chem. Mater.}, 2011,
  \textbf{23}, 4464--4468\relax
\mciteBstWouldAddEndPuncttrue
\mciteSetBstMidEndSepPunct{\mcitedefaultmidpunct}
{\mcitedefaultendpunct}{\mcitedefaultseppunct}\relax
\EndOfBibitem
\bibitem[Iv{\'{a}}dy \emph{et~al.}(2018)Iv{\'{a}}dy, Abrikosov, and
  Gali]{Ivdy2018}
V.~Iv{\'{a}}dy, I.~A. Abrikosov and A.~Gali, \emph{npj Comput Mater}, 2018,
  \textbf{4}, 76\relax
\mciteBstWouldAddEndPuncttrue
\mciteSetBstMidEndSepPunct{\mcitedefaultmidpunct}
{\mcitedefaultendpunct}{\mcitedefaultseppunct}\relax
\EndOfBibitem
\bibitem[Weber \emph{et~al.}(2010)Weber, Koehl, Varley, Janotti, Buckley,
  de~Walle, and Awschalom]{Weber2010}
J.~R. Weber, W.~F. Koehl, J.~B. Varley, A.~Janotti, B.~B. Buckley, C.~G.~V.
  de~Walle and D.~D. Awschalom, \emph{Proc. Natl. Acad. Sci. U.S.A.}, 2010,
  \textbf{107}, 8513--8518\relax
\mciteBstWouldAddEndPuncttrue
\mciteSetBstMidEndSepPunct{\mcitedefaultmidpunct}
{\mcitedefaultendpunct}{\mcitedefaultseppunct}\relax
\EndOfBibitem
\bibitem[Thomas \emph{et~al.}(2023)Thomas, Chen, Xiong, Barker, Zhou, Chen,
  Rossi, Kelly, Yu, Zhou, Kumari, Barnard, Robinson, Terrones, Schwartzberg,
  Ogletree, Rotenberg, Noack, Griffin, Raja, Strubbe, Rignanese,
  Weber-Bargioni, and Hautier]{Thomas2023}
J.~Thomas, W.~Chen, Y.~Xiong, B.~Barker, J.~Zhou, W.~Chen, A.~Rossi, N.~Kelly,
  Z.~Yu, D.~Zhou, S.~Kumari, E.~Barnard, J.~Robinson, M.~Terrones,
  A.~Schwartzberg, D.~F. Ogletree, E.~Rotenberg, M.~Noack, S.~Griffin, A.~Raja,
  D.~Strubbe, G.-M. Rignanese, A.~Weber-Bargioni and G.~Hautier, 2023,
  arXiv:cond-mat/2309.08032\relax
\mciteBstWouldAddEndPuncttrue
\mciteSetBstMidEndSepPunct{\mcitedefaultmidpunct}
{\mcitedefaultendpunct}{\mcitedefaultseppunct}\relax
\EndOfBibitem
\bibitem[Dreyer \emph{et~al.}(2018)Dreyer, Alkauskas, Lyons, Janotti, and
  Van~de Walle]{Dreyer_2018}
C.~E. Dreyer, A.~Alkauskas, J.~L. Lyons, A.~Janotti and C.~G. Van~de Walle,
  \emph{Annu. Rev. Mater. Res.}, 2018, \textbf{48}, 1--26\relax
\mciteBstWouldAddEndPuncttrue
\mciteSetBstMidEndSepPunct{\mcitedefaultmidpunct}
{\mcitedefaultendpunct}{\mcitedefaultseppunct}\relax
\EndOfBibitem
\bibitem[Yan \emph{et~al.}(2024)Yan, Kar, Chowdhury, and Bansil]{Yan_2024}
Q.~Yan, S.~Kar, S.~Chowdhury and A.~Bansil, \emph{Adv Mater}, 2024,
  2303098\relax
\mciteBstWouldAddEndPuncttrue
\mciteSetBstMidEndSepPunct{\mcitedefaultmidpunct}
{\mcitedefaultendpunct}{\mcitedefaultseppunct}\relax
\EndOfBibitem
\bibitem[Davidsson \emph{et~al.}(2023)Davidsson, Bertoldo, Thygesen, and
  Armiento]{Davidsson2023}
J.~Davidsson, F.~Bertoldo, K.~S. Thygesen and R.~Armiento, \emph{npj 2D Mater.
  Appl.}, 2023, \textbf{7}, 26\relax
\mciteBstWouldAddEndPuncttrue
\mciteSetBstMidEndSepPunct{\mcitedefaultmidpunct}
{\mcitedefaultendpunct}{\mcitedefaultseppunct}\relax
\EndOfBibitem
\bibitem[Sluydts \emph{et~al.}(2016)Sluydts, Pieters, Vanhellemont, Speybroeck,
  and Cottenier]{Sluydts2016}
M.~Sluydts, M.~Pieters, J.~Vanhellemont, V.~V. Speybroeck and S.~Cottenier,
  \emph{Chem. Mater.}, 2016, \textbf{29}, 975--984\relax
\mciteBstWouldAddEndPuncttrue
\mciteSetBstMidEndSepPunct{\mcitedefaultmidpunct}
{\mcitedefaultendpunct}{\mcitedefaultseppunct}\relax
\EndOfBibitem
\bibitem[Bertoldo \emph{et~al.}(2022)Bertoldo, Ali, Manti, and
  Thygesen]{Bertoldo2022}
F.~Bertoldo, S.~Ali, S.~Manti and K.~S. Thygesen, \emph{npj Comput Mater},
  2022, \textbf{8}, 56\relax
\mciteBstWouldAddEndPuncttrue
\mciteSetBstMidEndSepPunct{\mcitedefaultmidpunct}
{\mcitedefaultendpunct}{\mcitedefaultseppunct}\relax
\EndOfBibitem
\bibitem[Huang \emph{et~al.}(2023)Huang, Lukin, Faleev, Kazeev, Al-Maeeni,
  Andreeva, Ustyuzhanin, Tormasov, Castro~Neto, and
  Novoselov]{huang_unveiling_2023}
P.~Huang, R.~Lukin, M.~Faleev, N.~Kazeev, A.~R. Al-Maeeni, D.~V. Andreeva,
  A.~Ustyuzhanin, A.~Tormasov, A.~H. Castro~Neto and K.~S. Novoselov, \emph{npj
  2D Mater Appl}, 2023, \textbf{7}, 1--10\relax
\mciteBstWouldAddEndPuncttrue
\mciteSetBstMidEndSepPunct{\mcitedefaultmidpunct}
{\mcitedefaultendpunct}{\mcitedefaultseppunct}\relax
\EndOfBibitem
\bibitem[Medasani \emph{et~al.}(2016)Medasani, Gamst, Ding, Chen, Persson,
  Asta, Canning, and Haranczyk]{medasani_predicting_2016}
B.~Medasani, A.~Gamst, H.~Ding, W.~Chen, K.~A. Persson, M.~Asta, A.~Canning and
  M.~Haranczyk, \emph{npj Comput Mater}, 2016, \textbf{2}, 1--10\relax
\mciteBstWouldAddEndPuncttrue
\mciteSetBstMidEndSepPunct{\mcitedefaultmidpunct}
{\mcitedefaultendpunct}{\mcitedefaultseppunct}\relax
\EndOfBibitem
\bibitem[Rahman \emph{et~al.}(2023)Rahman, Gollapalli, Manganaris, Yadav,
  Pilania, DeCost, Choudhary, and Mannodi-Kanakkithodi]{Rahman_23}
M.~H. Rahman, P.~Gollapalli, P.~Manganaris, S.~K. Yadav, G.~Pilania, B.~DeCost,
  K.~Choudhary and A.~Mannodi-Kanakkithodi, 2023,
  arXiv:cond-mat/2309.06423\relax
\mciteBstWouldAddEndPuncttrue
\mciteSetBstMidEndSepPunct{\mcitedefaultmidpunct}
{\mcitedefaultendpunct}{\mcitedefaultseppunct}\relax
\EndOfBibitem
\bibitem[Ivanov \emph{et~al.}(2023)Ivanov, Ivanov, Simoni, Parajuli, Kanté,
  Schenkel, and Tan]{ivanov2023database}
V.~Ivanov, A.~Ivanov, J.~Simoni, P.~Parajuli, B.~Kanté, T.~Schenkel and
  L.~Tan, 2023, arXiv:quant-ph/2303.16283\relax
\mciteBstWouldAddEndPuncttrue
\mciteSetBstMidEndSepPunct{\mcitedefaultmidpunct}
{\mcitedefaultendpunct}{\mcitedefaultseppunct}\relax
\EndOfBibitem
\bibitem[Kumagai \emph{et~al.}(2021)Kumagai, Tsunoda, Takahashi, and
  Oba]{Kumagai_2021}
Y.~Kumagai, N.~Tsunoda, A.~Takahashi and F.~Oba, \emph{Phys. Rev. Mater.},
  2021, \textbf{5}, 123803\relax
\mciteBstWouldAddEndPuncttrue
\mciteSetBstMidEndSepPunct{\mcitedefaultmidpunct}
{\mcitedefaultendpunct}{\mcitedefaultseppunct}\relax
\EndOfBibitem
\bibitem[Deml \emph{et~al.}(2015)Deml, Holder, O'Hayre, Musgrave, and
  Stevanovi{\'{c}}]{Deml2015}
A.~M. Deml, A.~M. Holder, R.~P. O'Hayre, C.~B. Musgrave and
  V.~Stevanovi{\'{c}}, \emph{J. Phys. Chem. Lett.}, 2015, \textbf{6},
  1948--1953\relax
\mciteBstWouldAddEndPuncttrue
\mciteSetBstMidEndSepPunct{\mcitedefaultmidpunct}
{\mcitedefaultendpunct}{\mcitedefaultseppunct}\relax
\EndOfBibitem
\bibitem[Broberg \emph{et~al.}(2023)Broberg, Bystrom, Srivastava, Dahliah,
  Williamson, Weston, Scanlon, Rignanese, Dwaraknath, Varley, Persson, Asta,
  and Hautier]{Broberg2023}
D.~Broberg, K.~Bystrom, S.~Srivastava, D.~Dahliah, B.~A.~D. Williamson,
  L.~Weston, D.~O. Scanlon, G.-M. Rignanese, S.~Dwaraknath, J.~Varley, K.~A.
  Persson, M.~Asta and G.~Hautier, \emph{npj Comput Mater}, 2023, \textbf{9},
  72\relax
\mciteBstWouldAddEndPuncttrue
\mciteSetBstMidEndSepPunct{\mcitedefaultmidpunct}
{\mcitedefaultendpunct}{\mcitedefaultseppunct}\relax
\EndOfBibitem
\bibitem[Mannodi-Kanakkithodi \emph{et~al.}(2022)Mannodi-Kanakkithodi, Xiang,
  Jacoby, Biegaj, Dunham, Gamelin, and Chan]{MannodiKanakkithodi2022}
A.~Mannodi-Kanakkithodi, X.~Xiang, L.~Jacoby, R.~Biegaj, S.~T. Dunham, D.~R.
  Gamelin and M.~K. Chan, \emph{Patterns}, 2022, \textbf{3}, 100450\relax
\mciteBstWouldAddEndPuncttrue
\mciteSetBstMidEndSepPunct{\mcitedefaultmidpunct}
{\mcitedefaultendpunct}{\mcitedefaultseppunct}\relax
\EndOfBibitem
\bibitem[Varley \emph{et~al.}(2017)Varley, Samanta, and Lordi]{Varley2017}
J.~B. Varley, A.~Samanta and V.~Lordi, \emph{J. Phys. Chem. Lett.}, 2017,
  \textbf{8}, 5059--5063\relax
\mciteBstWouldAddEndPuncttrue
\mciteSetBstMidEndSepPunct{\mcitedefaultmidpunct}
{\mcitedefaultendpunct}{\mcitedefaultseppunct}\relax
\EndOfBibitem
\bibitem[Wan \emph{et~al.}(2021)Wan, Wang, Liu, and Liang]{Wan2021}
Z.~Wan, Q.-D. Wang, D.~Liu and J.~Liang, \emph{Phys. Chem. Chem. Phys.}, 2021,
  \textbf{23}, 15675--15684\relax
\mciteBstWouldAddEndPuncttrue
\mciteSetBstMidEndSepPunct{\mcitedefaultmidpunct}
{\mcitedefaultendpunct}{\mcitedefaultseppunct}\relax
\EndOfBibitem
\bibitem[Wexler \emph{et~al.}(2021)Wexler, Gautam, Stechel, and
  Carter]{Wexler2021}
R.~B. Wexler, G.~S. Gautam, E.~B. Stechel and E.~A. Carter, \emph{J. Am. Chem.
  Soc.}, 2021, \textbf{143}, 13212--13227\relax
\mciteBstWouldAddEndPuncttrue
\mciteSetBstMidEndSepPunct{\mcitedefaultmidpunct}
{\mcitedefaultendpunct}{\mcitedefaultseppunct}\relax
\EndOfBibitem
\bibitem[Frey \emph{et~al.}(2020)Frey, Akinwande, Jariwala, and
  Shenoy]{Frey2020}
N.~C. Frey, D.~Akinwande, D.~Jariwala and V.~B. Shenoy, \emph{{ACS} Nano},
  2020, \textbf{14}, 13406--13417\relax
\mciteBstWouldAddEndPuncttrue
\mciteSetBstMidEndSepPunct{\mcitedefaultmidpunct}
{\mcitedefaultendpunct}{\mcitedefaultseppunct}\relax
\EndOfBibitem
\bibitem[Sharma \emph{et~al.}(2020)Sharma, Kumar, Dev, and Pilania]{Sharma_20}
V.~Sharma, P.~Kumar, P.~Dev and G.~Pilania, \emph{J. Appl. Phys.}, 2020,
  \textbf{128}, 034902\relax
\mciteBstWouldAddEndPuncttrue
\mciteSetBstMidEndSepPunct{\mcitedefaultmidpunct}
{\mcitedefaultendpunct}{\mcitedefaultseppunct}\relax
\EndOfBibitem
\bibitem[Baldassarri \emph{et~al.}(2023)Baldassarri, He, Gopakumar, Griesemer,
  Salgado-Casanova, Liu, Torrisi, and Wolverton]{Baldassarri_23}
B.~Baldassarri, J.~He, A.~Gopakumar, S.~Griesemer, A.~J.~A. Salgado-Casanova,
  T.-C. Liu, S.~B. Torrisi and C.~Wolverton, \emph{Chem. Mater.}, 2023,
  \textbf{35}, 10619–10634\relax
\mciteBstWouldAddEndPuncttrue
\mciteSetBstMidEndSepPunct{\mcitedefaultmidpunct}
{\mcitedefaultendpunct}{\mcitedefaultseppunct}\relax
\EndOfBibitem
\bibitem[Park \emph{et~al.}(0)Park, Lee, Park, Park, Heo, and
  Lee]{Park_23_exploring}
S.~Park, N.~Lee, J.~O. Park, J.~Park, Y.~S. Heo and J.~Lee, \emph{ACS Mater.
  Lett.}, 0, \textbf{0}, 66--72\relax
\mciteBstWouldAddEndPuncttrue
\mciteSetBstMidEndSepPunct{\mcitedefaultmidpunct}
{\mcitedefaultendpunct}{\mcitedefaultseppunct}\relax
\EndOfBibitem
\bibitem[Kazeev \emph{et~al.}(2023)Kazeev, Al-Maeeni, Romanov, Faleev, Lukin,
  Tormasov, Castro~Neto, Novoselov, Huang, and Ustyuzhanin]{kazeev_sparse_2023}
N.~Kazeev, A.~R. Al-Maeeni, I.~Romanov, M.~Faleev, R.~Lukin, A.~Tormasov, A.~H.
  Castro~Neto, K.~S. Novoselov, P.~Huang and A.~Ustyuzhanin, \emph{npj Comput
  Mater}, 2023, \textbf{9}, 113\relax
\mciteBstWouldAddEndPuncttrue
\mciteSetBstMidEndSepPunct{\mcitedefaultmidpunct}
{\mcitedefaultendpunct}{\mcitedefaultseppunct}\relax
\EndOfBibitem
\bibitem[Choudhary and Sumpter(2023)]{Choudhary_2023_can}
K.~Choudhary and B.~G. Sumpter, \emph{AIP Adv.}, 2023, \textbf{13},
  095109\relax
\mciteBstWouldAddEndPuncttrue
\mciteSetBstMidEndSepPunct{\mcitedefaultmidpunct}
{\mcitedefaultendpunct}{\mcitedefaultseppunct}\relax
\EndOfBibitem
\bibitem[Zhao \emph{et~al.}(2023)Zhao, Yu, Zheng, Reece, and
  Zhang]{zhao_machine_2023}
X.~Zhao, S.~Yu, J.~Zheng, M.~J. Reece and R.-Z. Zhang, \emph{J. Eur. Ceram.
  Soc.}, 2023, \textbf{43}, 1315--1321\relax
\mciteBstWouldAddEndPuncttrue
\mciteSetBstMidEndSepPunct{\mcitedefaultmidpunct}
{\mcitedefaultendpunct}{\mcitedefaultseppunct}\relax
\EndOfBibitem
\bibitem[Manzoor \emph{et~al.}(2021)Manzoor, Arora, Jerome, Linton, Norman, and
  Aidhy]{Manzoor2021}
A.~Manzoor, G.~Arora, B.~Jerome, N.~Linton, B.~Norman and D.~S. Aidhy,
  \emph{Front. Mater.}, 2021, \textbf{8}, 673574\relax
\mciteBstWouldAddEndPuncttrue
\mciteSetBstMidEndSepPunct{\mcitedefaultmidpunct}
{\mcitedefaultendpunct}{\mcitedefaultseppunct}\relax
\EndOfBibitem
\bibitem[Polak \emph{et~al.}(2022)Polak, Jacobs, Mannodi-Kanakkithodi, Chan,
  and Morgan]{Polak2022}
M.~P. Polak, R.~Jacobs, A.~Mannodi-Kanakkithodi, M.~K.~Y. Chan and D.~Morgan,
  \emph{J. Chem. Phys.}, 2022, \textbf{156}, 114110\relax
\mciteBstWouldAddEndPuncttrue
\mciteSetBstMidEndSepPunct{\mcitedefaultmidpunct}
{\mcitedefaultendpunct}{\mcitedefaultseppunct}\relax
\EndOfBibitem
\bibitem[Witman \emph{et~al.}(2023)Witman, Goyal, Ogitsu, McDaniel, and
  Lany]{Witman2023}
M.~D. Witman, A.~Goyal, T.~Ogitsu, A.~H. McDaniel and S.~Lany, \emph{Nat Comput
  Sci}, 2023, \textbf{3}, 675--686\relax
\mciteBstWouldAddEndPuncttrue
\mciteSetBstMidEndSepPunct{\mcitedefaultmidpunct}
{\mcitedefaultendpunct}{\mcitedefaultseppunct}\relax
\EndOfBibitem
\bibitem[Arrigoni and Madsen(2021)]{Arrigoni2021}
M.~Arrigoni and G.~K.~H. Madsen, \emph{npj Comput. Mater.}, 2021, \textbf{7},
  1--13\relax
\mciteBstWouldAddEndPuncttrue
\mciteSetBstMidEndSepPunct{\mcitedefaultmidpunct}
{\mcitedefaultendpunct}{\mcitedefaultseppunct}\relax
\EndOfBibitem
\bibitem[Kavanagh \emph{et~al.}(2021)Kavanagh, Walsh, and
  Scanlon]{Kavanagh_rapid}
S.~R. Kavanagh, A.~Walsh and D.~O. Scanlon, \emph{ACS Energy Lett.}, 2021,
  \textbf{6}, 1392--1398\relax
\mciteBstWouldAddEndPuncttrue
\mciteSetBstMidEndSepPunct{\mcitedefaultmidpunct}
{\mcitedefaultendpunct}{\mcitedefaultseppunct}\relax
\EndOfBibitem
\bibitem[Mosquera-Lois and Kavanagh(2021)]{mosquera-lois_search_2021}
I.~Mosquera-Lois and S.~R. Kavanagh, \emph{Matter}, 2021, \textbf{4},
  2602--2605\relax
\mciteBstWouldAddEndPuncttrue
\mciteSetBstMidEndSepPunct{\mcitedefaultmidpunct}
{\mcitedefaultendpunct}{\mcitedefaultseppunct}\relax
\EndOfBibitem
\bibitem[Mosquera-Lois \emph{et~al.}(2023)Mosquera-Lois, Kavanagh, Walsh, and
  Scanlon]{Mosquera-Lois2023}
I.~Mosquera-Lois, S.~R. Kavanagh, A.~Walsh and D.~O. Scanlon, \emph{npj Comput.
  Mater.}, 2023, \textbf{9}, 1--11\relax
\mciteBstWouldAddEndPuncttrue
\mciteSetBstMidEndSepPunct{\mcitedefaultmidpunct}
{\mcitedefaultendpunct}{\mcitedefaultseppunct}\relax
\EndOfBibitem
\bibitem[Wang \emph{et~al.}(2023)Wang, Kavanagh, Scanlon, and
  Walsh]{wang2023fourelectron}
X.~Wang, S.~R. Kavanagh, D.~O. Scanlon and A.~Walsh, \emph{Phys. Rev. B}, 2023,
  \textbf{108}, 134102\relax
\mciteBstWouldAddEndPuncttrue
\mciteSetBstMidEndSepPunct{\mcitedefaultmidpunct}
{\mcitedefaultendpunct}{\mcitedefaultseppunct}\relax
\EndOfBibitem
\bibitem[Morris \emph{et~al.}(2009)Morris, Pickard, and Needs]{Morris2009}
A.~J. Morris, C.~J. Pickard and R.~J. Needs, \emph{Phys. Rev. B}, 2009,
  \textbf{80}, 144112\relax
\mciteBstWouldAddEndPuncttrue
\mciteSetBstMidEndSepPunct{\mcitedefaultmidpunct}
{\mcitedefaultendpunct}{\mcitedefaultseppunct}\relax
\EndOfBibitem
\bibitem[Mulroue \emph{et~al.}(2011)Mulroue, Morris, and Duffy]{Mulroue2011}
J.~Mulroue, A.~J. Morris and D.~M. Duffy, \emph{Phys. Rev. B}, 2011,
  \textbf{84}, 094118\relax
\mciteBstWouldAddEndPuncttrue
\mciteSetBstMidEndSepPunct{\mcitedefaultmidpunct}
{\mcitedefaultendpunct}{\mcitedefaultseppunct}\relax
\EndOfBibitem
\bibitem[Al-Mushadani and Needs(2003)]{al-mushadani_free-energy_2003}
O.~K. Al-Mushadani and R.~J. Needs, \emph{Phys. Rev. B}, 2003, \textbf{68},
  235205\relax
\mciteBstWouldAddEndPuncttrue
\mciteSetBstMidEndSepPunct{\mcitedefaultmidpunct}
{\mcitedefaultendpunct}{\mcitedefaultseppunct}\relax
\EndOfBibitem
\bibitem[Kononov \emph{et~al.}(2023)Kononov, Lee, Shapera, and
  Schleife]{kononov_2023}
A.~Kononov, C.-W. Lee, E.~Shapera and A.~Schleife, \emph{J. Phys. Condens.},
  2023, \textbf{35}, 334002\relax
\mciteBstWouldAddEndPuncttrue
\mciteSetBstMidEndSepPunct{\mcitedefaultmidpunct}
{\mcitedefaultendpunct}{\mcitedefaultseppunct}\relax
\EndOfBibitem
\bibitem[Schaarschmidt \emph{et~al.}(2022)Schaarschmidt, Riviere, Ganose,
  Spencer, Gaunt, Kirkpatrick, Axelrod, Battaglia, and
  Godwin]{Schaarschmidt_22}
M.~Schaarschmidt, M.~Riviere, A.~M. Ganose, J.~S. Spencer, A.~L. Gaunt,
  J.~Kirkpatrick, S.~Axelrod, P.~W. Battaglia and J.~Godwin, 2022,
  arXiv:cond-mat/2209.12466\relax
\mciteBstWouldAddEndPuncttrue
\mciteSetBstMidEndSepPunct{\mcitedefaultmidpunct}
{\mcitedefaultendpunct}{\mcitedefaultseppunct}\relax
\EndOfBibitem
\bibitem[Lan \emph{et~al.}(2023)Lan, Palizhati, Shuaibi, Wood, Wander, Das,
  Uyttendaele, Zitnick, and Ulissi]{Lan2023}
J.~Lan, A.~Palizhati, M.~Shuaibi, B.~M. Wood, B.~Wander, A.~Das,
  M.~Uyttendaele, C.~L. Zitnick and Z.~W. Ulissi, \emph{npj Comput Mater},
  2023, \textbf{9}, 172\relax
\mciteBstWouldAddEndPuncttrue
\mciteSetBstMidEndSepPunct{\mcitedefaultmidpunct}
{\mcitedefaultendpunct}{\mcitedefaultseppunct}\relax
\EndOfBibitem
\bibitem[Lany and Zunger(2004)]{lany_metal-dimer_2004}
S.~Lany and A.~Zunger, \emph{Phys. Rev. Lett.}, 2004, \textbf{93}, 156404\relax
\mciteBstWouldAddEndPuncttrue
\mciteSetBstMidEndSepPunct{\mcitedefaultmidpunct}
{\mcitedefaultendpunct}{\mcitedefaultseppunct}\relax
\EndOfBibitem
\bibitem[Kang and Wang(2017)]{Kang2017}
J.~Kang and L.-W. Wang, \emph{J. Phys. Chem. Lett.}, 2017, \textbf{8},
  489--493\relax
\mciteBstWouldAddEndPuncttrue
\mciteSetBstMidEndSepPunct{\mcitedefaultmidpunct}
{\mcitedefaultendpunct}{\mcitedefaultseppunct}\relax
\EndOfBibitem
\bibitem[Wilson \emph{et~al.}(2008)Wilson, Sokol, French, and
  Catlow]{wilson_defect_2008}
D.~J. Wilson, A.~A. Sokol, S.~A. French and C.~R.~A. Catlow, \emph{Phys. Rev.
  B}, 2008, \textbf{77}, 064115\relax
\mciteBstWouldAddEndPuncttrue
\mciteSetBstMidEndSepPunct{\mcitedefaultmidpunct}
{\mcitedefaultendpunct}{\mcitedefaultseppunct}\relax
\EndOfBibitem
\bibitem[Zhao \emph{et~al.}(2016)Zhao, Zhou, Ma, Meng, Li, Wei, Fu, Liu, Yu,
  and Zhao]{zhao_correlations_2016}
Y.~Zhao, W.~Zhou, W.~Ma, S.~Meng, H.~Li, J.~Wei, R.~Fu, K.~Liu, D.~Yu and
  Q.~Zhao, \emph{ACS Energy Lett.}, 2016, \textbf{1}, 266--272\relax
\mciteBstWouldAddEndPuncttrue
\mciteSetBstMidEndSepPunct{\mcitedefaultmidpunct}
{\mcitedefaultendpunct}{\mcitedefaultseppunct}\relax
\EndOfBibitem
\bibitem[Ágoston \emph{et~al.}(2009)Ágoston, Erhart, Klein, and
  Albe]{Agoston_2009}
P.~Ágoston, P.~Erhart, A.~Klein and K.~Albe, \emph{J. Phys. Condens. Matter},
  2009, \textbf{21}, 455801\relax
\mciteBstWouldAddEndPuncttrue
\mciteSetBstMidEndSepPunct{\mcitedefaultmidpunct}
{\mcitedefaultendpunct}{\mcitedefaultseppunct}\relax
\EndOfBibitem
\bibitem[Han \emph{et~al.}(2017)Han, Du, Dai, Sun, and Chen]{Han2017}
D.~Han, M.-H. Du, C.-M. Dai, D.~Sun and S.~Chen, \emph{J. Mater. Chem. A},
  2017, \textbf{5}, 6200--6210\relax
\mciteBstWouldAddEndPuncttrue
\mciteSetBstMidEndSepPunct{\mcitedefaultmidpunct}
{\mcitedefaultendpunct}{\mcitedefaultseppunct}\relax
\EndOfBibitem
\bibitem[Meggiolaro \emph{et~al.}(2020)Meggiolaro, Ricciarelli, Alasmari,
  Alasmary, and De~Angelis]{meggiolaro_tin_2020}
D.~Meggiolaro, D.~Ricciarelli, A.~A. Alasmari, F.~A.~S. Alasmary and
  F.~De~Angelis, \emph{J. Phys. Chem. Lett.}, 2020, \textbf{11},
  3546--3556\relax
\mciteBstWouldAddEndPuncttrue
\mciteSetBstMidEndSepPunct{\mcitedefaultmidpunct}
{\mcitedefaultendpunct}{\mcitedefaultseppunct}\relax
\EndOfBibitem
\bibitem[Erhart \emph{et~al.}(2005)Erhart, Klein, and
  Albe]{erhart_first-principles_2005}
P.~Erhart, A.~Klein and K.~Albe, \emph{Phys. Rev. B}, 2005, \textbf{72},
  085213\relax
\mciteBstWouldAddEndPuncttrue
\mciteSetBstMidEndSepPunct{\mcitedefaultmidpunct}
{\mcitedefaultendpunct}{\mcitedefaultseppunct}\relax
\EndOfBibitem
\bibitem[Sokol \emph{et~al.}(2010)Sokol, Walsh, and Catlow]{sokol_oxygen_2010}
A.~A. Sokol, A.~Walsh and C.~R.~A. Catlow, \emph{Chem. Phys. Lett.}, 2010,
  \textbf{492}, 44--48\relax
\mciteBstWouldAddEndPuncttrue
\mciteSetBstMidEndSepPunct{\mcitedefaultmidpunct}
{\mcitedefaultendpunct}{\mcitedefaultseppunct}\relax
\EndOfBibitem
\bibitem[Evarestov \emph{et~al.}(1996)Evarestov, Jacobs, and
  Leko]{evarestov_oxygen_1996}
R.~A. Evarestov, P.~W.~M. Jacobs and A.~V. Leko, \emph{Phys. Rev. B}, 1996,
  \textbf{54}, 8969--8972\relax
\mciteBstWouldAddEndPuncttrue
\mciteSetBstMidEndSepPunct{\mcitedefaultmidpunct}
{\mcitedefaultendpunct}{\mcitedefaultseppunct}\relax
\EndOfBibitem
\bibitem[Kotomin and Popov(1998)]{kotomin_radiation-induced_1998}
E.~A. Kotomin and A.~I. Popov, \emph{Nucl. Instrum. Methods Phys. Res. B},
  1998, \textbf{141}, 1--15\relax
\mciteBstWouldAddEndPuncttrue
\mciteSetBstMidEndSepPunct{\mcitedefaultmidpunct}
{\mcitedefaultendpunct}{\mcitedefaultseppunct}\relax
\EndOfBibitem
\bibitem[Burbano \emph{et~al.}(2011)Burbano, Scanlon, and
  Watson]{burbano_sources_2011}
M.~Burbano, D.~O. Scanlon and G.~W. Watson, \emph{J. Am. Chem. Soc.}, 2011,
  \textbf{133}, 15065--15072\relax
\mciteBstWouldAddEndPuncttrue
\mciteSetBstMidEndSepPunct{\mcitedefaultmidpunct}
{\mcitedefaultendpunct}{\mcitedefaultseppunct}\relax
\EndOfBibitem
\bibitem[Scanlon and Watson(2012)]{scanlon_possibility_2012}
D.~O. Scanlon and G.~W. Watson, \emph{J. Mater. Chem.}, 2012, \textbf{22},
  25236--25245\relax
\mciteBstWouldAddEndPuncttrue
\mciteSetBstMidEndSepPunct{\mcitedefaultmidpunct}
{\mcitedefaultendpunct}{\mcitedefaultseppunct}\relax
\EndOfBibitem
\bibitem[Godinho \emph{et~al.}(2009)Godinho, Walsh, and
  Watson]{godinho_energetic_2009}
K.~G. Godinho, A.~Walsh and G.~W. Watson, \emph{J. Phys. Chem. C}, 2009,
  \textbf{113}, 439--448\relax
\mciteBstWouldAddEndPuncttrue
\mciteSetBstMidEndSepPunct{\mcitedefaultmidpunct}
{\mcitedefaultendpunct}{\mcitedefaultseppunct}\relax
\EndOfBibitem
\bibitem[Scanlon \emph{et~al.}(2011)Scanlon, Kehoe, Watson, Jones, David,
  Payne, Egdell, Edwards, and Walsh]{scanlon_nature_2011}
D.~O. Scanlon, A.~B. Kehoe, G.~W. Watson, M.~O. Jones, W.~I.~F. David, D.~J.
  Payne, R.~G. Egdell, P.~P. Edwards and A.~Walsh, \emph{Phys. Rev. Lett.},
  2011, \textbf{107}, 246402\relax
\mciteBstWouldAddEndPuncttrue
\mciteSetBstMidEndSepPunct{\mcitedefaultmidpunct}
{\mcitedefaultendpunct}{\mcitedefaultseppunct}\relax
\EndOfBibitem
\bibitem[Keating \emph{et~al.}(2012)Keating, Scanlon, Morgan, Galea, and
  Watson]{keating_analysis_2012}
P.~R.~L. Keating, D.~O. Scanlon, B.~J. Morgan, N.~M. Galea and G.~W. Watson,
  \emph{J. Phys. Chem. C}, 2012, \textbf{116}, 2443--2452\relax
\mciteBstWouldAddEndPuncttrue
\mciteSetBstMidEndSepPunct{\mcitedefaultmidpunct}
{\mcitedefaultendpunct}{\mcitedefaultseppunct}\relax
\EndOfBibitem
\bibitem[Walsh \emph{et~al.}(2009)Walsh, Da~Silva, and
  Wei]{walsh_interplay_2009}
A.~Walsh, J.~L.~F. Da~Silva and S.-H. Wei, \emph{Chem. Mater.}, 2009,
  \textbf{21}, 5119--5124\relax
\mciteBstWouldAddEndPuncttrue
\mciteSetBstMidEndSepPunct{\mcitedefaultmidpunct}
{\mcitedefaultendpunct}{\mcitedefaultseppunct}\relax
\EndOfBibitem
\bibitem[Whalley \emph{et~al.}(2017)Whalley, Crespo-Otero, and
  Walsh]{whalley_h-center_2017}
L.~D. Whalley, R.~Crespo-Otero and A.~Walsh, \emph{ACS Energy Lett.}, 2017,
  \textbf{2}, 2713--2714\relax
\mciteBstWouldAddEndPuncttrue
\mciteSetBstMidEndSepPunct{\mcitedefaultmidpunct}
{\mcitedefaultendpunct}{\mcitedefaultseppunct}\relax
\EndOfBibitem
\bibitem[Agiorgousis \emph{et~al.}(2014)Agiorgousis, Sun, Zeng, and
  Zhang]{agiorgousis_strong_2014}
M.~L. Agiorgousis, Y.-Y. Sun, H.~Zeng and S.~Zhang, \emph{J. Am. Chem. Soc.},
  2014, \textbf{136}, 14570--14575\relax
\mciteBstWouldAddEndPuncttrue
\mciteSetBstMidEndSepPunct{\mcitedefaultmidpunct}
{\mcitedefaultendpunct}{\mcitedefaultseppunct}\relax
\EndOfBibitem
\bibitem[Whalley \emph{et~al.}(2021)Whalley, van Gerwen, Frost, Kim, Hood, and
  Walsh]{whalley_giant_2021}
L.~D. Whalley, P.~van Gerwen, J.~M. Frost, S.~Kim, S.~N. Hood and A.~Walsh,
  \emph{J. Am. Chem. Soc.}, 2021, \textbf{143}, 9123--9128\relax
\mciteBstWouldAddEndPuncttrue
\mciteSetBstMidEndSepPunct{\mcitedefaultmidpunct}
{\mcitedefaultendpunct}{\mcitedefaultseppunct}\relax
\EndOfBibitem
\bibitem[Motti \emph{et~al.}(2019)Motti, Meggiolaro, Martani, Sorrentino,
  Barker, De~Angelis, and Petrozza]{motti_defect_2019}
S.~G. Motti, D.~Meggiolaro, S.~Martani, R.~Sorrentino, A.~J. Barker,
  F.~De~Angelis and A.~Petrozza, \emph{Adv Mater}, 2019, \textbf{31},
  1901183\relax
\mciteBstWouldAddEndPuncttrue
\mciteSetBstMidEndSepPunct{\mcitedefaultmidpunct}
{\mcitedefaultendpunct}{\mcitedefaultseppunct}\relax
\EndOfBibitem
\bibitem[Xiao \emph{et~al.}(2016)Xiao, Meng, Wang, and Yan]{xiao_defect_2016}
Z.~Xiao, W.~Meng, J.~Wang and Y.~Yan, \emph{Phys. Chem. Chem. Phys.}, 2016,
  \textbf{18}, 25786--25790\relax
\mciteBstWouldAddEndPuncttrue
\mciteSetBstMidEndSepPunct{\mcitedefaultmidpunct}
{\mcitedefaultendpunct}{\mcitedefaultseppunct}\relax
\EndOfBibitem
\bibitem[Na-Phattalung \emph{et~al.}(2006)Na-Phattalung, Smith, Kim, Du, Wei,
  Zhang, and Limpijumnong]{Na-Phattalung_2006}
S.~Na-Phattalung, M.~F. Smith, K.~Kim, M.-H. Du, S.-H. Wei, S.~B. Zhang and
  S.~Limpijumnong, \emph{Phys. Rev. B}, 2006, \textbf{73}, 125205\relax
\mciteBstWouldAddEndPuncttrue
\mciteSetBstMidEndSepPunct{\mcitedefaultmidpunct}
{\mcitedefaultendpunct}{\mcitedefaultseppunct}\relax
\EndOfBibitem
\bibitem[Li \emph{et~al.}(2023)Li, Willis, Kavanagh, and Scanlon]{Li_2023}
K.~Li, J.~Willis, S.~R. Kavanagh and D.~O. Scanlon, 2023,
  chemrxiv-2023-8l7pb\relax
\mciteBstWouldAddEndPuncttrue
\mciteSetBstMidEndSepPunct{\mcitedefaultmidpunct}
{\mcitedefaultendpunct}{\mcitedefaultseppunct}\relax
\EndOfBibitem
\bibitem[Scanlon(2013)]{Scanlon_2013_basno3}
D.~O. Scanlon, \emph{Phys. Rev. B}, 2013, \textbf{87}, 161201\relax
\mciteBstWouldAddEndPuncttrue
\mciteSetBstMidEndSepPunct{\mcitedefaultmidpunct}
{\mcitedefaultendpunct}{\mcitedefaultseppunct}\relax
\EndOfBibitem
\bibitem[Cen \emph{et~al.}(2023)Cen, Zhu, Kavanagh, Squires, and
  Scanlon]{Cen2023}
J.~Cen, B.~Zhu, S.~R. Kavanagh, A.~G. Squires and D.~O. Scanlon, \emph{J.
  Mater. Chem. A}, 2023, \textbf{11}, 13353–13370\relax
\mciteBstWouldAddEndPuncttrue
\mciteSetBstMidEndSepPunct{\mcitedefaultmidpunct}
{\mcitedefaultendpunct}{\mcitedefaultseppunct}\relax
\EndOfBibitem
\bibitem[Mosquera-Lois \emph{et~al.}(2022)Mosquera-Lois, Kavanagh, Walsh, and
  Scanlon]{shakenbreak2022}
I.~Mosquera-Lois, S.~R. Kavanagh, A.~Walsh and D.~O. Scanlon, \emph{J. Open
  Source Softw.}, 2022, \textbf{7}, 4817\relax
\mciteBstWouldAddEndPuncttrue
\mciteSetBstMidEndSepPunct{\mcitedefaultmidpunct}
{\mcitedefaultendpunct}{\mcitedefaultseppunct}\relax
\EndOfBibitem
\bibitem[Chen and Ong(2022)]{Chen2022}
C.~Chen and S.~P. Ong, \emph{Nat Comput Sci}, 2022, \textbf{2}, 718--728\relax
\mciteBstWouldAddEndPuncttrue
\mciteSetBstMidEndSepPunct{\mcitedefaultmidpunct}
{\mcitedefaultendpunct}{\mcitedefaultseppunct}\relax
\EndOfBibitem
\bibitem[Qi \emph{et~al.}(2023)Qi, Ko, Wood, Pham, and Ong]{Qi_2023}
J.~Qi, T.~W. Ko, B.~C. Wood, T.~A. Pham and S.~P. Ong, 2023,
  arXiv:cond-mat/22307.13710\relax
\mciteBstWouldAddEndPuncttrue
\mciteSetBstMidEndSepPunct{\mcitedefaultmidpunct}
{\mcitedefaultendpunct}{\mcitedefaultseppunct}\relax
\EndOfBibitem
\bibitem[van~der Maaten and Hinton(2008)]{vanDerMaaten2008}
L.~van~der Maaten and G.~Hinton, \emph{J. Mach. Learn. Res.}, 2008, \textbf{9},
  2579--2605\relax
\mciteBstWouldAddEndPuncttrue
\mciteSetBstMidEndSepPunct{\mcitedefaultmidpunct}
{\mcitedefaultendpunct}{\mcitedefaultseppunct}\relax
\EndOfBibitem
\bibitem[Hinton and Roweis(2002)]{hinton2002stochastic}
G.~E. Hinton and S.~Roweis, \emph{Adv Neural Inf Process Syst}, 2002,
  \textbf{15}, 857--864\relax
\mciteBstWouldAddEndPuncttrue
\mciteSetBstMidEndSepPunct{\mcitedefaultmidpunct}
{\mcitedefaultendpunct}{\mcitedefaultseppunct}\relax
\EndOfBibitem
\bibitem[Chen \emph{et~al.}(2019)Chen, Ye, Zuo, Zheng, and Ong]{Chen2019}
C.~Chen, W.~Ye, Y.~Zuo, C.~Zheng and S.~P. Ong, \emph{Chem Mater}, 2019,
  \textbf{31}, 3564–3572\relax
\mciteBstWouldAddEndPuncttrue
\mciteSetBstMidEndSepPunct{\mcitedefaultmidpunct}
{\mcitedefaultendpunct}{\mcitedefaultseppunct}\relax
\EndOfBibitem
\bibitem[nis()]{nist}
\emph{NIST Chemistry WebBook}, https://doi.org/10.18434/M32147, (accessed May
  2023)\relax
\mciteBstWouldAddEndPuncttrue
\mciteSetBstMidEndSepPunct{\mcitedefaultmidpunct}
{\mcitedefaultendpunct}{\mcitedefaultseppunct}\relax
\EndOfBibitem
\bibitem[Deng \emph{et~al.}(2023)Deng, Zhong, Jun, Riebesell, Han, Bartel, and
  Ceder]{Deng2023}
B.~Deng, P.~Zhong, K.~Jun, J.~Riebesell, K.~Han, C.~J. Bartel and G.~Ceder,
  \emph{Nat. Mach. Intell.}, 2023, \textbf{5}, 1031–1041\relax
\mciteBstWouldAddEndPuncttrue
\mciteSetBstMidEndSepPunct{\mcitedefaultmidpunct}
{\mcitedefaultendpunct}{\mcitedefaultseppunct}\relax
\EndOfBibitem
\bibitem[Merchant \emph{et~al.}(2023)Merchant, Batzner, Schoenholz, Aykol,
  Cheon, and Cubuk]{Merchant2023}
A.~Merchant, S.~Batzner, S.~S. Schoenholz, M.~Aykol, G.~Cheon and E.~D. Cubuk,
  \emph{Nature}, 2023, \textbf{624}, 80–85\relax
\mciteBstWouldAddEndPuncttrue
\mciteSetBstMidEndSepPunct{\mcitedefaultmidpunct}
{\mcitedefaultendpunct}{\mcitedefaultseppunct}\relax
\EndOfBibitem
\bibitem[Batatia \emph{et~al.}(2024)Batatia, Benner, Chiang, Elena, Kovács,
  Riebesell, Advincula, Asta, Baldwin, Bernstein, Bhowmik, Blau, Cărare,
  Darby, De, Della~Pia, Deringer, Elijošius, El-Machachi, Fako, Ferrari,
  Genreith-Schriever, George, Goodall, Grey, Han, Handley, Heenen, Hermansson,
  Holm, Jaafar, Hofmann, Jakob, Jung, Kapil, Kaplan, Karimitari, Kroupa,
  Kullgren, Kuner, Kuryla, Liepuoniute, Margraf, Magdău, Michaelides, Moore,
  Naik, Niblett, Norwood, O'Neill, Ortner, Persson, Reuter, Rosen, Schaaf,
  Schran, Sivonxay, Stenczel, Svahn, Sutton, van~der Oord, Varga-Umbrich,
  Vegge, Vondrák, Wang, Witt, Zills, and Csányi]{Batatia_2024}
I.~Batatia, P.~Benner, Y.~Chiang, A.~M. Elena, D.~P. Kovács, J.~Riebesell,
  X.~R. Advincula, M.~Asta, W.~J. Baldwin, N.~Bernstein, A.~Bhowmik, S.~M.
  Blau, V.~Cărare, J.~P. Darby, S.~De, F.~Della~Pia, V.~L. Deringer,
  R.~Elijošius, Z.~El-Machachi, E.~Fako, A.~C. Ferrari, A.~Genreith-Schriever,
  J.~George, R.~E.~A. Goodall, C.~P. Grey, S.~Han, W.~Handley, H.~H. Heenen,
  K.~Hermansson, C.~Holm, J.~Jaafar, S.~Hofmann, K.~S. Jakob, H.~Jung,
  V.~Kapil, A.~D. Kaplan, N.~Karimitari, N.~Kroupa, J.~Kullgren, M.~C. Kuner,
  D.~Kuryla, G.~Liepuoniute, J.~T. Margraf, I.-B. Magdău, A.~Michaelides,
  J.~H. Moore, A.~A. Naik, S.~P. Niblett, S.~W. Norwood, N.~O'Neill, C.~Ortner,
  K.~A. Persson, K.~Reuter, A.~S. Rosen, L.~L. Schaaf, C.~Schran, E.~Sivonxay,
  T.~K. Stenczel, V.~Svahn, C.~Sutton, C.~van~der Oord, E.~Varga-Umbrich,
  T.~Vegge, M.~Vondrák, Y.~Wang, W.~C. Witt, F.~Zills and G.~Csányi, 2024,
  arXiv:physics.chem-ph/2401.00096\relax
\mciteBstWouldAddEndPuncttrue
\mciteSetBstMidEndSepPunct{\mcitedefaultmidpunct}
{\mcitedefaultendpunct}{\mcitedefaultseppunct}\relax
\EndOfBibitem
\bibitem[Batatia \emph{et~al.}(2022)Batatia, Kovács, Simm, Ortner, and
  Csányi]{Batatia_2022}
I.~Batatia, D.~P. Kovács, G.~N.~C. Simm, C.~Ortner and G.~Csányi, 2022,
  arXiv:stat.ML/2206.07697\relax
\mciteBstWouldAddEndPuncttrue
\mciteSetBstMidEndSepPunct{\mcitedefaultmidpunct}
{\mcitedefaultendpunct}{\mcitedefaultseppunct}\relax
\EndOfBibitem
\bibitem[Chen and Ong()]{m3gnet_repo}
C.~Chen and S.~P. Ong, M3GNet (version 0.2.4), 2023\relax
\mciteBstWouldAddEndPuncttrue
\mciteSetBstMidEndSepPunct{\mcitedefaultmidpunct}
{\mcitedefaultendpunct}{\mcitedefaultseppunct}\relax
\EndOfBibitem
\bibitem[Salzbrenner \emph{et~al.}(2023)Salzbrenner, Joo, Conway, Cooke, Zhu,
  Matraszek, Witt, and Pickard]{Salzbrenner_2023}
P.~T. Salzbrenner, S.~H. Joo, L.~J. Conway, P.~I.~C. Cooke, B.~Zhu, M.~P.
  Matraszek, W.~C. Witt and C.~J. Pickard, \emph{J. Chem. Phys.}, 2023,
  \textbf{159}, 144801\relax
\mciteBstWouldAddEndPuncttrue
\mciteSetBstMidEndSepPunct{\mcitedefaultmidpunct}
{\mcitedefaultendpunct}{\mcitedefaultseppunct}\relax
\EndOfBibitem
\bibitem[Pickard(2022)]{Pickard2022}
C.~J. Pickard, \emph{Phys. Rev. B}, 2022, \textbf{106}, 014102\relax
\mciteBstWouldAddEndPuncttrue
\mciteSetBstMidEndSepPunct{\mcitedefaultmidpunct}
{\mcitedefaultendpunct}{\mcitedefaultseppunct}\relax
\EndOfBibitem
\bibitem[Musielewicz \emph{et~al.}(2022)Musielewicz, Wang, Tian, and
  Ulissi]{Musielewicz2022}
J.~Musielewicz, X.~Wang, T.~Tian and Z.~Ulissi, \emph{Mach. Learn. Technol.},
  2022, \textbf{3}, 03LT01\relax
\mciteBstWouldAddEndPuncttrue
\mciteSetBstMidEndSepPunct{\mcitedefaultmidpunct}
{\mcitedefaultendpunct}{\mcitedefaultseppunct}\relax
\EndOfBibitem
\bibitem[Jung \emph{et~al.}(2023)Jung, Sauerland, Stocker, Reuter, and
  Margraf]{Jung2023}
H.~Jung, L.~Sauerland, S.~Stocker, K.~Reuter and J.~T. Margraf, \emph{npj
  Comput Mater}, 2023, \textbf{9}, 114\relax
\mciteBstWouldAddEndPuncttrue
\mciteSetBstMidEndSepPunct{\mcitedefaultmidpunct}
{\mcitedefaultendpunct}{\mcitedefaultseppunct}\relax
\EndOfBibitem
\bibitem[Bart\'ok \emph{et~al.}(2013)Bart\'ok, Kondor, and
  Cs\'anyi]{bartok_2013}
A.~P. Bart\'ok, R.~Kondor and G.~Cs\'anyi, \emph{Phys. Rev. B}, 2013,
  \textbf{87}, 184115\relax
\mciteBstWouldAddEndPuncttrue
\mciteSetBstMidEndSepPunct{\mcitedefaultmidpunct}
{\mcitedefaultendpunct}{\mcitedefaultseppunct}\relax
\EndOfBibitem
\bibitem[Hu(2023)]{hu_first-principles_2023}
Y.-J. Hu, \emph{Comput. Mater. Sci.}, 2023, \textbf{216}, 111831\relax
\mciteBstWouldAddEndPuncttrue
\mciteSetBstMidEndSepPunct{\mcitedefaultmidpunct}
{\mcitedefaultendpunct}{\mcitedefaultseppunct}\relax
\EndOfBibitem
\bibitem[Piochaud \emph{et~al.}(2014)Piochaud, Klaver, Adjanor, Olsson, Domain,
  and Becquart]{Piochaud_2014}
J.~B. Piochaud, T.~P.~C. Klaver, G.~Adjanor, P.~Olsson, C.~Domain and C.~S.
  Becquart, \emph{Phys. Rev. B}, 2014, \textbf{89}, 024101\relax
\mciteBstWouldAddEndPuncttrue
\mciteSetBstMidEndSepPunct{\mcitedefaultmidpunct}
{\mcitedefaultendpunct}{\mcitedefaultseppunct}\relax
\EndOfBibitem
\bibitem[Guan \emph{et~al.}(2020)Guan, Huang, Ding, Tian, Xu, and
  Zhao]{Guan_2020}
H.~Guan, S.~Huang, J.~Ding, F.~Tian, Q.~Xu and J.~Zhao, \emph{Acta Mater.},
  2020, \textbf{187}, 122--134\relax
\mciteBstWouldAddEndPuncttrue
\mciteSetBstMidEndSepPunct{\mcitedefaultmidpunct}
{\mcitedefaultendpunct}{\mcitedefaultseppunct}\relax
\EndOfBibitem
\bibitem[Rio \emph{et~al.}(2011)Rio, Sampedro, Dogo, Caturla, Caro, Caro, and
  Perlado]{rio_formation_2011}
E.~d. Rio, J.~M. Sampedro, H.~Dogo, M.~J. Caturla, M.~Caro, A.~Caro and J.~M.
  Perlado, \emph{J. Nucl. Mater.}, 2011, \textbf{408}, 18--24\relax
\mciteBstWouldAddEndPuncttrue
\mciteSetBstMidEndSepPunct{\mcitedefaultmidpunct}
{\mcitedefaultendpunct}{\mcitedefaultseppunct}\relax
\EndOfBibitem
\bibitem[Zhang \emph{et~al.}(2015)Zhang, Stocks, Jin, Lu, Bei, Sales, Wang,
  Béland, Stoller, Samolyuk, Caro, Caro, and Weber]{zhang_influence_2015}
Y.~Zhang, G.~M. Stocks, K.~Jin, C.~Lu, H.~Bei, B.~C. Sales, L.~Wang, L.~K.
  Béland, R.~E. Stoller, G.~D. Samolyuk, M.~Caro, A.~Caro and W.~J. Weber,
  \emph{Nat Commun}, 2015, \textbf{6}, 8736\relax
\mciteBstWouldAddEndPuncttrue
\mciteSetBstMidEndSepPunct{\mcitedefaultmidpunct}
{\mcitedefaultendpunct}{\mcitedefaultseppunct}\relax
\EndOfBibitem
\bibitem[Zhang \emph{et~al.}(2017)Zhang, Zhao, Weber, Nordlund, Granberg, and
  Djurabekova]{zhang_atomic-level_2017}
Y.~Zhang, S.~Zhao, W.~J. Weber, K.~Nordlund, F.~Granberg and F.~Djurabekova,
  \emph{Curr. Opin. Solid State Mater. Sci.}, 2017, \textbf{21}, 221--237\relax
\mciteBstWouldAddEndPuncttrue
\mciteSetBstMidEndSepPunct{\mcitedefaultmidpunct}
{\mcitedefaultendpunct}{\mcitedefaultseppunct}\relax
\EndOfBibitem
\bibitem[Arora \emph{et~al.}(2021)Arora, Bonny, Castin, and
  Aidhy]{arora_effect_2021}
G.~Arora, G.~Bonny, N.~Castin and D.~S. Aidhy, \emph{Acta Mater}, 2021,
  \textbf{15}, 100974\relax
\mciteBstWouldAddEndPuncttrue
\mciteSetBstMidEndSepPunct{\mcitedefaultmidpunct}
{\mcitedefaultendpunct}{\mcitedefaultseppunct}\relax
\EndOfBibitem
\bibitem[Zhao \emph{et~al.}(2016)Zhao, Stocks, and Zhang]{zhao_defect_2016}
S.~Zhao, G.~M. Stocks and Y.~Zhang, \emph{Phys. Chem. Chem. Phys.}, 2016,
  \textbf{18}, 24043--24056\relax
\mciteBstWouldAddEndPuncttrue
\mciteSetBstMidEndSepPunct{\mcitedefaultmidpunct}
{\mcitedefaultendpunct}{\mcitedefaultseppunct}\relax
\EndOfBibitem
\bibitem[Manzoor and Zhang(2022)]{manzoor_influence_2022}
A.~Manzoor and Y.~Zhang, \emph{JOM}, 2022, \textbf{74}, 4107--4120\relax
\mciteBstWouldAddEndPuncttrue
\mciteSetBstMidEndSepPunct{\mcitedefaultmidpunct}
{\mcitedefaultendpunct}{\mcitedefaultseppunct}\relax
\EndOfBibitem
\bibitem[Zhao \emph{et~al.}(2018)Zhao, Egami, Stocks, and
  Zhang]{zhao_effect_2018}
S.~Zhao, T.~Egami, G.~M. Stocks and Y.~Zhang, \emph{Phys. Rev. Mater.}, 2018,
  \textbf{2}, 013602\relax
\mciteBstWouldAddEndPuncttrue
\mciteSetBstMidEndSepPunct{\mcitedefaultmidpunct}
{\mcitedefaultendpunct}{\mcitedefaultseppunct}\relax
\EndOfBibitem
\bibitem[Li \emph{et~al.}(2019)Li, Yin, Odbadrakh, Sales, Zinkle, Stocks, and
  Wirth]{li_first_2019}
C.~Li, J.~Yin, K.~Odbadrakh, B.~C. Sales, S.~J. Zinkle, G.~M. Stocks and B.~D.
  Wirth, \emph{J. Appl. Phys}, 2019, \textbf{125}, 155103\relax
\mciteBstWouldAddEndPuncttrue
\mciteSetBstMidEndSepPunct{\mcitedefaultmidpunct}
{\mcitedefaultendpunct}{\mcitedefaultseppunct}\relax
\EndOfBibitem
\bibitem[Manzoor \emph{et~al.}(2021)Manzoor, Zhang, and
  Aidhy]{manzoor_factors_2021}
A.~Manzoor, Y.~Zhang and D.~S. Aidhy, \emph{Comput. Mater. Sci.}, 2021,
  \textbf{198}, 110669\relax
\mciteBstWouldAddEndPuncttrue
\mciteSetBstMidEndSepPunct{\mcitedefaultmidpunct}
{\mcitedefaultendpunct}{\mcitedefaultseppunct}\relax
\EndOfBibitem
\bibitem[Muzyk \emph{et~al.}(2011)Muzyk, Nguyen-Manh, Kurzyd\l{}owski, Baluc,
  and Dudarev]{Muzyk_2011}
M.~Muzyk, D.~Nguyen-Manh, K.~J. Kurzyd\l{}owski, N.~L. Baluc and S.~L. Dudarev,
  \emph{Phys. Rev. B}, 2011, \textbf{84}, 104115\relax
\mciteBstWouldAddEndPuncttrue
\mciteSetBstMidEndSepPunct{\mcitedefaultmidpunct}
{\mcitedefaultendpunct}{\mcitedefaultseppunct}\relax
\EndOfBibitem
\bibitem[Wang \emph{et~al.}(2022)Wang, Kavanagh, Burgués-Ceballos, Walsh,
  Scanlon, and Konstantatos]{wang_cation_2022}
Y.~Wang, S.~R. Kavanagh, I.~Burgués-Ceballos, A.~Walsh, D.~O. Scanlon and
  G.~Konstantatos, \emph{Nat. Photon.}, 2022, \textbf{16}, 235--241\relax
\mciteBstWouldAddEndPuncttrue
\mciteSetBstMidEndSepPunct{\mcitedefaultmidpunct}
{\mcitedefaultendpunct}{\mcitedefaultseppunct}\relax
\EndOfBibitem
\bibitem[Williford \emph{et~al.}(1999)Williford, Weber, Devanathan, and
  Gale]{williford_effects_1999}
R.~Williford, W.~Weber, R.~Devanathan and J.~Gale, \emph{J. Electroceramics},
  1999, \textbf{3}, 409--424\relax
\mciteBstWouldAddEndPuncttrue
\mciteSetBstMidEndSepPunct{\mcitedefaultmidpunct}
{\mcitedefaultendpunct}{\mcitedefaultseppunct}\relax
\EndOfBibitem
\bibitem[Quadir \emph{et~al.}(2022)Quadir, Qorbani, Sabbah, Wu, Anbalagan,
  Chen, Meledath~Valiyaveettil, Thong, Wang, Chen, Lee, Chen, and
  Chen]{Quadir_2022}
S.~Quadir, M.~Qorbani, A.~Sabbah, T.-S. Wu, A.~k. Anbalagan, W.-T. Chen,
  S.~Meledath~Valiyaveettil, H.-T. Thong, C.-W. Wang, C.-Y. Chen, C.-H. Lee,
  K.-H. Chen and L.-C. Chen, \emph{Chem. Mater.}, 2022, \textbf{34},
  7058--7068\relax
\mciteBstWouldAddEndPuncttrue
\mciteSetBstMidEndSepPunct{\mcitedefaultmidpunct}
{\mcitedefaultendpunct}{\mcitedefaultseppunct}\relax
\EndOfBibitem
\bibitem[Morrow \emph{et~al.}(2023)Morrow, Ugwumadu, Drabold, Elliott, Goodwin,
  and Deringer]{morrow_understanding_2023}
J.~D. Morrow, C.~Ugwumadu, D.~A. Drabold, S.~R. Elliott, A.~L. Goodwin and
  V.~L. Deringer, 2023, arXiv:cond-mat/2308.16868\relax
\mciteBstWouldAddEndPuncttrue
\mciteSetBstMidEndSepPunct{\mcitedefaultmidpunct}
{\mcitedefaultendpunct}{\mcitedefaultseppunct}\relax
\EndOfBibitem
\bibitem[Riebesell \emph{et~al.}(2023)Riebesell, Goodall, Jain, Benner,
  Persson, and Lee]{riebesell2023matbench}
J.~Riebesell, R.~E.~A. Goodall, A.~Jain, P.~Benner, K.~A. Persson and A.~A.
  Lee, 2023, arXiv:cond-mat/2308.14920\relax
\mciteBstWouldAddEndPuncttrue
\mciteSetBstMidEndSepPunct{\mcitedefaultmidpunct}
{\mcitedefaultendpunct}{\mcitedefaultseppunct}\relax
\EndOfBibitem
\bibitem[Shimizu \emph{et~al.}(2022)Shimizu, Dou, Arguelles, Moriya,
  Minamitani, and Watanabe]{shimizu_using_2022}
K.~Shimizu, Y.~Dou, E.~F. Arguelles, T.~Moriya, E.~Minamitani and S.~Watanabe,
  \emph{Phys. Rev. B}, 2022, \textbf{106}, 054108\relax
\mciteBstWouldAddEndPuncttrue
\mciteSetBstMidEndSepPunct{\mcitedefaultmidpunct}
{\mcitedefaultendpunct}{\mcitedefaultseppunct}\relax
\EndOfBibitem
\bibitem[Ko \emph{et~al.}(2021)Ko, Finkler, Goedecker, and
  Behler]{ko_general-purpose_2021}
T.~W. Ko, J.~A. Finkler, S.~Goedecker and J.~Behler, \emph{Acc. Chem. Res.},
  2021, \textbf{54}, 808--817\relax
\mciteBstWouldAddEndPuncttrue
\mciteSetBstMidEndSepPunct{\mcitedefaultmidpunct}
{\mcitedefaultendpunct}{\mcitedefaultseppunct}\relax
\EndOfBibitem
\bibitem[Kavanagh \emph{et~al.}(2022)Kavanagh, Scanlon, Walsh, and
  Freysoldt]{Kavanagh2022}
S.~R. Kavanagh, D.~O. Scanlon, A.~Walsh and C.~Freysoldt, \emph{Faraday
  Discuss.}, 2022, \textbf{239}, 339–356\relax
\mciteBstWouldAddEndPuncttrue
\mciteSetBstMidEndSepPunct{\mcitedefaultmidpunct}
{\mcitedefaultendpunct}{\mcitedefaultseppunct}\relax
\EndOfBibitem
\bibitem[Mosquera-Lois \emph{et~al.}(2023)Mosquera-Lois, Kavanagh, Klarbring,
  Tolborg, and Walsh]{MosqueraLois2023}
I.~Mosquera-Lois, S.~R. Kavanagh, J.~Klarbring, K.~Tolborg and A.~Walsh,
  \emph{Chem. Soc. Rev.}, 2023, \textbf{52}, 5812–5826\relax
\mciteBstWouldAddEndPuncttrue
\mciteSetBstMidEndSepPunct{\mcitedefaultmidpunct}
{\mcitedefaultendpunct}{\mcitedefaultseppunct}\relax
\EndOfBibitem
\bibitem[Pols \emph{et~al.}(2023)Pols, Brouwers, Calero, and
  Tao]{pols_how_2023}
M.~Pols, V.~Brouwers, S.~Calero and S.~Tao, \emph{Chem. Commun.}, 2023,
  \textbf{59}, 4660--4663\relax
\mciteBstWouldAddEndPuncttrue
\mciteSetBstMidEndSepPunct{\mcitedefaultmidpunct}
{\mcitedefaultendpunct}{\mcitedefaultseppunct}\relax
\EndOfBibitem
\bibitem[Freysoldt \emph{et~al.}(2014)Freysoldt, Grabowski, Hickel, Neugebauer,
  Kresse, Janotti, and Van~de Walle]{Freysoldt_2014}
C.~Freysoldt, B.~Grabowski, T.~Hickel, J.~Neugebauer, G.~Kresse, A.~Janotti and
  C.~G. Van~de Walle, \emph{Rev. Mod. Phys.}, 2014, \textbf{86}, 253--305\relax
\mciteBstWouldAddEndPuncttrue
\mciteSetBstMidEndSepPunct{\mcitedefaultmidpunct}
{\mcitedefaultendpunct}{\mcitedefaultseppunct}\relax
\EndOfBibitem
\bibitem[Lany and Zunger(2008)]{Zunger_2008_assessment}
S.~Lany and A.~Zunger, \emph{Phys. Rev. B}, 2008, \textbf{78}, 235104\relax
\mciteBstWouldAddEndPuncttrue
\mciteSetBstMidEndSepPunct{\mcitedefaultmidpunct}
{\mcitedefaultendpunct}{\mcitedefaultseppunct}\relax
\EndOfBibitem
\bibitem[Heyd \emph{et~al.}(2003)Heyd, Scuseria, and Ernzerhof]{Heyd2003}
J.~Heyd, G.~E. Scuseria and M.~Ernzerhof, \emph{J. Chem. Phys.}, 2003,
  \textbf{118}, 8207--8215\relax
\mciteBstWouldAddEndPuncttrue
\mciteSetBstMidEndSepPunct{\mcitedefaultmidpunct}
{\mcitedefaultendpunct}{\mcitedefaultseppunct}\relax
\EndOfBibitem
\bibitem[Kresse and Furthmüller(1996)]{Kresse_1996}
G.~Kresse and J.~Furthmüller, \emph{Comput. Mater. Sci.}, 1996, \textbf{6},
  15--50\relax
\mciteBstWouldAddEndPuncttrue
\mciteSetBstMidEndSepPunct{\mcitedefaultmidpunct}
{\mcitedefaultendpunct}{\mcitedefaultseppunct}\relax
\EndOfBibitem
\bibitem[Kresse and Hafner(1993)]{Kresse_1993}
G.~Kresse and J.~Hafner, \emph{Phys. Rev. B}, 1993, \textbf{47}, 558--561\relax
\mciteBstWouldAddEndPuncttrue
\mciteSetBstMidEndSepPunct{\mcitedefaultmidpunct}
{\mcitedefaultendpunct}{\mcitedefaultseppunct}\relax
\EndOfBibitem
\bibitem[Kresse and Hafner(1994)]{Kresse_1994}
G.~Kresse and J.~Hafner, \emph{Phys. Rev. B}, 1994, \textbf{49},
  14251--14269\relax
\mciteBstWouldAddEndPuncttrue
\mciteSetBstMidEndSepPunct{\mcitedefaultmidpunct}
{\mcitedefaultendpunct}{\mcitedefaultseppunct}\relax
\EndOfBibitem
\bibitem[Pizzi \emph{et~al.}(2016)Pizzi, Cepellotti, Sabatini, Marzari, and
  Kozinsky]{Pizzi2016}
G.~Pizzi, A.~Cepellotti, R.~Sabatini, N.~Marzari and B.~Kozinsky, \emph{Comput.
  Mater. Sci.}, 2016, \textbf{111}, 218--230\relax
\mciteBstWouldAddEndPuncttrue
\mciteSetBstMidEndSepPunct{\mcitedefaultmidpunct}
{\mcitedefaultendpunct}{\mcitedefaultseppunct}\relax
\EndOfBibitem
\bibitem[Uhrin \emph{et~al.}(2021)Uhrin, Huber, Yu, Marzari, and
  Pizzi]{Uhrin2021}
M.~Uhrin, S.~P. Huber, J.~Yu, N.~Marzari and G.~Pizzi, \emph{Comput. Mater.
  Sci.}, 2021, \textbf{187}, 110086\relax
\mciteBstWouldAddEndPuncttrue
\mciteSetBstMidEndSepPunct{\mcitedefaultmidpunct}
{\mcitedefaultendpunct}{\mcitedefaultseppunct}\relax
\EndOfBibitem
\bibitem[Huber \emph{et~al.}(2020)Huber, Zoupanos, Uhrin, Talirz, Kahle,
  H\"{a}uselmann, Gresch, M\"{u}ller, Yakutovich, Andersen, Ramirez, Adorf,
  Gargiulo, Kumbhar, Passaro, Johnston, Merkys, Cepellotti, Mounet, Marzari,
  Kozinsky, and Pizzi]{Huber2020}
S.~P. Huber, S.~Zoupanos, M.~Uhrin, L.~Talirz, L.~Kahle, R.~H\"{a}uselmann,
  D.~Gresch, T.~M\"{u}ller, A.~V. Yakutovich, C.~W. Andersen, F.~F. Ramirez,
  C.~S. Adorf, F.~Gargiulo, S.~Kumbhar, E.~Passaro, C.~Johnston, A.~Merkys,
  A.~Cepellotti, N.~Mounet, N.~Marzari, B.~Kozinsky and G.~Pizzi, \emph{Sci.
  Data}, 2020, \textbf{7}, 300\relax
\mciteBstWouldAddEndPuncttrue
\mciteSetBstMidEndSepPunct{\mcitedefaultmidpunct}
{\mcitedefaultendpunct}{\mcitedefaultseppunct}\relax
\EndOfBibitem
\bibitem[Ong \emph{et~al.}(2013)Ong, Richards, Jain, Hautier, Kocher, Cholia,
  Gunter, Chevrier, Persson, and Ceder]{Ong2013}
S.~P. Ong, W.~D. Richards, A.~Jain, G.~Hautier, M.~Kocher, S.~Cholia,
  D.~Gunter, V.~L. Chevrier, K.~A. Persson and G.~Ceder, \emph{Comput. Mater.
  Sci.}, 2013, \textbf{68}, 314--319\relax
\mciteBstWouldAddEndPuncttrue
\mciteSetBstMidEndSepPunct{\mcitedefaultmidpunct}
{\mcitedefaultendpunct}{\mcitedefaultseppunct}\relax
\EndOfBibitem
\bibitem[Jain \emph{et~al.}(2013)Jain, Ong, Hautier, Chen, Richards, Dacek,
  Cholia, Gunter, Skinner, Ceder, and Persson]{Jain2013}
A.~Jain, S.~P. Ong, G.~Hautier, W.~Chen, W.~D. Richards, S.~Dacek, S.~Cholia,
  D.~Gunter, D.~Skinner, G.~Ceder and K.~A. Persson, \emph{{APL} Materials},
  2013, \textbf{1}, 011002\relax
\mciteBstWouldAddEndPuncttrue
\mciteSetBstMidEndSepPunct{\mcitedefaultmidpunct}
{\mcitedefaultendpunct}{\mcitedefaultseppunct}\relax
\EndOfBibitem
\bibitem[Ong \emph{et~al.}(2015)Ong, Cholia, Jain, Brafman, Gunter, Ceder, and
  Persson]{Ong2015}
S.~P. Ong, S.~Cholia, A.~Jain, M.~Brafman, D.~Gunter, G.~Ceder and K.~A.
  Persson, \emph{Comput. Mater. Sci.}, 2015, \textbf{97}, 209--215\relax
\mciteBstWouldAddEndPuncttrue
\mciteSetBstMidEndSepPunct{\mcitedefaultmidpunct}
{\mcitedefaultendpunct}{\mcitedefaultseppunct}\relax
\EndOfBibitem
\bibitem[Shen and Varley()]{pymatgen_analysis_defects}
J.-X. Shen and J.~B. Varley, pymatgen-analysis-defects (version 2023.10.19 ),
  2023\relax
\mciteBstWouldAddEndPuncttrue
\mciteSetBstMidEndSepPunct{\mcitedefaultmidpunct}
{\mcitedefaultendpunct}{\mcitedefaultseppunct}\relax
\EndOfBibitem
\bibitem[Larsen \emph{et~al.}(2017)Larsen, Mortensen, Blomqvist, Castelli,
  Christensen, Dułak, Friis, Groves, Hammer, Hargus, Hermes, Jennings, Jensen,
  Kermode, Kitchin, Kolsbjerg, Kubal, Kaasbjerg, Lysgaard, Maronsson, Maxson,
  Olsen, Pastewka, Peterson, Rostgaard, Schiøtz, Schütt, Strange, Thygesen,
  Vegge, Vilhelmsen, Walter, Zeng, and Jacobsen]{ase-paper}
A.~H. Larsen, J.~J. Mortensen, J.~Blomqvist, I.~E. Castelli, R.~Christensen,
  M.~Dułak, J.~Friis, M.~N. Groves, B.~Hammer, C.~Hargus, E.~D. Hermes, P.~C.
  Jennings, P.~B. Jensen, J.~Kermode, J.~R. Kitchin, E.~L. Kolsbjerg, J.~Kubal,
  K.~Kaasbjerg, S.~Lysgaard, J.~B. Maronsson, T.~Maxson, T.~Olsen, L.~Pastewka,
  A.~Peterson, C.~Rostgaard, J.~Schiøtz, O.~Schütt, M.~Strange, K.~S.
  Thygesen, T.~Vegge, L.~Vilhelmsen, M.~Walter, Z.~Zeng and K.~W. Jacobsen,
  \emph{J. Condens. Matter Phys.}, 2017, \textbf{29}, 273002\relax
\mciteBstWouldAddEndPuncttrue
\mciteSetBstMidEndSepPunct{\mcitedefaultmidpunct}
{\mcitedefaultendpunct}{\mcitedefaultseppunct}\relax
\EndOfBibitem
\bibitem[Kavanagh \emph{et~al.}()Kavanagh, Brlec, Zhu, Nicolson, Hachmioune,
  Aggarwal, Walsh, and Scanlon]{doped}
S.~R. Kavanagh, K.~Brlec, B.~Zhu, A.~Nicolson, S.~Hachmioune, S.~Aggarwal,
  A.~Walsh and D.~O. Scanlon, Defect Oriented Python Environment Distribution
  ({DOPED}) (version 1.1.2), 2023\relax
\mciteBstWouldAddEndPuncttrue
\mciteSetBstMidEndSepPunct{\mcitedefaultmidpunct}
{\mcitedefaultendpunct}{\mcitedefaultseppunct}\relax
\EndOfBibitem
\bibitem[Bitzek \emph{et~al.}(2006)Bitzek, Koskinen, G\"ahler, Moseler, and
  Gumbsch]{Bitzek_2006}
E.~Bitzek, P.~Koskinen, F.~G\"ahler, M.~Moseler and P.~Gumbsch, \emph{Phys.
  Rev. Lett.}, 2006, \textbf{97}, 170201\relax
\mciteBstWouldAddEndPuncttrue
\mciteSetBstMidEndSepPunct{\mcitedefaultmidpunct}
{\mcitedefaultendpunct}{\mcitedefaultseppunct}\relax
\EndOfBibitem
\bibitem[Morrow \emph{et~al.}(2023)Morrow, Gardner, and
  Deringer]{Morrow2023_how}
J.~D. Morrow, J.~L.~A. Gardner and V.~L. Deringer, \emph{J. Chem. Phys.}, 2023,
  \textbf{158}, 121501\relax
\mciteBstWouldAddEndPuncttrue
\mciteSetBstMidEndSepPunct{\mcitedefaultmidpunct}
{\mcitedefaultendpunct}{\mcitedefaultseppunct}\relax
\EndOfBibitem
\end{mcitethebibliography}

% Include SI after main
\clearpage
\begin{center}
\textbf{Supplementary Information for `Machine-learning structural reconstructions for accelerated point defect calculations'}
\end{center}

% Figure with S
\renewcommand\thefigure{S\arabic{figure}}    
\setcounter{figure}{0}
\renewcommand\thetable{S\arabic{table}}  
\setcounter{table}{0}
\renewcommand\thesection{S\arabic{section}}  
\setcounter{section}{0}

\section{Dataset analysis}

\begin{table}[ht]
\caption{Percentage of defects that lead to energy-lowering reconstructions (compared to a typical unperturbed relaxation) higher than the specified energy threshold. The column `Reconstructions found with rattling' indicates the percentage of reconstructions identified if we only apply randomised distortions to the unperturbed, ideal defect structure (i.e. no bond distortions), demonstrating the need for proper structure searching. }\label{stab:energy_lowerings}
    \begin{tabular}{ccc}
    \hline
    \thead{Threshold \\(eV)} & 
    \thead{Significant \\reconstructions (\%)} &
    \thead{Reconstructions\\ found with rattling (\%)}\\
    \hline
    -0.05 & 40.3 & 3.7 \\
    -0.10 & 33.6 & 0 \\
    -0.50 & 29.9 & 0 \\
    -1.00 & 20.9 & 0 \\
    -2.00 & 6.7  & 0 \\
    \hline
    \end{tabular}
\end{table}

\begin{figure}[ht]
\centering\includegraphics[width=0.7\textwidth]{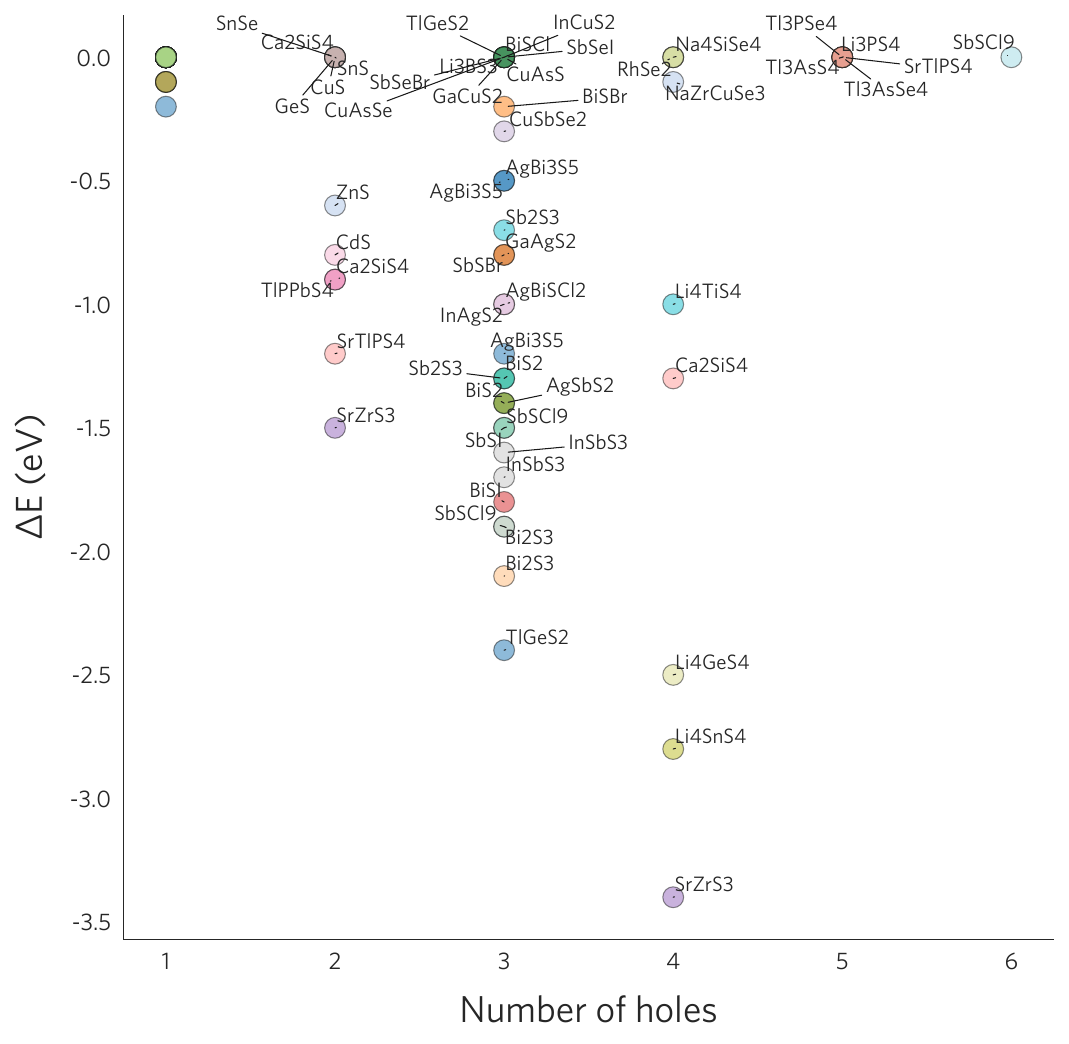}
    \caption{Energy difference between the ground state structure identified with ShakeNBreak and the final structure obtained by relaxing the ideal, high-symmetry defect structure for cation vacancies across each chalcogenide. Note that all defects with more than 3 missing electrons form dimer reconstructions, but some appear with $\Delta E = 0~\mathrm{eV}$ since the ground state structure is already identified from the high-symmetry relaxation.}
    \label{sfig:delta_E_vs_holes}
\end{figure}

\FloatBarrier

%\clearpage
\subsection{Defect reconstructions}

\begin{figure}[h!]
\centering\includegraphics[width=0.75\textwidth]{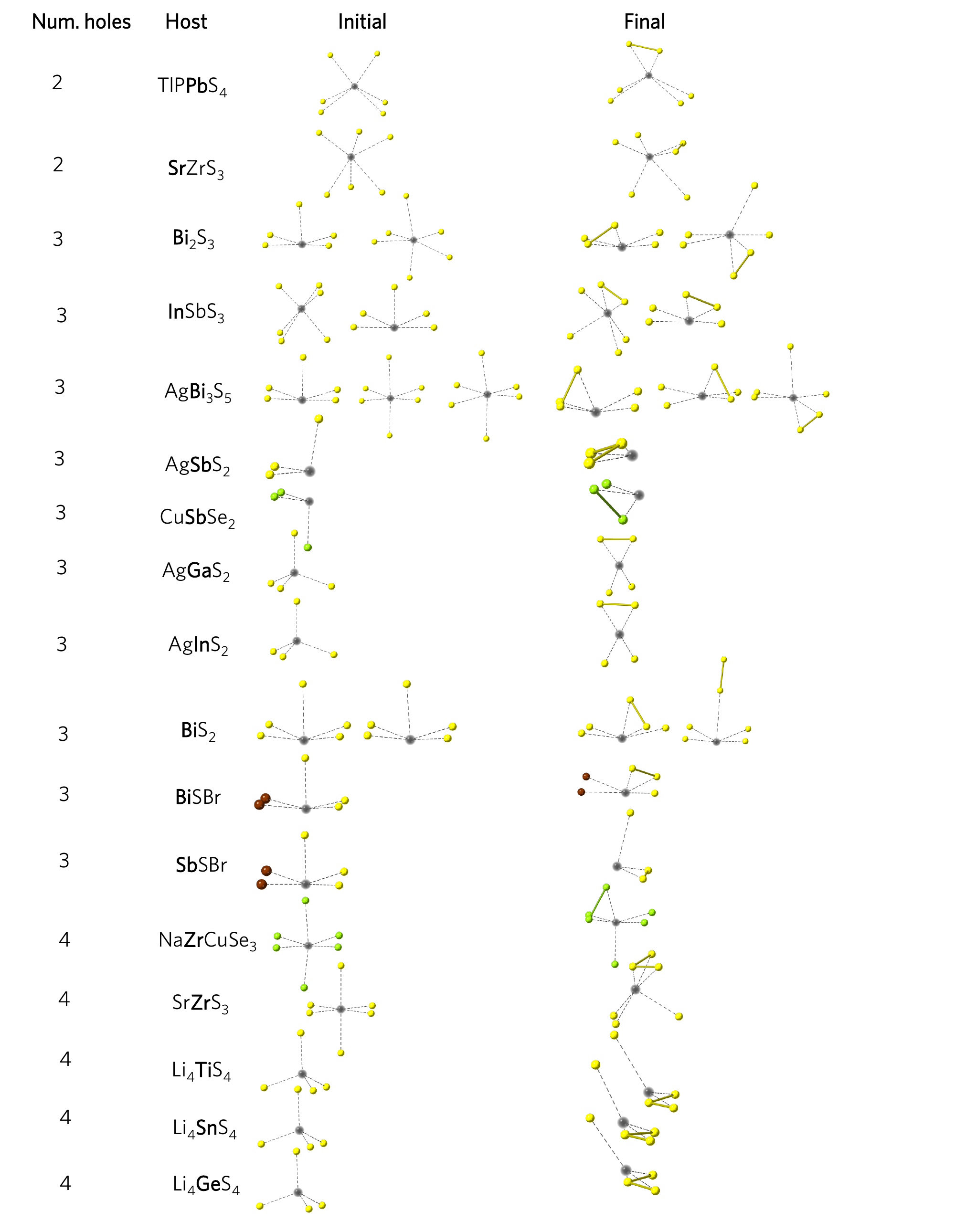}
    \caption{Illustration of some of the reconstructions identified, showing the similar motifs undergone by different hosts. The initial (high-symmetry) and the ground state structures are shown. The defect element is shown in bold in the host composition.}
    \label{sfig:reconstructions}
\end{figure}
\begin{figure}[ht]
\centering\includegraphics[width=0.95\textwidth]{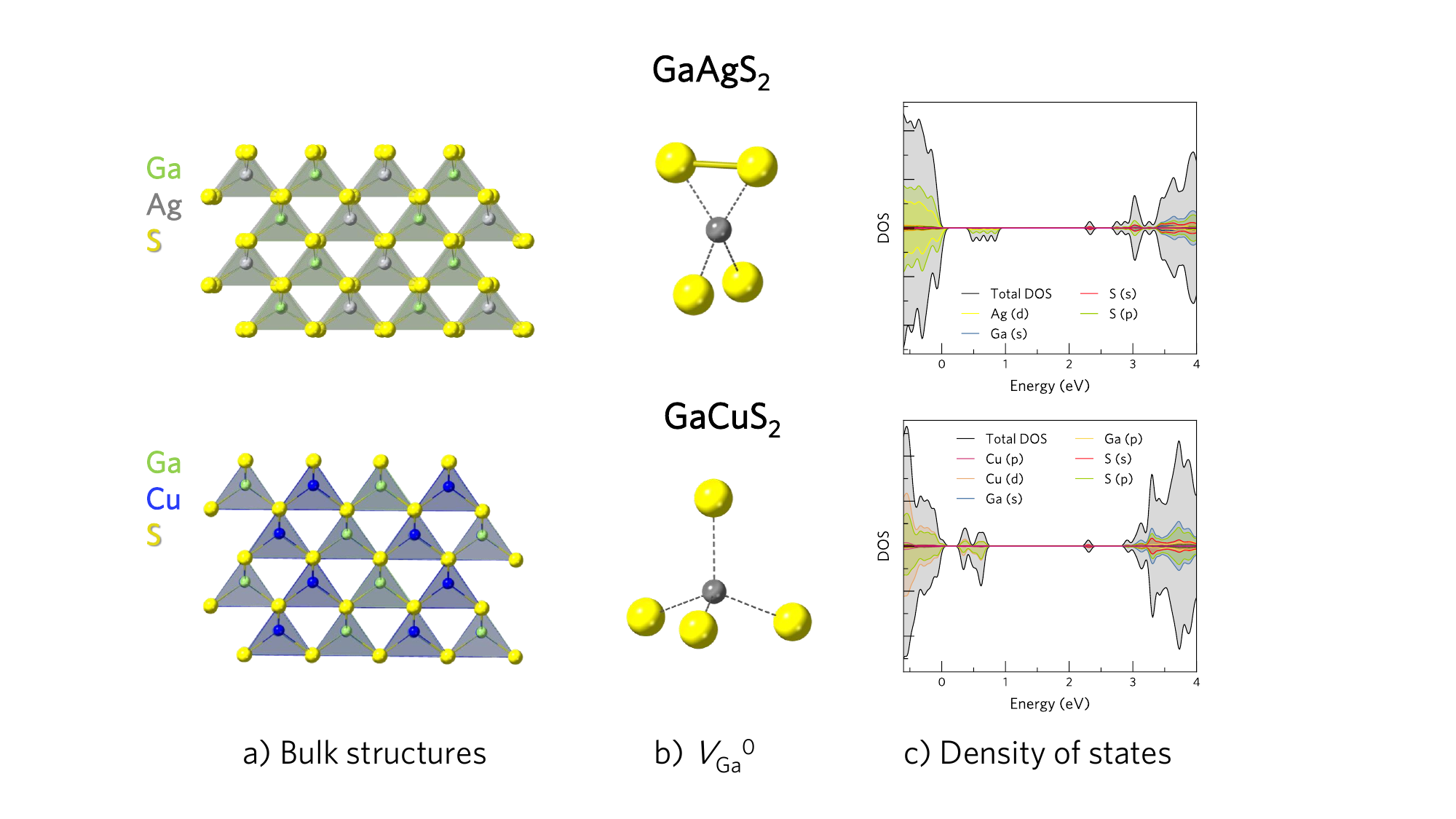}
    \caption{Reconstructions undergone by \kvc{V}{Ga}{0} in \ce{GaAgS2} and \ce{GaCuS2}. a) Bulk structures for \ce{GaAgS2} (top) and \ce{GaCuS2} (bottom). b) \kvc{V}{Ga}{0}, showing the dimerisation undergone in \ce{GaAgS2} while in \ce{GaCuS2} \kvc{V}{Ga}{0} maintains the ideal tetrahedral coordination. Similar behaviour is observed for \kvc{V}{In}{0} in \ce{InAg/CuS2}. c) Density of states for \kvc{V}{Ga}{0}, showing the localised defect states within the gap.}
    \label{sfig:V_Ga_GaAgS2}
\end{figure}
%

%
%\begin{figure}[ht]
%\centering\includegraphics[width=0.5\textwidth]{Figures/Cu_Ag_bandgaps.pdf}
%    \caption{Diagram of the electronic structure of bulk \ce{GaAgS2} and \ce{GaCuS2} illustrating the positions of the valence and conduction bands (VB and CB, respectively), as well as the state associated to a S--S dimer. The higher position of the VBM in \ce{GaCuS2} reduces the energy gain of localising a hole at the dimer state, making the S--S dimer distortion unfavourable for this system.}
%    \label{sfig:V_Ga_GaAgS2_bandgaps}
%\end{figure}
%

%
\begin{figure}[ht]
\centering\includegraphics[width=0.8\textwidth]{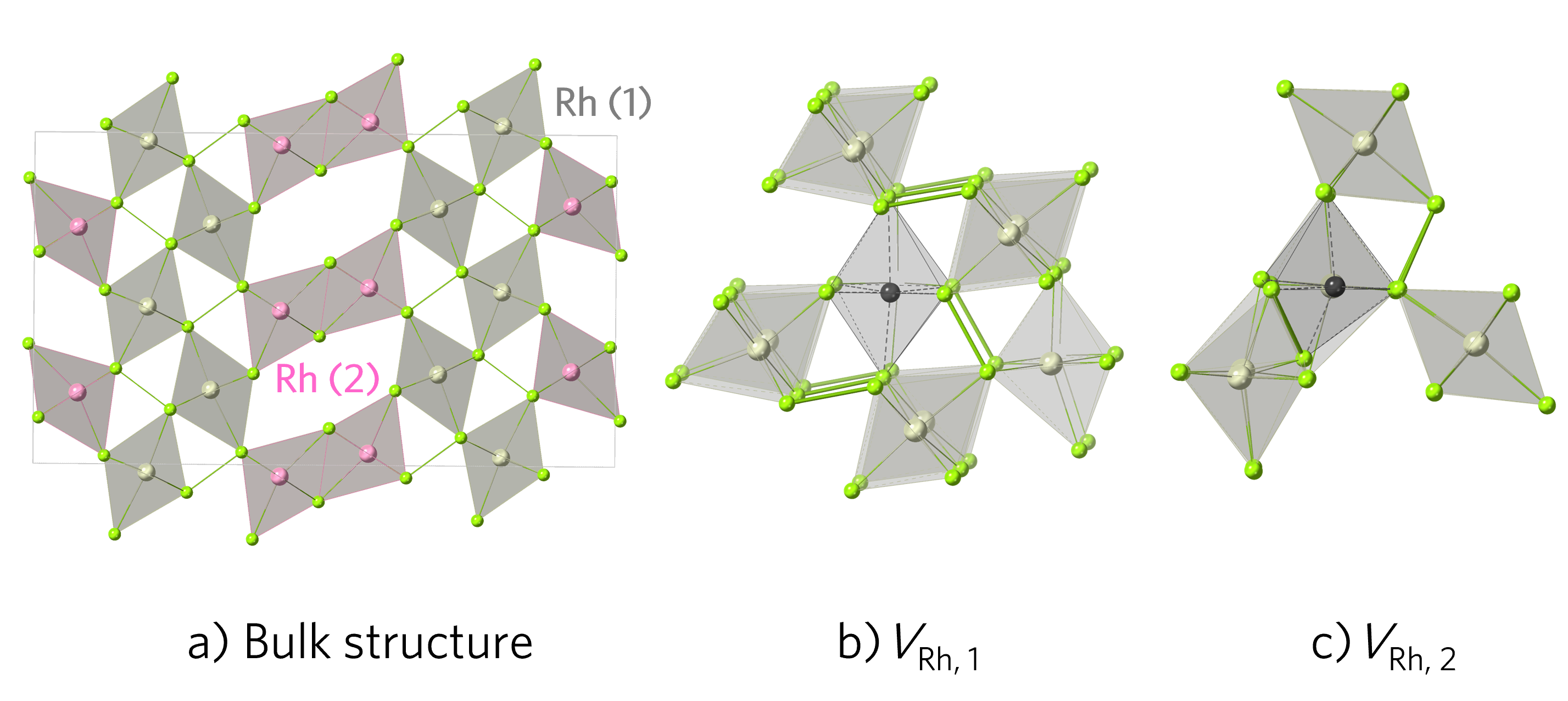}
    \caption{Reconstructions undergone by the symmetry inequivalent Rh vacancies in \ce{RhSe2}. a) Bulk structure showing the two symmetry inequivalent Rh sites, in grey (Rh(1)) and pink (Rh(2)). b) \kv{V}{Rh,1} does not form a Se dimer since the holes can localise in the Se-Se bonds already neighbouring this site. c) \kv{V}{Rh,2} forms an additional Se-Se bond as it lacks enough neighbouring Se-Se bonds to localise the four holes.}
    \label{sfig:RhSe2_reconstructions}
\end{figure}
\begin{figure}[ht]
\centering\includegraphics[width=0.65\textwidth]{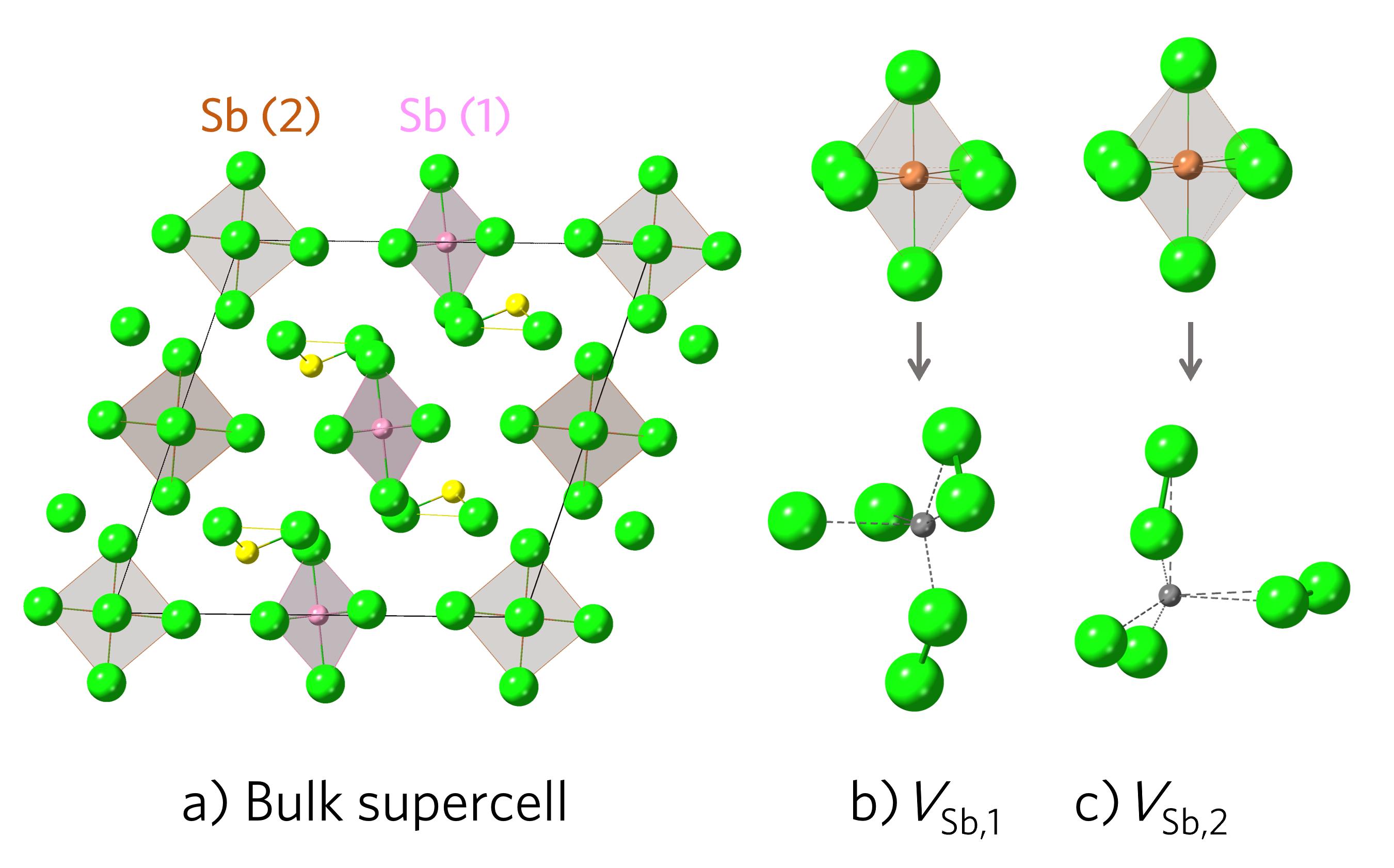}
    \caption{Reconstructions undergone by the symmetry inequivalent Sb vacancies in \ce{SbSCl9}. a) Bulk structure showing the two symmetry inequivalent Sb sites, in pink (Sb(1)) and brown (Sb(2)). b, c) \kv{V}{Sb,1} and \kv{V}{Sb,2} forming two Cl-Cl dimers.}
    \label{sfig:SbSCl9_reconstructions}
\end{figure}

\FloatBarrier
\section{Model training}\label{ssec:training}
In this section, we compare different training strategies and parameters. We note that errors are higher than in the final model since here most of the parameters were set to default values except for the parameter being benchmarked. Further, in these experiments, the validation set only contained unseen compositions (and not unseen configurations from compositions included in the training set).
% \FloatBarrier
\subsection{Splits}
\begin{figure}[ht]
\centering\includegraphics[width=0.55\textwidth]{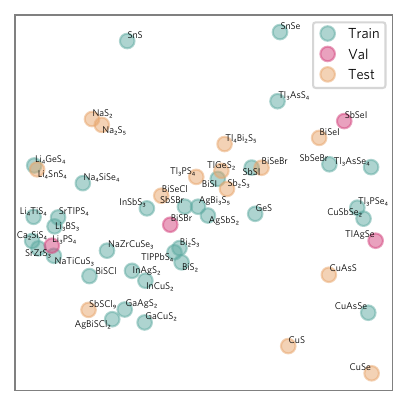}
    \caption{Plot of the first two principal components of the feature space for the pristine structures, showing the division into training, validation, and test sets. We note that the validation set was then augmented with some configurations from the compositions in the training set. The structures were encoded with the 128-element vector outputs from the M3GNet model trained on the formation energies of  bulk materials in the Materials Project database, as done in Ref.~\citenum{Qi_2023}.}
    \label{sfig:split_2d_proj}
\end{figure}

The dataset was first split by composition into training, composition, and test sets, as shown in \ref{sfig:split_2d_proj}.

\subsection{Reference energies}
Before fitting most MLFFs, it is common to subtract the reference elemental energies from the total energies to improve training stability\cite{Chen2022}. In M3GNet\cite{Chen2022}, the reference elemental energies can be calculated from the training or other user-specified data using linear regression\cite{Chen2022}. 
To compare the effect of the reference elemental energies, we considered three different approaches to calculate them: i) energies of the isolated atoms,  ii) linear regression\footnote{The linear regression was performed using the \texttt{AtomRef} class from Ref.~\citenum{Chen2022}.} to the bulk energies, and iii) linear regression to the energies of all training defect structures. As expected, the last case decreases performance since the regressed reference energies include defect formation contributions. On the other hand, when the reference elemental energies are calculated from the bulk systems, these constitute a robust reference that simplifies learning the energies of different defect structures.

\begin{table}[ht]
\caption{Comparison of model performance depending on the elemental reference energies used. The same training and validation data and training parameters are used for all cases and the performance is measured with the mean absolute errors of the energies, forces, and stresses.}\label{stab:ref_energies}
\vspace{10pt}
\begin{tabular}{lrrrrrrr}
%\toprule
\hline
 Method & 
 $\rm Loss_{val}$ & 
 \makecell{$\rm E_{val}$ \\(meV/atom)} & 
 \makecell{$\rm F_{val}$ \\(meV/\AA)} & 
 \makecell{$\rm S_{val}$ \\(GPa)} & 
 \makecell{$\rm E_{train}$ \\(meV/atom)} & 
 \makecell{$\rm F_{train}$ \\(meV/\AA)} & 
 \makecell{$\rm S_{train}$ \\(GPa)} \\
% \midrule
\hline
Bulk & 0.22 & 41.8 & 153.8 & 0.22 & 50.50 & 118.0 & 0.19 \\
Isolated atoms & 0.31 & 72.4 & 191.6 & 0.43 & 134.5 & 154.3 & 0.30 \\
Defect & 0.34 & 130.3 & 179.3 & 0.34 & 109.6 & 158.2 & 0.28 \\
\hline
%\bottomrule
\end{tabular}
\end{table}

\FloatBarrier
\subsection{Sampling methods}
\begin{figure}[ht]
    %\centering
    \begin{subfigure}[c]{0.40\textwidth}
        \includegraphics[width=\textwidth]{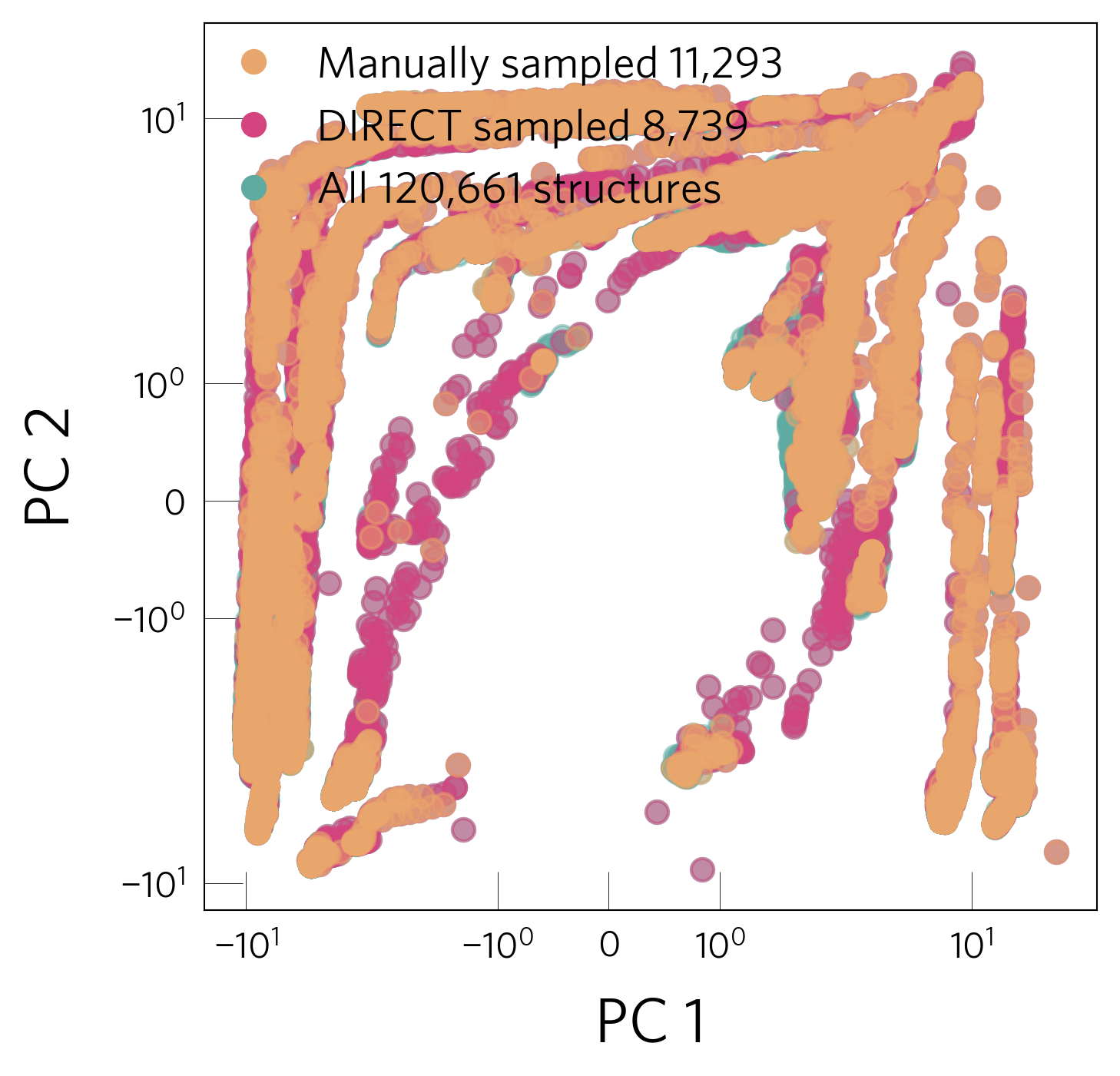}
        \caption{DIRECT and manual coverage}
    \end{subfigure}
    \hspace{5pt}
    \begin{subfigure}[c]{0.4\textwidth}
        \includegraphics[width=\textwidth]{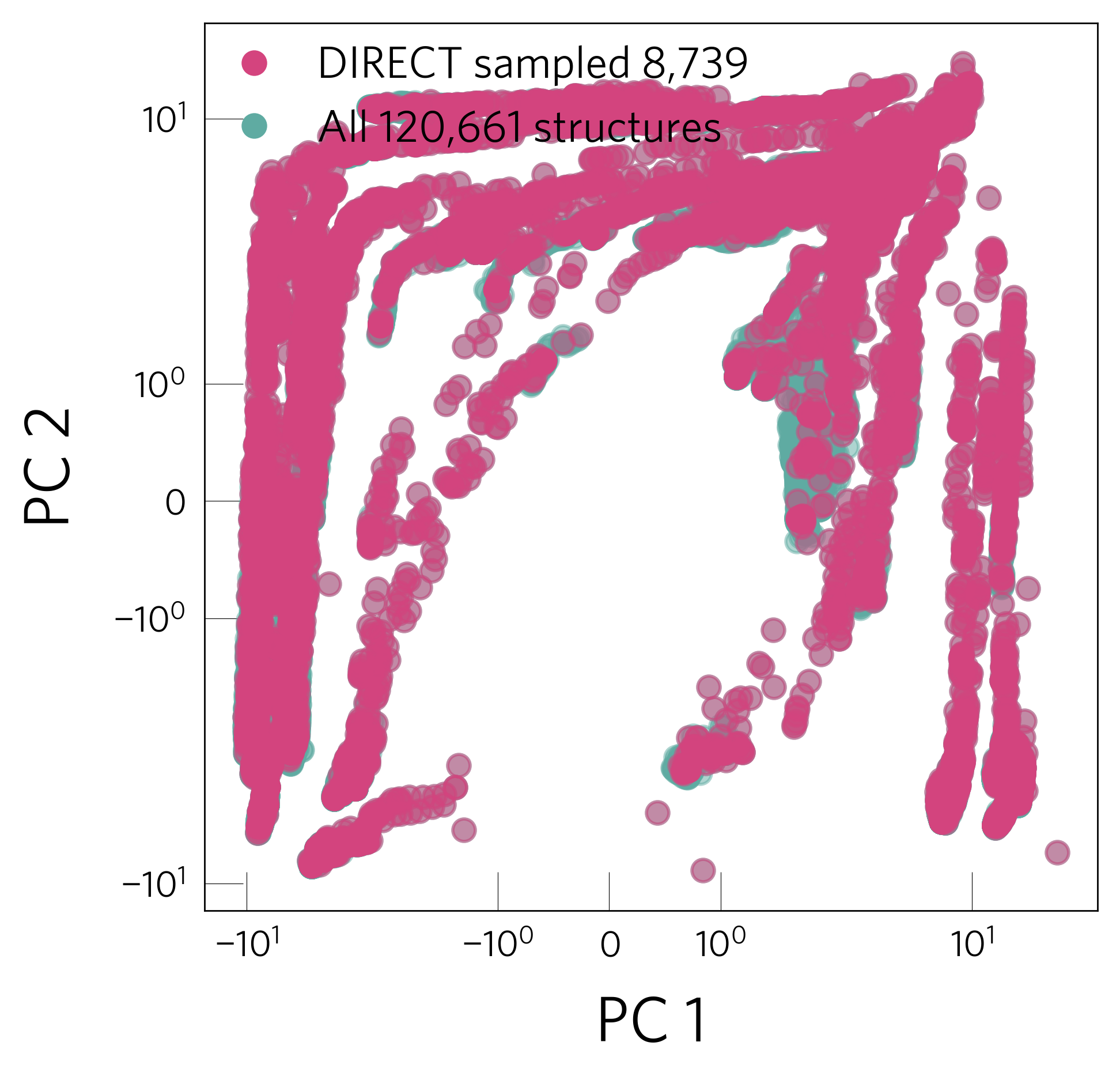}
        \caption{DIRECT only (for clarity)}
    \end{subfigure}
    \caption{Sampling coverage of manual (orange) and DIRECT sampling (pink) of the full dataset (green), illustrating that the DIRECT approach covers better the configurational landscape. For the DIRECT method, we considered the parameters t=0.02, n=4000, k=20, but similar behaviour is observed for the other cases of \ref{stab:sampling_methods}. 
    Note that the dataset is only plotted along two dimensions (see \ref{sfig:pca_explained}). 
    The structures were encoded with the 128-element vector outputs from the M3GNet model trained on the formation energies of bulk materials in the Materials Project database\cite{Qi_2023}. For further details of the method used to featurise structures, dimensionality reduction, clustering, and stratified sampling see the original publication of Qi \emph{et al}\cite{Qi_2023}.}\label{sfig:direct_coverage}
\end{figure}

\begin{figure}[ht]
\includegraphics[width=0.75\textwidth]{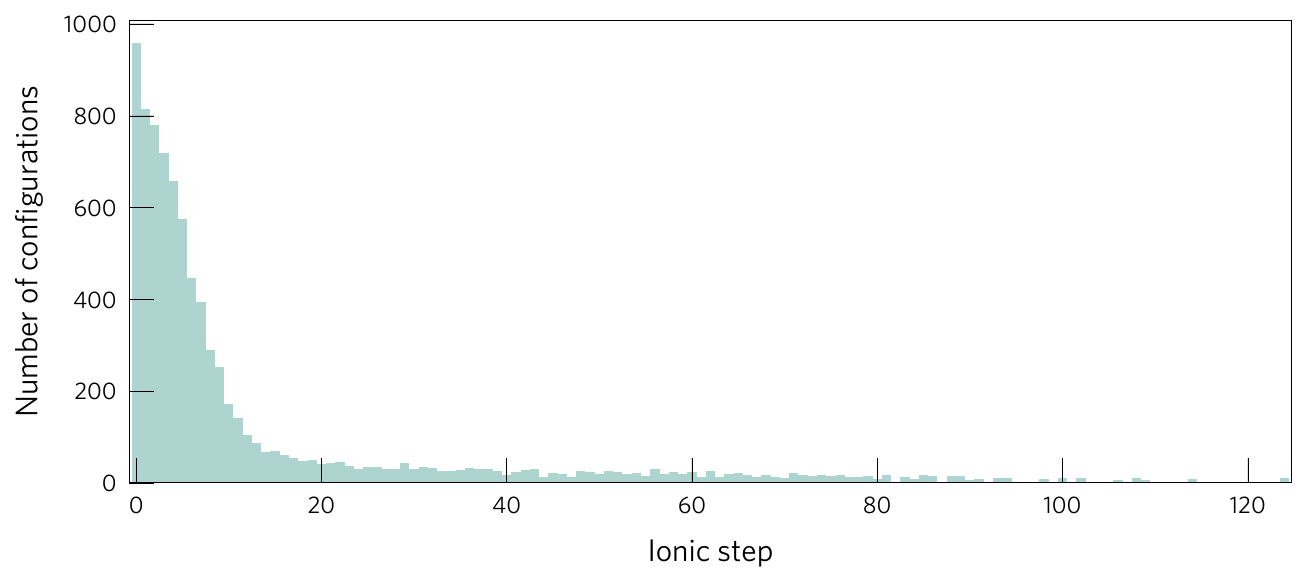}
\caption{Distribution of DIRECT (t=0.02, n=4000, k=20) sampled configurations, showing that structures are mainly selected from the initial ionic steps, which correspond to highly distorted structures.}\label{sfig:direct_ionic_step}
\end{figure}

To investigate how to best sample configurations from the relaxation trajectories to generate the training set we compared two approaches: i) a manual method where we sample 10 evenly spaced frames from each relaxation (MS) and ii) the Dimensionality-Reduced Encoded Clusters approach (DIRECT)\cite{Qi_2023}, which aims to select a robust training set from a complex configurational space (\ref{sfig:direct_coverage}). This comparison was performed without including configurations of the bulk (pristine) host structures. After generating the different training sets, four M3GNet models\cite{Chen2022} with default parameters were trained and evaluated. As shown in \ref{stab:sampling_methods}, when using training sets of similar sizes, the manual model performs better. This results from the DIRECT approach mostly sampling highly distorted configurations from the initial ionic steps (\ref{sfig:direct_ionic_step}), hindering the learning of the low-energy region of the potential energy surface (PES).

\begin{figure}[ht]
\includegraphics[width=0.55\textwidth]{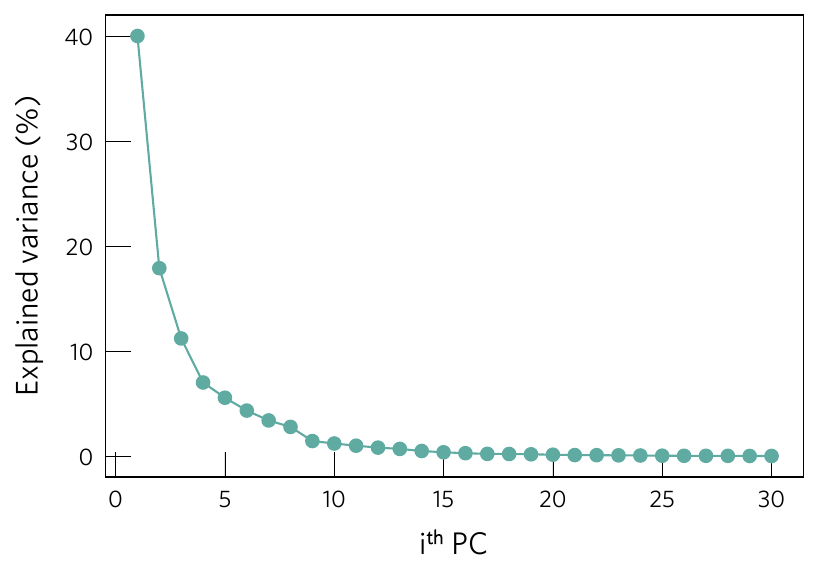}
\caption{Explained variance versus the number of PCA dimensions. PCA was used to reduce the dimensions of the feature vector used to describe the defect configurations (see Ref.~\citenum{Qi_2023} for details).}\label{sfig:pca_explained}
\end{figure}
\begin{table}[ht]
\caption{Comparison of the mean absolute errors in the validation and training sets obtained with the different methods of sampling the training data. The same validation data and training parameters are used for all cases. The column `Params.' lists the sampling parameters used with the BIRCH clustering method, as detailed in Ref.~\citenum{Qi_2023}.}\label{stab:sampling_methods}
\vspace{10pt}
\begin{tabular}{lrcrrrrrrrr}
%\toprule
\hline
 Method & Size & \makecell{Params.\\ (t, n, k)} &$\rm Loss_{val}$ & \thead{$\rm E_{val}$ \\(meV/atom)} & \makecell{$\rm F_{val}$ \\(meV/\AA)} & \thead{$\rm S_{val}$ \\(GPa)} & \thead{$\rm E_{train}$ \\(meV/atom)} & \thead{$\rm F_{train}$ \\(meV/\AA)} & \thead{$\rm S_{train}$ \\(GPa)} \\
 %\toprule
\hline
Direct & 21149 & 0.02, 4000, 20 &0.20 & 28.6 & 148.6 & 0.19 & 38.1 & 147.6 & 0.19 \\
% \hline
Manual & 11335 & - & 0.22 & 51.4 & 146.7 & 0.26 & 62.0 & 123.3 & 0.23 \\
% \hline
Direct & 8705 & 0.1, 1000, 20 & 0.25 & 47.0	& 174.7	& 0.33	&42.2	&177.3	&0.23 \\
% \hline
Direct & 13144 & 0.03, 2000, 20 &0.28 & 93.4 & 165.3 & 0.26 & 54.9 & 192.3 & 0.26 \\
\hline
%\bottomrule
\end{tabular}
\end{table}
\vspace{5pt}

\begin{figure}[ht]
    %\centering
    \includegraphics[width=\textwidth]{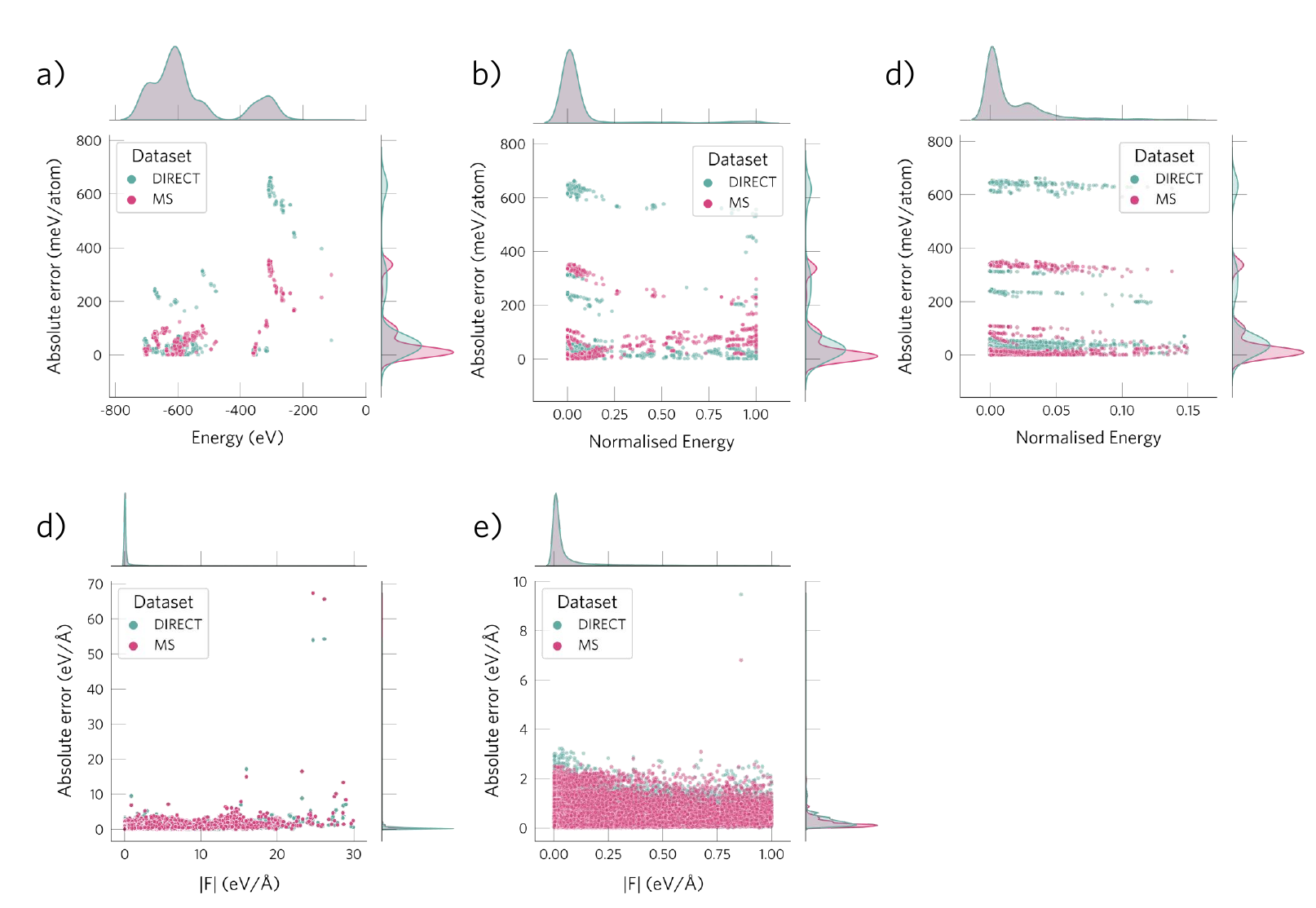}
    \caption{
    Distribution of the energy (a, b, c) and force (d, e) absolute errors for the test set. The MS sampling results in lower errors for the low-energy structures while the DIRECT approach leads to higher errors for these structures. a) Distribution of absolute energy errors versus the absolute DFT energy for all test compositions. b) Same as (a) but with the DFT energies normalised for each defect, so that energies of different defects/compositions are comparable. c) Same as (b) but removing very high energy configurations to aid visualisation. d) Distribution of absolute force errors versus the magnitude of the DFT forces. Same as (c) but removing configurations with very high forces. We note that the configurations with very high errors correspond to the most challenging compositions (more different from the training set, like the quasi-one-dimensional system \ce{SbSCl9}).
    }\label{sfig:distribution_mae_sampling}
\end{figure}

\FloatBarrier
\subsection{Model architecture}\label{ssubsec:models}
To compare different model architectures, we trained from scratch (i.e. initialising the weights with random values) three models: M3GNet\cite{Chen2022}, MACE\cite{Batatia_2022} and CHGNet\cite{Deng2023}. Again, for simplicity, this comparison was performed with default parameters for each model. The same training, validation, and test sets were used to compare the different models. The worse performance of CHGNet compared to the other models seems to arise from two reasons: i) the smaller value of the 3-body cutoff that is used by default (\SI{3}{\angstrom} compared to \SI{4}{\angstrom} in M3GNet) and ii) its default readout or pooling layer, which involves an average function (with the attention layer performing better). Regarding M3GNet and MACE, MACE seems to perform better on the test set. However, due to the lack of a pre-trained MACE universal model at the time when this comparison was performed, we decided to use M3GNet. However, with the now available MACE universal model, we expect the MACE architecture to perform better. 

\begin{table}[ht]
\caption{Comparison of validation and test mean absolute errors for different models (M3GNet, CHGNet, MACE). For the test set, the Spearman coefficient for the energies ($\rho$) is also shown. The default parameters of each model were used and the models were trained from scratch (without fine-tuning/pretraining). 
In all cases, the training data included energies, forces and stresses. 
}\label{stab:model_comparison}
\vspace{10pt}
\begin{tabular}{lrrrrrr}
    \hline
    Model & \thead{$\rm E_{val}$ \\(meV/atom)} & \makecell{$\rm F_{val}$ \\(meV/\AA)} 
    %& \thead{$\rm S_{val}$ \\(GPa)} 
    & \thead{$\rm E_{test}$ \\(meV/atom)} & \makecell{$\rm F_{test}$ \\(meV/\AA)} & $\rho_{\rm test}$\\
    \hline
    M3GNet & 45.8 & 143.1 
        %& 0.19 
        & 129.2 & 263.0 & 0.44\\
    MACE\footnotemark & 47.8 & 173.9 
        %& 1.34 
        & 153.7 & 187.8 & 0.66\\
    CHGNet & 192.3 & 351.4 
        %& 3.0 
        & 376.0 &  333.8 & 0.37\\
    \hline
\end{tabular}
\vskip 1em
\footnotetext[1]{For MACE, the following parameters were used: hidden irreducible representations: 32x0e, radial cutoff: \SI{5}{\angstrom}; weight for energies, forces and stresses: 1, 1, 0.1; learning rate scheduler: exponential; batch size: 20; maximum number of epochs: 250.}
\end{table}

\FloatBarrier

\subsection{Training parameters for M3GNet model}\label{ssubsec:params}
To determine the best training parameters for the M3GNet model, we performed several benchmarks by comparing training and validation errors. From these comparisons, we determined the best learning rate, radial and 3-body cutoffs and readout function. We note that these comparisons were performed on the training and validation errors, rather than test ones, as they were intended as quick benchmarks to determine appropriate training parameters. Further, note that the errors are higher than for the final model since the default values were used for the parameters not being tested in each experiment.
\subsubsection{Learning rate}
\begin{table}[ht]
\caption{Comparison of validation mean absolute errors for different learning rates and learning rate schedulers. `Constant' denotes using a constant learning rate. When a scheduler is used, the initial learning rate is given in parentheses. Rows are ordered by ascending validation loss. The same training data (manual sampling) and training parameters were used for all experiments.}\label{stab:learning_rate}
\vspace{10pt}
\begin{tabular}{llrrrrrr}
    \hline
     Method &$\rm Loss_{val}$ & \thead{$\rm E_{val}$ \\(meV/atom)} & \makecell{$\rm F_{val}$ \\(meV/\AA)} & \thead{$\rm S_{val}$ \\(GPa)} & \thead{$\rm E_{train}$ \\(meV/atom)} & \thead{$\rm F_{train}$ \\(meV/\AA)} & \thead{$\rm S_{train}$ \\(GPa)} \\
    \hline
    Exp. scheduler ($5 \cdot10^{-4}$) & 0.19 & 35.7 & 128.0 & 0.22 & 30.9 & 93.0 & 0.16 \\
    Cos. scheduler ($10^{-3}$) & 0.20 & 41.8 & 135.4 & 0.26 & 32.2 & 106.8 & 0.17 \\
    Constant ($5 \cdot10^{-4}$) &  0.21 & 45.0 & 142.6 & 0.19 & 34.1 & 96.2 & 0.16 \\
    Constant ($10^{-3}$) &  0.21 & 45.8 & 143.1 & 0.19 & 36.7 & 97.9 & 0.17 \\
    Exp. scheduler ($10^{-3}$) & 0.21 & 25.8 & 152.9 & 0.36 & 21.8 & 85.70 & 0.12 \\
    Constant ($10^{-4}$) & 0.24 & 58.8 & 152.1 & 0.29 & 56.7 & 131.4 & 0.22 \\
    Time. scheduler ($10^{-3}$) & 0.25 & 64.3 & 152.0 & 0.31 & 35.2 & 101.4 & 0.16 \\
    Constant ($10^{-5}$) & 0.29 & 70.9 & 191.2 & 0.32 & 84.8 & 174.1 & 0.41 \\
    \hline
\end{tabular}
\end{table}

\FloatBarrier
\subsubsection{Structure featurisation}
We performed benchmarks to identify the optimal values for the radial and 3-body cutoff and the optimal function for the pooling layer. As shown in \ref{stab:cutoff_3body}, \ref{stab:cutoff_radial}, \ref{stab:readout}, we found a 3-body cutoff of \SI{4}{\angstrom}, a radial cutoff of \SI{5}{\angstrom}, and the weighted atom layer from the M3GNet model to work best. 

\begin{table}[ht]
\caption{Comparison of cutoff values for the 3-body radius in the M3GNet model. Mean absolute errors for energies, forces, and stresses for the training and validation sets.
Rows are ordered by ascending validation loss.}\label{stab:cutoff_3body}
\vspace{10pt}
\begin{tabular}{crrrrrrr}
\hline
 \makecell{cutoff \\(\AA)} &$\rm Loss_{val}$ & \thead{$\rm E_{val}$ \\(meV/atom)} & \makecell{$\rm F_{val}$ \\(meV/\AA)} & \thead{$\rm S_{val}$ \\(GPa)} & \thead{$\rm E_{train}$ \\(meV/atom)} & \thead{$\rm F_{train}$ \\(meV/\AA)} & \thead{$\rm S_{train}$ \\(GPa)} \\
\hline
4 &  0.19 & 35.7 & 128.0 & 0.22 & 30.9 & 93.0 & 0.16 \\
3 &  0.26 & 59.2 & 164.9 & 0.32 & 50.6 & 123.5 & 0.20 \\
\hline
\end{tabular}
\end{table}
\begin{table}[ht]
\caption{Performance comparison for different radial cutoffs in the M3GNet model. Mean absolute errors for energies, forces, and stresses for the training and validation sets. Rows are ordered by ascending validation loss.}\label{stab:cutoff_radial}
\vspace{10pt}
\begin{tabular}{crrrrrrr}
\hline
 \makecell{cutoff \\(\AA)} &$\rm Loss_{val}$ & \thead{$\rm E_{val}$ \\(meV/atom)} & \makecell{$\rm F_{val}$ \\(meV/\AA)} & \thead{$\rm S_{val}$ \\(GPa)} & \thead{$\rm E_{train}$ \\(meV/atom)} & \thead{$\rm F_{train}$ \\(meV/\AA)} & \thead{$\rm S_{train}$ \\(GPa)} \\
\hline
5 & 0.19 & 35.7 & 128.0 & 0.22 & 30.9 & 93.0 & 0.16 \\
4.5 & 0.22 & 40.1	&142.0	& 0.34	&35.1	&97.4	&0.16\\
5.5 & 0.23 & 57.1 & 140.7 & 0.31 & 46.8 & 114.4 & 0.20 \\
6 & 0.23 & 73.9 & 133.0 & 0.22 & 45.9 & 103.4 & 0.18 \\
\hline
\end{tabular}
\end{table}
\begin{table}[ht]
\caption{Performance comparison for different readout functions in the M3GNet model. Mean absolute errors for energies, forces, and stresses for the training and validation sets. Rows are ordered by ascending validation loss.}\label{stab:readout}
\vspace{10pt}
\begin{tabular}{crrrrrrr}
\hline
 Readout &$\rm Loss_{val}$ & \thead{$\rm E_{val}$ \\(meV/atom)} & \makecell{$\rm F_{val}$ \\(meV/\AA)} & \thead{$\rm S_{val}$ \\(GPa)} & \thead{$\rm E_{train}$ \\(meV/atom)} & \thead{$\rm F_{train}$ \\(meV/\AA)} & \thead{$\rm S_{train}$ \\(GPa)} \\
\hline
Weighted atom &  0.19 & 35.7 & 128.0 & 0.22 & 30.9 & 93.0 & 0.16 \\
Reduce readout &  0.23 & 54.6 & 143.1 & 0.30 & 65.0 & 135.1 & 0.24 \\
Set2Set & 0.26 & 57.7 & 163.9 & 0.39 & 44.3 & 130.9 & 0.2 \\
\hline
\end{tabular}
\end{table}

\FloatBarrier
\subsubsection{Fine-tuning}
To assess the effect of fine-tuning the model from a model trained on the bulk relaxations of the Materials Project database, we compared the performance of four different models: training from scratch (i.e. initialising the weights with random values), or fine-tuning the bulk model (either training all layers or only the last final layers). When comparing their performance on the validation set, we found that fine-tuning the bulk model and training all layers resulted in the best performance, with the main benefit observed for the forces.

\begin{table}[ht]
\caption{Performance comparison between different training strategies: training from scratch (initialising the weights with random values), or fine-tuning the model previously trained on the Materials Project dataset of bulk structures\cite{Chen2022} (with either re-training all layers (all), only the last two layers (2) or only the last layer (1)). The performance is measured with the mean absolute errors in the validation and training sets.
}
\label{stab:fine-tuning}
\begin{tabular}{llrrrrrrr}
\hline
 Training &$\rm Loss_{val}$ & \thead{$\rm E_{val}$ \\(meV/atom)} & \makecell{$\rm F_{val}$ \\(meV/\AA)} & \thead{$\rm S_{val}$ \\(GPa)} & \thead{$\rm E_{train}$ \\(meV/atom)} & \thead{$\rm F_{train}$ \\(meV/\AA)} & \thead{$\rm S_{train}$ \\(GPa)} \\
\hline
Fine-tuning (all) & 0.157 & 42.6 & 101.1 & 0.1 & 18.7 & 68.1 & 0.1 \\
Fine-tuning (1) & 0.176 & 37.6 & 118.8 & 0.2 & 18.3 & 76.0 & 0.1 \\
Fine-tuning (2) & 0.183 & 36.5 & 104.9 & 0.4 & 15.2 & 55.4 & 0.1 \\
From scratch & 0.187 & 46.4 & 121.5 & 0.2 & 27.7 & 86.1 & 0.1 \\
\hline
\end{tabular}
\end{table}

\FloatBarrier
\subsection{Adding bulk data to enhance defect dataset}\label{ssubsec:add_bulk}
\begin{figure}[ht]
    %\centering
    \begin{subfigure}[c]{0.70\textwidth}
        \includegraphics[width=\textwidth]{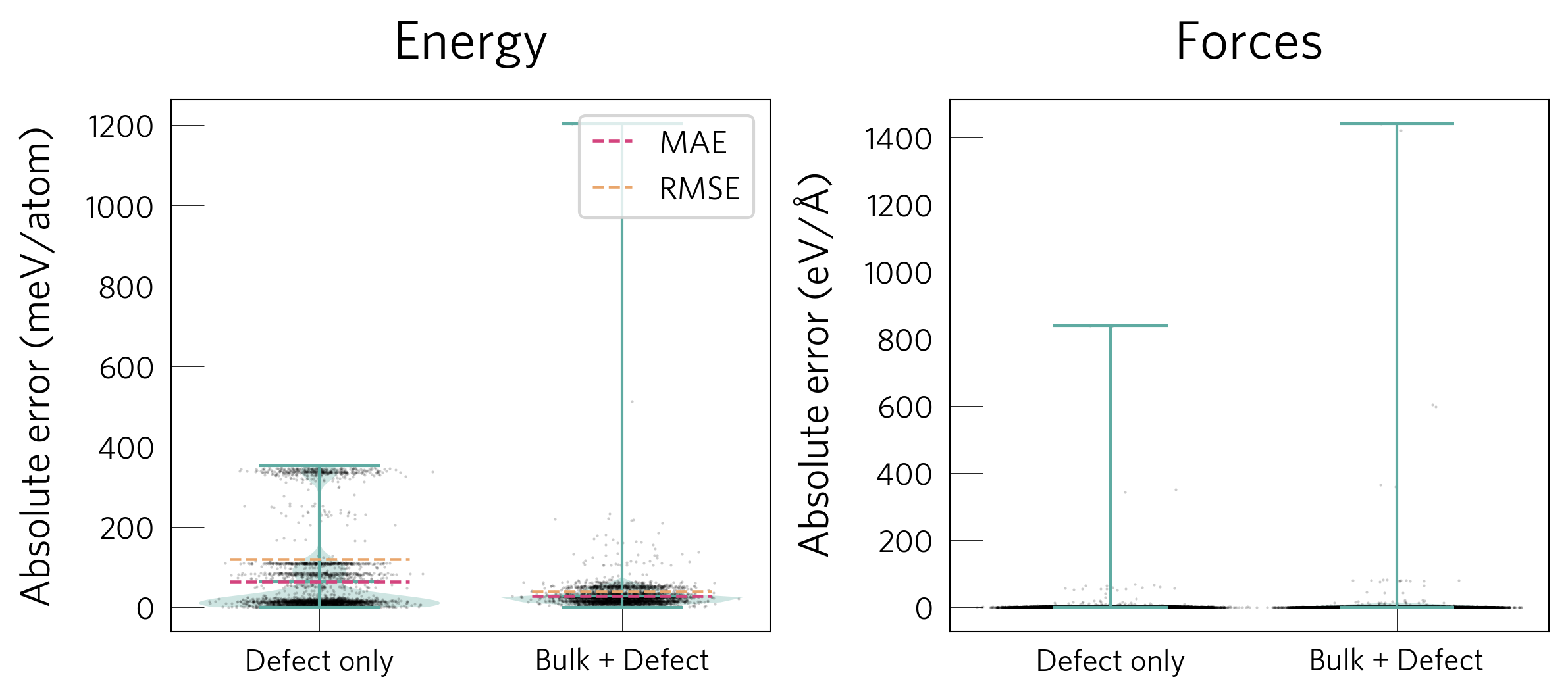}
        \caption{All test systems}
    \end{subfigure}
    \hspace{5pt}
    \begin{subfigure}[c]{0.7\textwidth}
        \includegraphics[width=\textwidth]{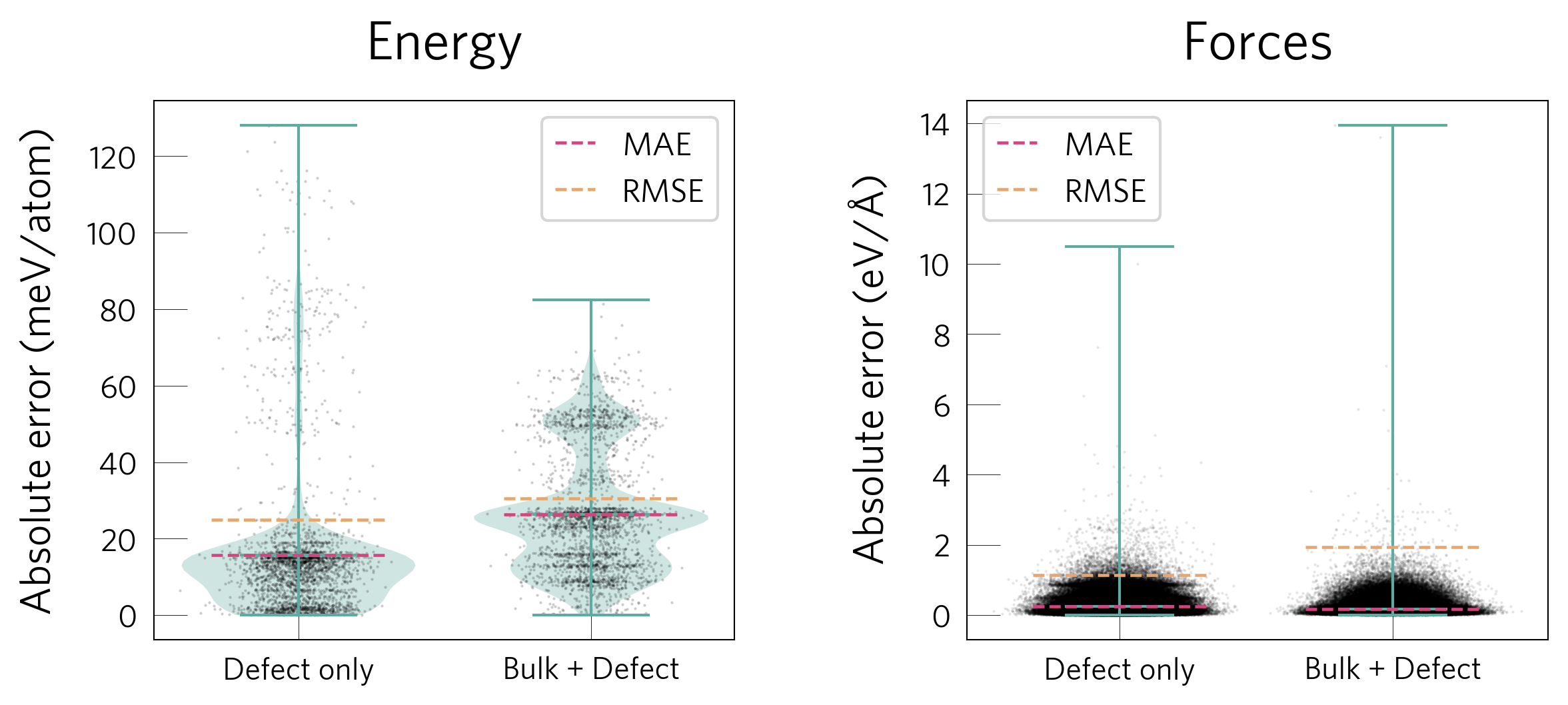}
        \caption{Removing difficult compositions from test set %(\ce{SbSCl9}, \ce{NaS2} and \ce{CuS})
        }
    \end{subfigure}
    \caption{Violin plots\cite{Morrow2023_how} and error metrics for energies and forces on the test set. a) When considering all test systems. b) When filtering out the systems that are hard to learn due to high structural differences to the training set (\ce{SbSCl9}, \ce{NaS2} and \ce{CuS}).}\label{sfig:bulk_data}
\end{figure}

To test whether adding data for pristine systems would improve model performance, we trained two models: one on just the defect dataset (using MS sampling) and another model using the same defect dataset but combined with a small fraction of pristine configurations for the same compositions (5 evenly spaced frames from the relaxation of each pristine structure). As shown in \ref{stab:errors_defect_with_bulk}, adding bulk data reduces the mean absolute errors for the energies and the forces of the test configurations. By analysing the error distributions in \ref{sfig:bulk_data}, we see that the main benefit of including bulk data is to reduce the errors for systems that are difficult due to their higher structural difference from the training set (i.e. including bulk data results in lower MAE and RMSE when considering all test systems (\ref{sfig:bulk_data}.a) but in a higher MAE and RMSE when not considering the harder compositions (\ref{sfig:bulk_data}.b)). 

\begin{table}[ht]
\caption{Performance comparison when adding bulk (pristine) configurations to the training dataset. Mean absolute errors (MAE) and root mean square errors (RMSE) for the (defect) test set when only training on defect data and when training on both defect and bulk data.}\label{stab:errors_defect_with_bulk}
\vspace{10pt}
\begin{tabular}{crrrrrrr}
\hline
Dataset & 
\thead{$\rm MAE_{E}$ \\(meV/atom)}  & 
\thead{$\rm RMSE_{E}$ \\(meV/atom)} &
$\rho$ &
\makecell{$\rm MAE_{F}$ \\(meV/\AA)} & 
\makecell{$\rm RMSE_{F}$ \\(meV/\AA)} & 
\thead{$\rm MAE_{S}$ \\(GPa)} &
\thead{$\rm RMSE_{S}$ \\(GPa)} \\
\hline
Defect + Bulk & 27.3 & 39.4  & 0.70 &  86.8  &  1943.5  & 0.19 & 0.54 \\
Defect only   & 63.4 & 119.5 &    0.63  &  135.0  & 1139.6  & 0.34 & 0.70 \\
\hline 
\end{tabular}
\end{table}

\begin{figure}[ht]
    %\centering
    \includegraphics[width=\linewidth]{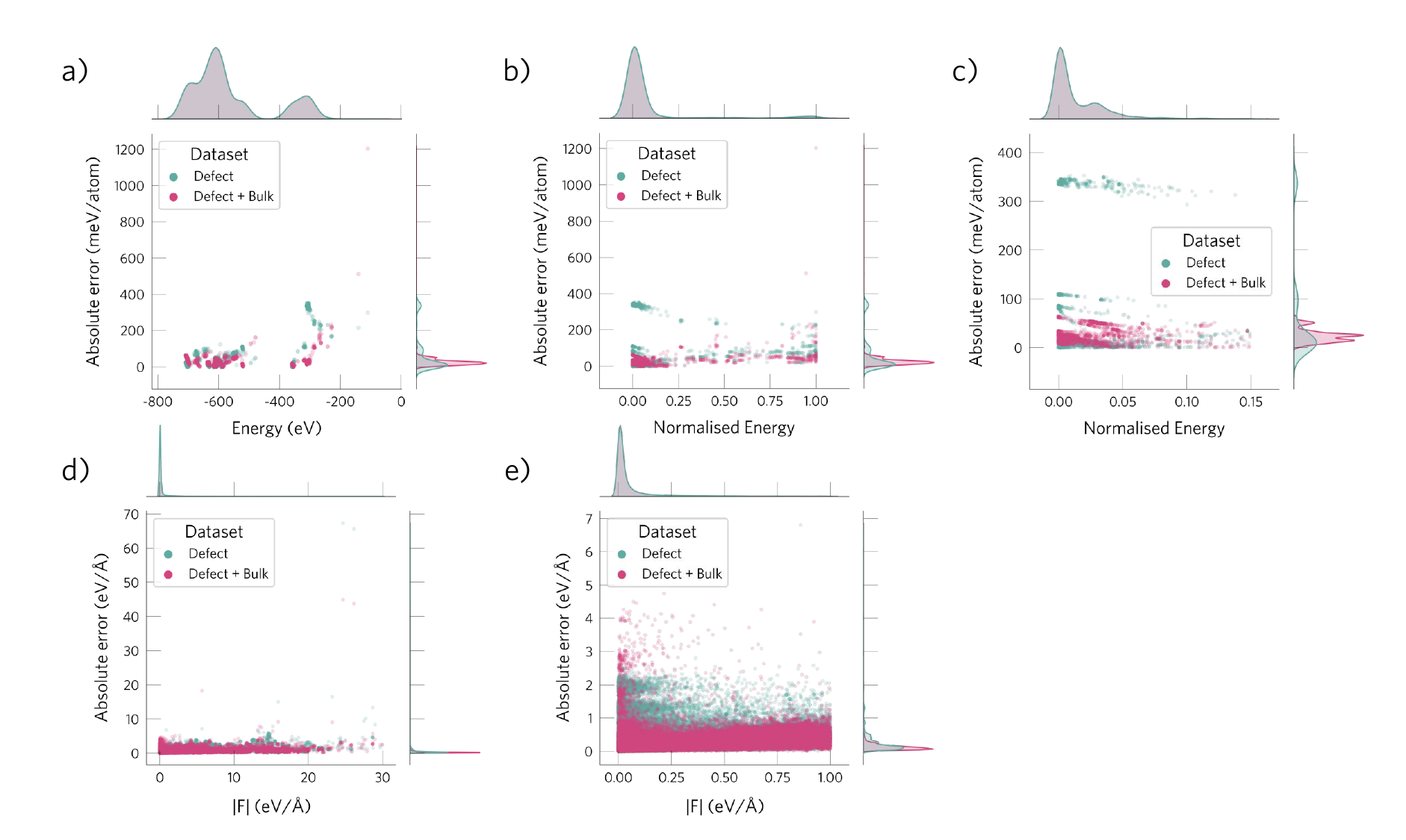}
    \caption{Distribution of absolute errors for energies (a, b, c) and forces (c, d) when training on only defect data (green) and both defect and bulk data (pink). a) Distribution of absolute energy errors versus the absolute DFT energy for all test compositions. b) Same as (a) but with the DFT energies normalised for each defect, so that energies of different defects/compositions are comparable. c) Same as (b) but removing very high energy configurations, illustrating how adding bulk data reduces the errors of many low-energy structures. %(with these \ce{SbSCl9} and \ce{CuS})
    d) Distribution of absolute force errors versus the magnitude of the DFT forces. Same as (c) but removing configurations with very high forces, illustrating how adding bulk data reduces the errors for low force configurations.
    }\label{sfig:bulk_data_error_distribution}
\end{figure}

\begin{figure}[ht]
    %\centering
    \includegraphics[width=\linewidth]{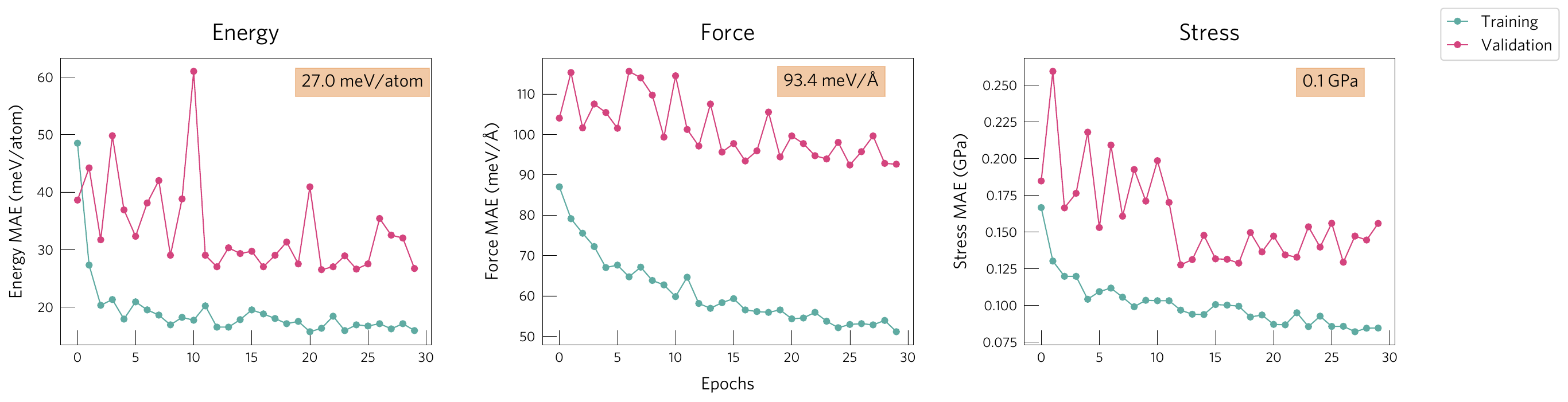}
    \caption{Evolution of training and validation errors with epoch number. The validation errors obtained for the epoch with the lowest validation loss are shown at the top right of each subplot.
    }\label{sfig:training_evolution}
\end{figure}

\FloatBarrier

\FloatBarrier

\section{Model performance}

\subsection{Metrics}

\begin{table}[ht!]
\caption{Performance of the fine-tuned surrogate model on the test set. Mean absolute errors (MAE, in eV and eV/atom) when predicting the relative energies of the low-energy configurations of a defect (i.e. structures that are less than \SI{5}{eV} above the defect ground state configuration). 
}
\setlength{\tabcolsep}{6.5pt} % Default value: 6pt
\renewcommand{\arraystretch}{0.98} % Default value: 1
\begin{tabular}{ccc}
\toprule[1pt]
         & MAE (eV) & MAE (meV/atom) \\ 
\midrule[0.2pt]
\ce{Li4SnS4}  & 1.6      & 11.2           \\
\ce{BiSeBr} & 0.7        & 1.4            \\
\ce{TlGeS2}   & 0.6      & 0.7            \\
\ce{Tl3PS4}   & 0.4      & 1.6            \\
\ce{SbSCl9}  & 0.2      & 9.6            \\
\ce{Tl4Bi2S5} & 0.2      & 5.0            \\
\ce{BiSeCl}  & 0.2      & 1.2            \\
\ce{CuS}      & 0.2      & 1.2            \\
\ce{Na2S5}    & 0.2      & 2.9            \\
\ce{CuAsS}    & 0.1      & 0.9            \\ 
\ce{NaS2} &    0.1 & 2.8 \\
\ce{CuSe} & 0.1 & 1.9 \\
\bottomrule[1pt]
\end{tabular}
\end{table}

\FloatBarrier
\subsection{New ground states: \ce{TlGeS2}}
\begin{figure}[ht]
    %\centering
    \includegraphics[width=0.95\linewidth]{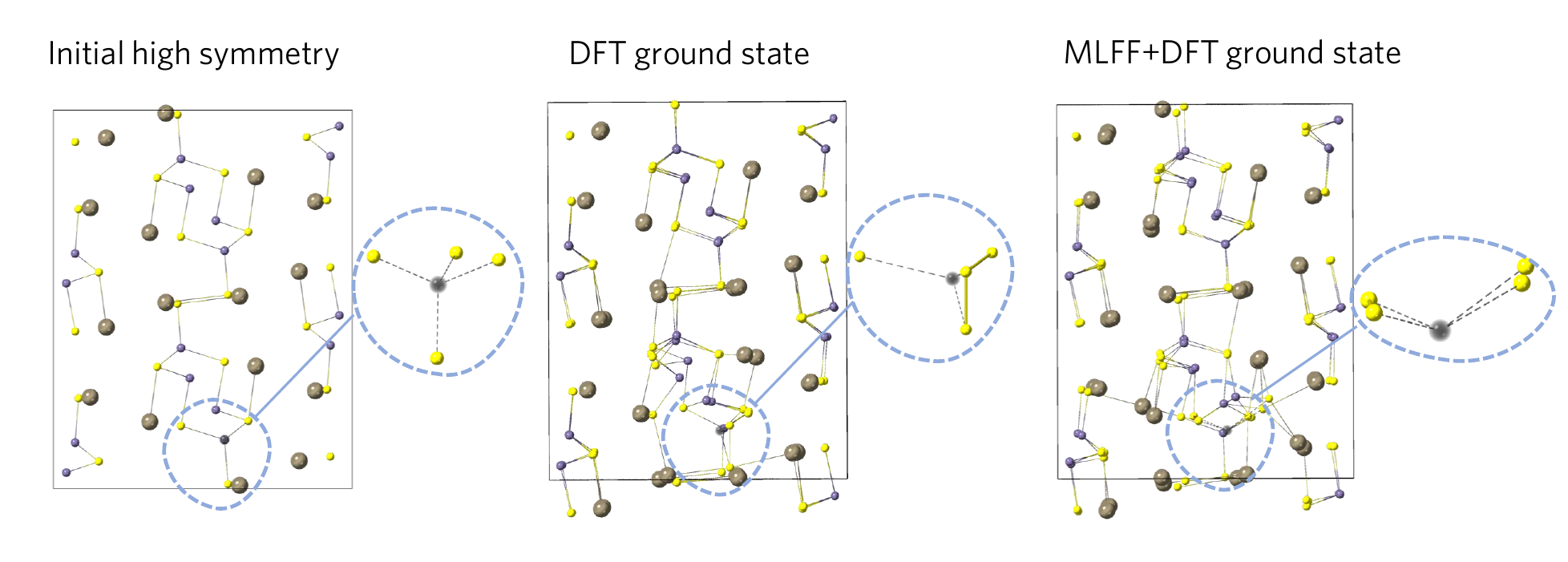}
    \caption{\kv{V}{Ge,9} in \ce{TlGeS2}. Ground state structures identified with the DFT search (S trimer) and with the MLFF+DFT approach (\emph{one} S dimer), with the second being \SI{0.5}{eV} lower in energy. The structure with only one S-S bond seems to be more favourable due to avoiding strain caused by the rearrangement of the atoms. For clarity, the initial high symmetry defect structure is shown on the left.
    }
    \label{sfig:v_Ge_TlGeS2}
\end{figure}
\FloatBarrier
\subsection{Failed systems}
\begin{figure}[ht]
    %\centering
    \includegraphics[width=0.95\linewidth]{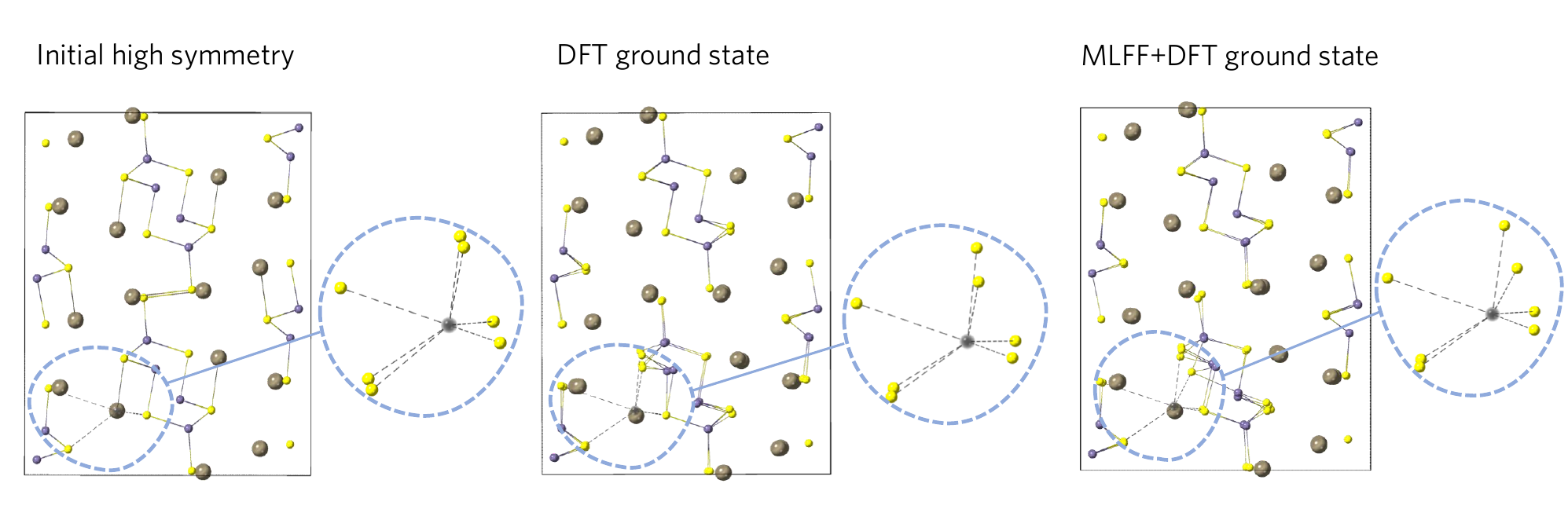}
    \caption{
    High symmetry (initial) and ground state structures of \kv{V}{Tl,0} in \ce{TlGeS2} identified with the DFT-only search and the MLFF+DFT approach. While the configuration identified with the DFT search is \SI{0.1}{eV} lower in energy than the MLFF+DFT structure, we note that the reconstruction motifs are very similar. 
    }
    \label{sfig:TlGeS2_v_Tl0}
\end{figure}
\begin{figure}[ht]
    %\centering
    \includegraphics[width=0.7\linewidth]{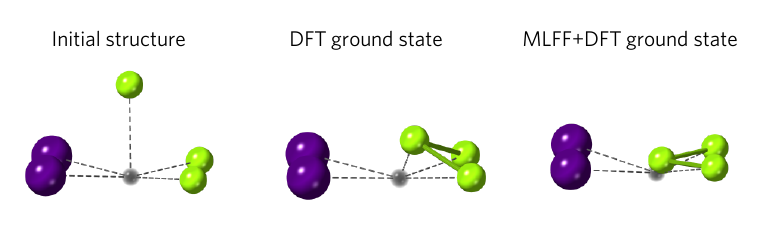}
    \caption{
    High symmetry (initial) and ground state structures of \kv{V}{Bi} in \ce{BiSeI} identified with the DFT-only search and the MLFF+DFT approach. While the configuration identified with the DFT search is \SI{0.2}{eV} lower in energy than the MLFF+DFT structure, we note that the reconstruction motifs are similar, involving a Se trimer in both cases. 
    }
    \label{sfig:BiSeI_v_Bi}
\end{figure}
\begin{figure}[ht]
    %\centering
    \includegraphics[width=0.9\linewidth]{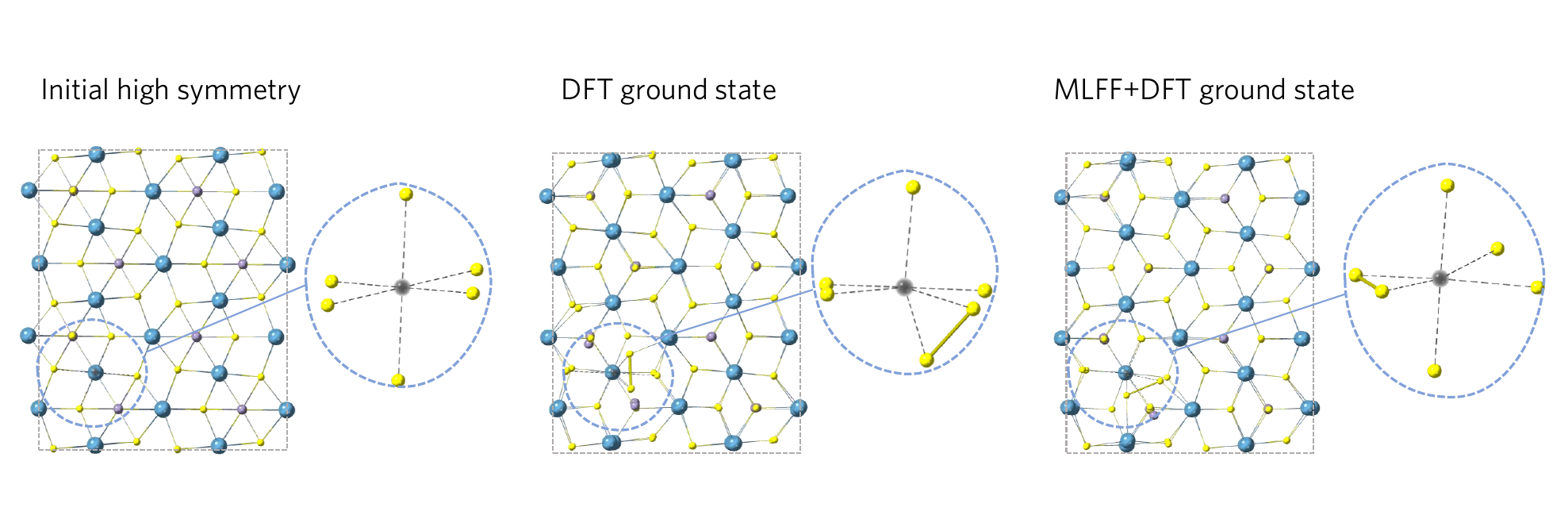}
    \caption{
    High symmetry (initial) and ground state structures of \kv{V}{Ca,0} in \ce{Ca2SnS4} identified with the DFT-only search and the MLFF+DFT approach. While the configuration identified with the DFT search is \SI{0.7}{eV} lower in energy than the MLFF+DFT structure, we note that the reconstruction motifs are similar, involving a S dimer in both cases.  
    }
    \label{sfig:Ca2SnS4_v_Ca}
\end{figure}
\begin{figure}[ht]
    %\centering
    \includegraphics[width=1.0\linewidth]{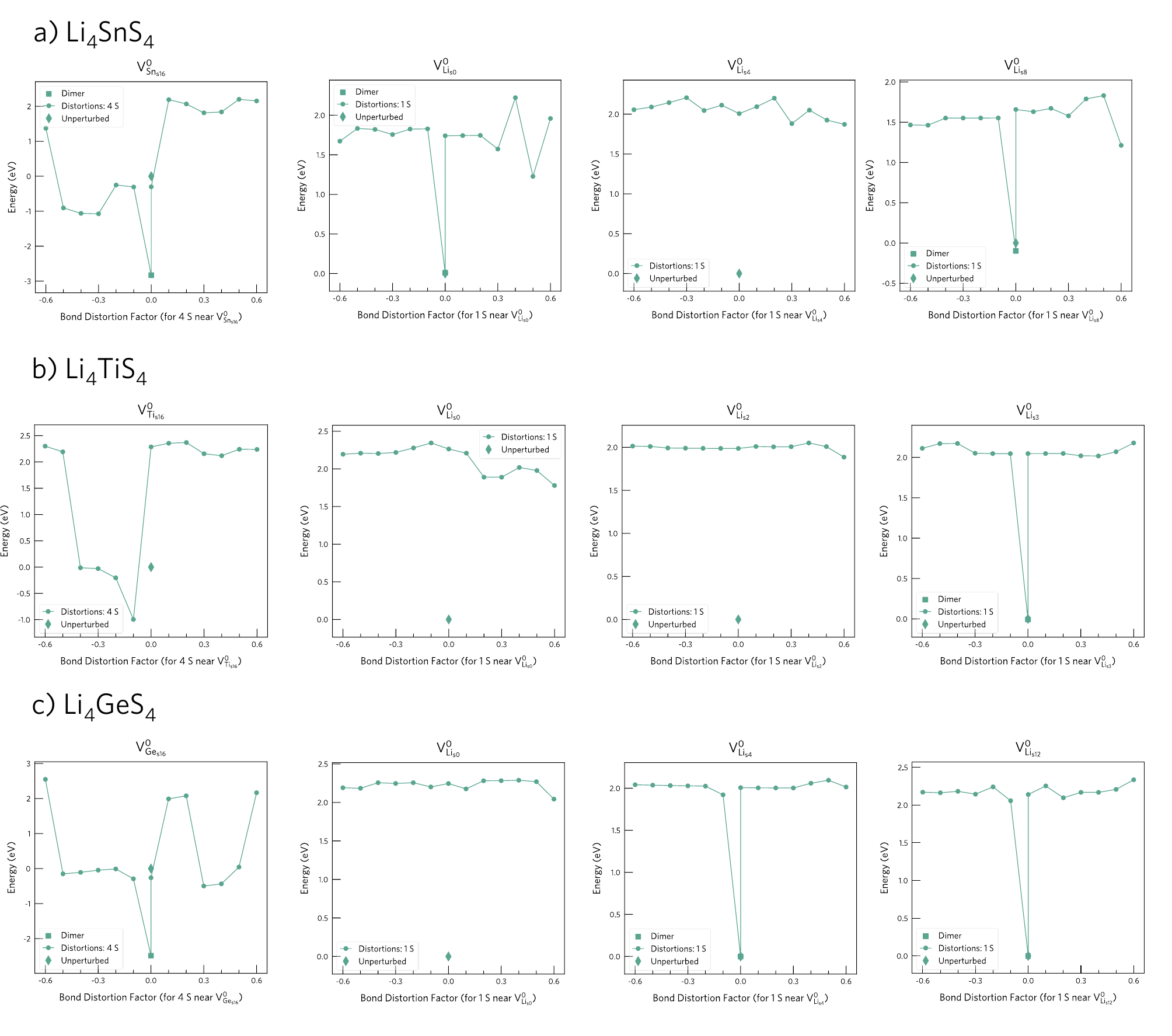}
    \caption{Energy vs distortion plots for all the cation vacancies in the test system \ce{Li4SnS4} (a) and the training compositions \ce{Li4TiS4} and \ce{Li4GeS4} (b, c). Note that most of the relaxations are trapped in high-energy minima, thus hindering learning the low-energy region of the PES for these systems. This trapping into high-energy basins can be avoided by reducing the magnitude of the rattle distortion that is applied with ShakeNBreak to the sampling structures. }
    \label{sfig:Li4SnS4_snb_plots}
\end{figure}

\FloatBarrier
\subsection{Acceleration factors}\label{ssubsec:speedup}
\begin{table}[ht!]
\caption{
Comparison of computing times for structure searching when using the full DFT approach and the MLFF+DFT approach. 
We compare the time required to run the structure searching relaxations when only using DFT (i.e. starting from the ShakeNBreak initial structures) and when pre-relaxing the structures with the MLFF. These timings are listed in the columns `DFT time' (full DFT relaxation of ShakeNBreak structures), `MLFF time' (DFT relaxation of the MLFF-relaxed ShakeNBreak structures) and `Inference time' (MLFF relaxation of ShakeNBreak structures).
}
\fontsize{10.75pt}{10.75pt}\selectfont
\setlength{\tabcolsep}{6.1pt} % Default value: 6pt
\renewcommand{\arraystretch}{0.97} % Default value: 1
\begin{tabular}{cccccccc}
\toprule[1pt]
Host 
& Defect 
& \thead{Num \\atoms} 
& \thead{Num \\electrons} 
& \thead{DFT time \\(CPU h)}
& \thead{MLFF time\\ (CPU h)} 
& \thead{Inference time \\(GPU h)} 
& Speedup\footnotemark \\
\midrule[0.1pt]
\ce{BiSeBr}  & \textit{V}$_{\rm Bi_0}$  & 71  & 657  & 4736.7   & 3742.7  & 0.05 & 1.3  \\
\ce{BiSeI}   & \textit{V}$_{\rm Bi_0}$  & 71  & 657  & 5778.6   & 3160.8  & 0.04 & 1.8  \\
\ce{Li4SnS4} & \textit{V}$_{\rm Sn_16}$ & 143 & 786  & 13976.7  & 18042.3 & 0.21 & 0.8  \\
\ce{Li4SnS4} & \textit{V}$_{\rm Li_0}$  & 143 & 797  & 11549.0  & 2576.1  & 0.18 & 4.5  \\
\ce{Li4SnS4} & \textit{V}$_{\rm Li_4}$  & 143 & 797  & 433083.1 & 6847.8  & 0.24 & 63.2 \\
\ce{Li4SnS4} & \textit{V}$_{\rm Li_8}$  & 143 & 797  & 41380.0  & 6205.6  & 0.15 & 6.7  \\
\ce{CuSe}    & \textit{V}$_{\rm Cu_0}$  & 107 & 907  & 21196.8  & 969.0   & 0.06 & 21.9 \\
\ce{CuSe}    & \textit{V}$_{\rm Cu_4}$  & 107 & 907  & 27734.9  & 1647.9  & 0.05 & 16.8 \\
\ce{CuS}     & \textit{V}$_{\rm Cu_0}$  & 143 & 1213 & 30292.0  & 7821.3  & 0.03 & 3.9  \\
\ce{CuS}     & \textit{V}$_{\rm Cu_3}$  & 143 & 1213 & 39892.4  & 3701.4  & 0.03 & 10.8 \\
\bottomrule[1pt]
\end{tabular}
\vskip 1em
\footnotetext[1]{The speedup factor is calculated as $\frac{t_{\rm relax, DFT}}{t_{\rm relax, MLFF} + t_{\rm infer, MLFF}}$, with a mean value of 13.2. The DFT relaxations of both the ShakeNBreak initial structures and the MLFF-relaxed structures are performed on 48 cores (node with two AMD EPYC 7742 processors). The MLFF relaxations of the ShakeNBreak initial structures are performed on a Quadro RTX6000 GPU.}
\end{table}

\section{Surrogate model for closely-related systems}\label{ssec:chalcohalides}
\begin{figure}[ht]
    %\centering
    \includegraphics[width=0.4\linewidth]{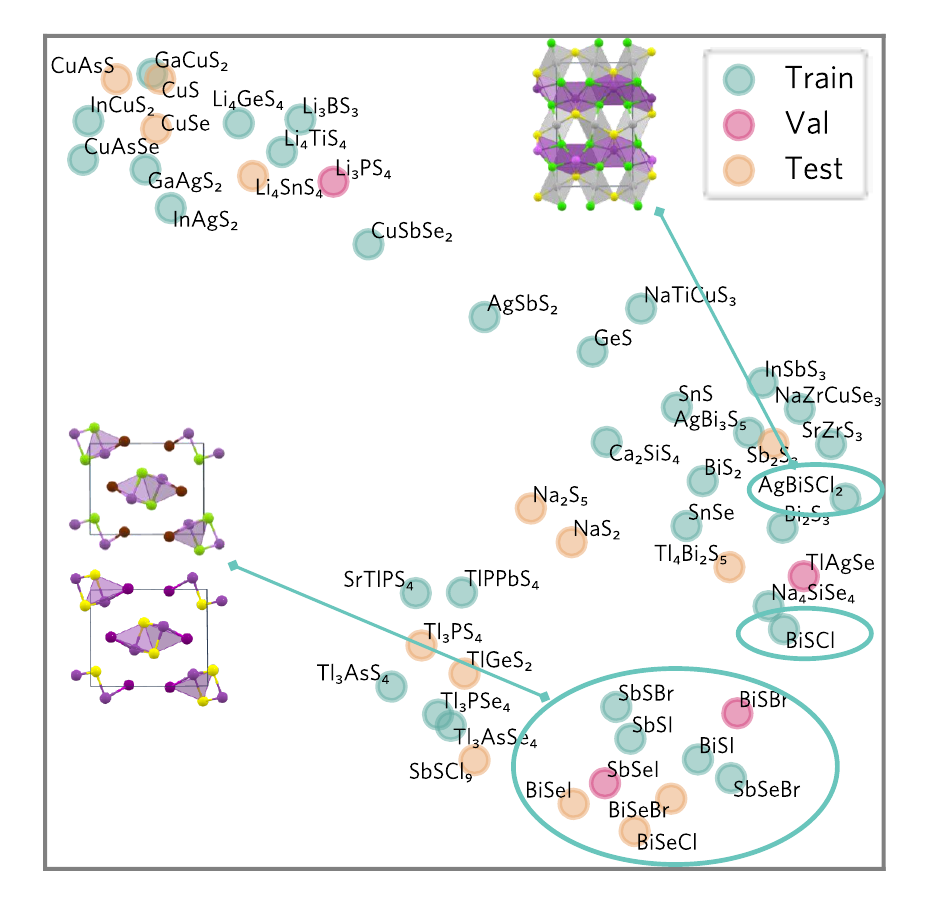}
    \caption{Plot of the first two principal components of the feature space for the pristine structures, showing the selection of the chalcohalide systems. The structures were encoded with the SOAP descriptor, whose dimensions were reduced for plotting with Principal Component Analysis.\cite{bartok_2013}.}
    \label{sfig:chalco_umap}
\end{figure}

\begin{figure}[ht]
    %\centering
    \includegraphics[width=\linewidth]{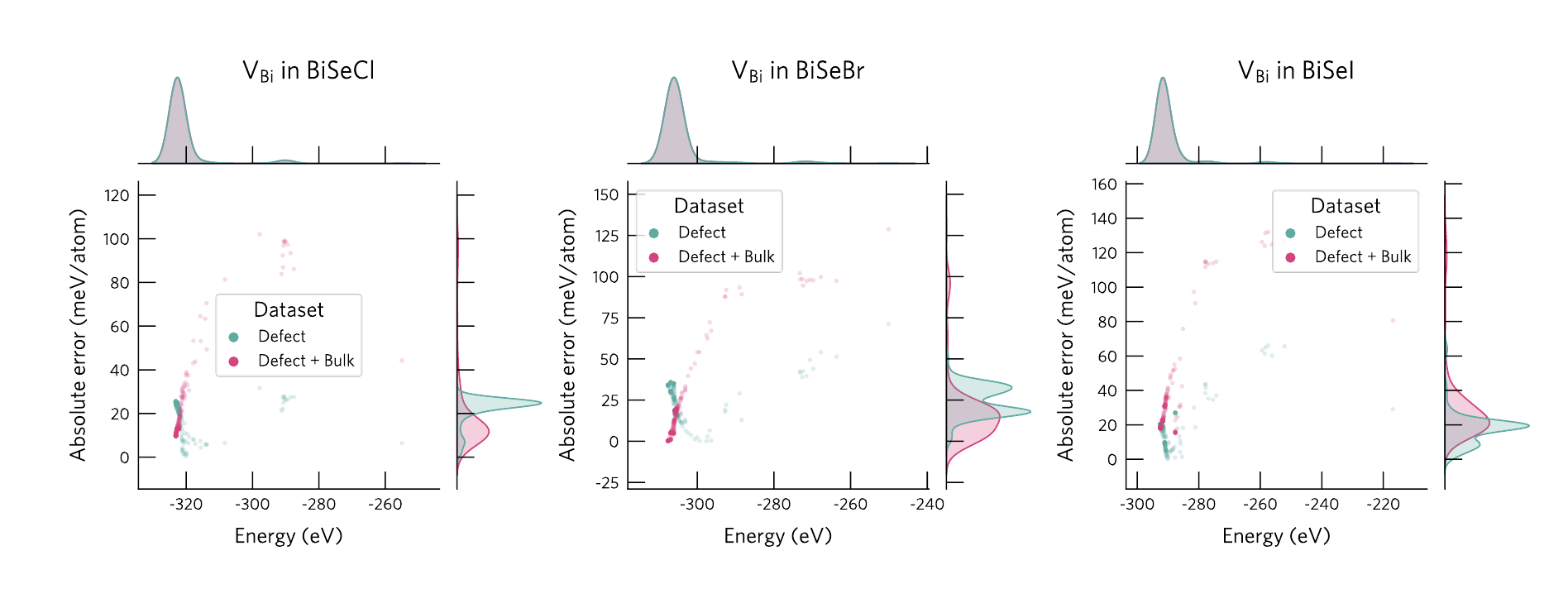}
    \caption{Comparison of absolute energy error distributions for the test systems when training on defect (green) and both defect and bulk data (pink). As seen for \ce{BiSeCl} and \ce{BiSeBr}, adding bulk data to the training set reduces the errors for the low-energy configurations.}\label{sfig:chalco_E_error_distribution}
\end{figure}

\begin{figure}[ht]
    \centering
    \includegraphics[width=\textwidth]{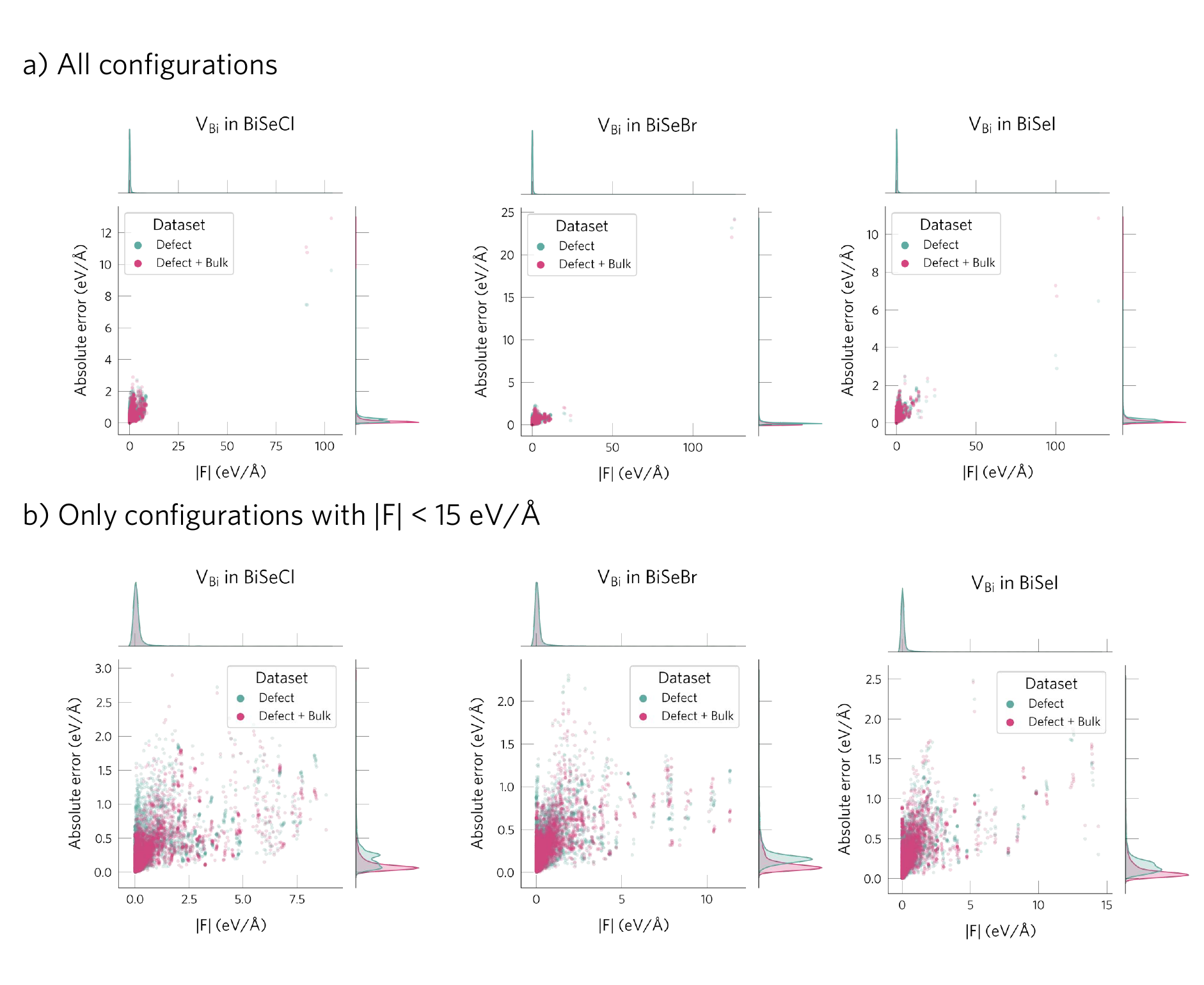}
    \caption{Comparison of absolute force error distributions for the test systems when training on defect (green) and both defect and bulk data (pink). a) All configurations and b) removing outliers with high forces ($|F| < 15~\mathrm{eV}/\mathrm{\angstrom}$). Adding bulk data to the training set reduces the force errors for the low-energy configurations. }\label{sfig:chalco_F_error_distribution}
\end{figure}

To investigate whether targeting more similar systems would reduce the dataset size required for similar model performance, we developed a model for chalcohalide systems. We selected the chalcohalides from our dataset, resulting in 11 compositions (BiSCl, BiSBr, BiSI, BiSeCl, BiSeBr, BiSeI,
SbSBr, SbSI, SbSeBr, SbSeI
\ce{AgBiSCl2}) (\ref{sfig:chalco_umap}). Three of these (27\%; BiSeCl, BiSeBr, BiSeI) were held out as the test set, and the remaining data was split into training and validation sets with 0.9 and 0.1 fractions \footnote{This split was done by selecting evenly spaced frames for the validation data, to ensure that both sets are representative of the original dataset.}, resulting in a training and validation sizes of 1476 and 164 configurations, respectively. 
The resulting training set was then increased by adding ten evenly spaced frames from the relaxation of each \emph{pristine} host structure (110 configurations) --- as this was observed to improve performance (\ref{stab:chalco_bulk}, \ref{sfig:chalco_E_error_distribution}, \ref{sfig:chalco_F_error_distribution}). 

\begin{table}[ht]
\caption{Mean absolute errors, root mean squared errors and Spearman coefficients on the test set for the chalcohalide model when only training on defect data and when training on both defect and bulk data. The distribution of the absolute errors is shown in \ref{sfig:chalco_E_error_distribution}, \ref{sfig:chalco_F_error_distribution}.}\label{stab:chalco_bulk}
\vspace{10pt}
\begin{tabular}{crrrrrrr}
\hline
Dataset & 
\thead{$\rm MAE_{E}$ \\(meV/atom)}  & 
\thead{$\rm RMSE_{E}$ \\(meV/atom)} &
$\rho$ &
\makecell{$\rm MAE_{F}$ \\(meV/\AA)} & 
\makecell{$\rm RMSE_{F}$ \\(meV/\AA)} & 
\thead{$\rm MAE_{S}$ \\(GPa)} &
\thead{$\rm RMSE_{S}$ \\(GPa)} \\
\hline
Bulk + Defect & 21.9   &  30.6  & 0.82 & 63.8 & 153.7 & 0.10 & 0.17 \\
Defect only   & 21.3   & 23.1   & 0.72 & 97.6 & 175.2 & 0.10 & 0.16 \\
\hline
\end{tabular}
\end{table}

\begin{figure}[ht]
\centering\includegraphics[width=0.98\textwidth]{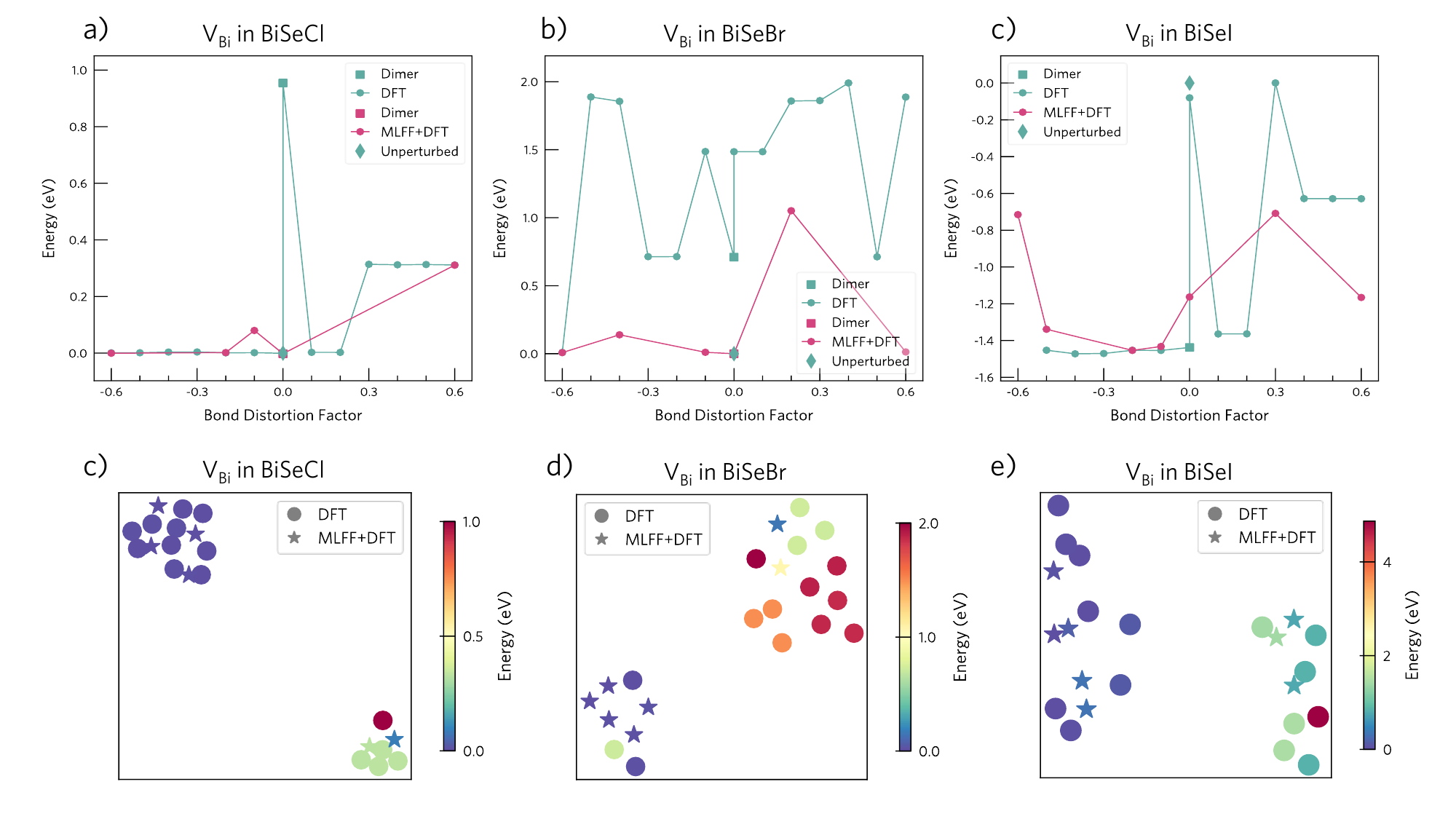}
    \caption{
    Comparison of DFT and MLFF+DFT approaches. (a, b, c) Plots of final (DFT) energy versus initial distortion for \kvc{V}{Bi}{0} in \ce{BiSeCl}, \ce{BiSeBr} and \ce{BiSeI}. The full DFT approach (green) is compared with our MLFF+DFT strategy (pink). Note that the MLFF+DFT successfully targets the low-energy regions of the PES. Further, we note that several low-energy metastable configurations are missed with the DFT-only approach but identified with the MLFF+DFT one. The label ``Unperturbed" denotes the configuration obtained by relaxing the high symmetry ideal structure while the label ``Dimer" denotes a targeted distortion that pushes two of the defect nearest neighbours towards each other. (c, d, e) 2D projection of structural similarity for final structures obtained with the full DFT (circles) and MLFF+DFT approach (stars), illustrating that the latter targets the low-energy regions of the PES (e.g. no red stars).}
    \label{sfig:snb_chalco}
\end{figure}

As shown in \ref{stab:mae_chalco}, the mean absolute errors are slightly lower than for the full model (trained on all compositions), confirming that smaller datasets can be used when targeting more similar host structures since their PES is easier to learn. After applying our MLFF+DFT approach to the test systems, it identifies the correct ground state for all three defects, while reducing the number of DFT calculations by 53\%. 
Further, we note that the candidate structures selected in our approach (by relaxing the initial sampling structures with the MLFF and then selecting the structures with a unique SOAP fingerprint\cite{bartok_2013} for the defect site) target the low-energy region of the PES, as demonstrated in \ref{sfig:snb_chalco}. 
Further, it also identifies two low-energy metastable structures that are missed with the DFT-only approach, validating that the model learns to suggest good candidate structures.

\begin{table}[ht]
\caption{Mean absolute errors and Spearman coefficients ($\rho$) for training, validation and test sets of the chalcohalide model (training on both defect and bulk data).}\label{stab:mae_chalco}
\vspace{10pt}
\begin{tabular}{crrrr}
\hline
Set & \thead{$\rm E$ \\(meV/atom)} & \makecell{$\rm F$ \\(meV/\AA)} & \thead{$\rm S$ \\(GPa)} & $\rho$ \\
\hline
Train & 19.2 & 51.0	& 0.06 & 0.92 \\
Val & 13.3 & 51.7 & 0.05 & 0.91 \\
Test & 21.9 & 63.7 & 0.10 & 0.82 \\
\hline
\end{tabular}
\end{table}

\begin{figure}[ht]
\centering\includegraphics[width=0.80\textwidth]{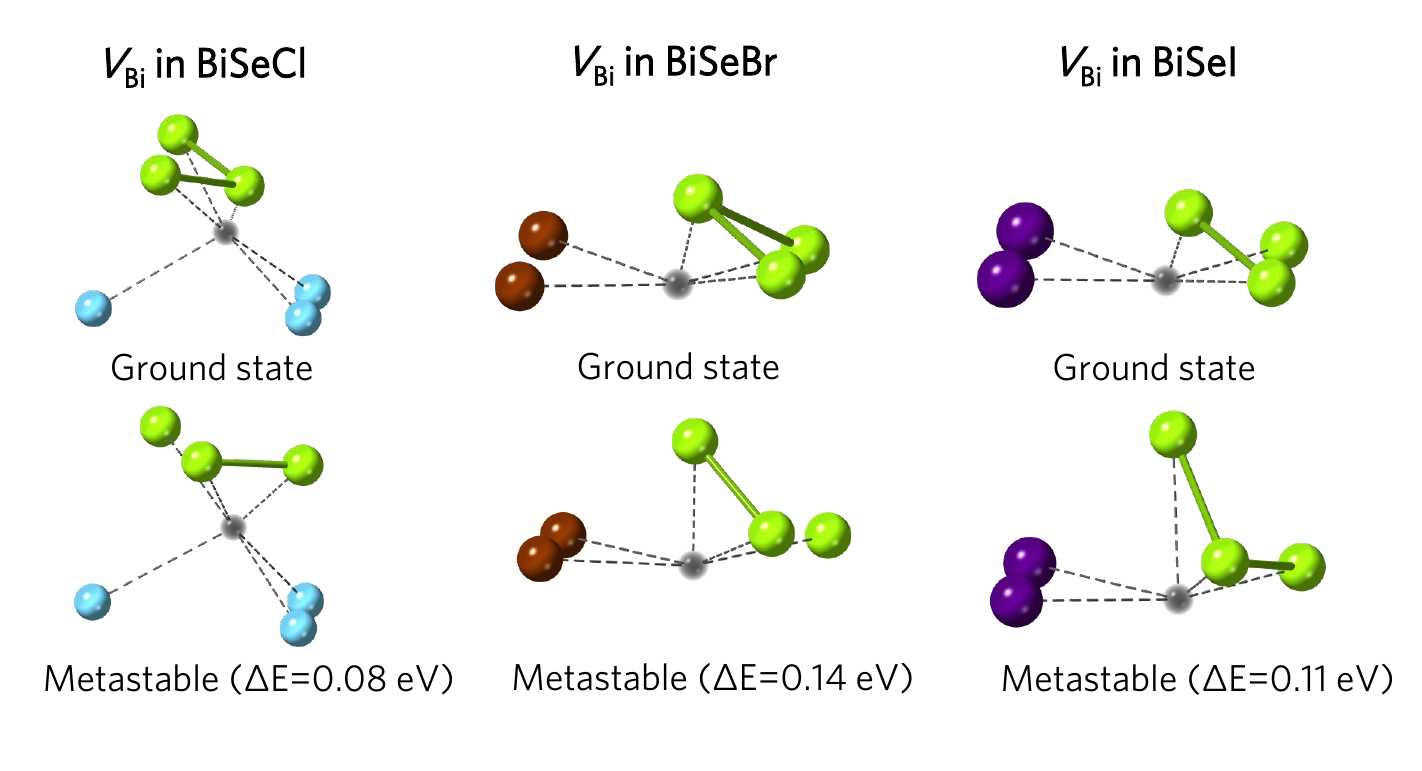}
    \caption{
    Ground state structures for \kv{V}{Bi} in \ce{BiSeCl}, \ce{BiSeBr} and \ce{BiSeI}. On the bottom, the low-energy metastable configurations that are only identified with the MLFF+DFT strategy are shown.}
    \label{sfig:snb_chalco_structs}
\end{figure}
%

% \subsection{Learning curve}
% To investigate the change in model performance with training set size, we train a set of models on increasingly large portions of the training set. These subsets are created so that each of them covers all the elements in the test set (\ref{fig:subsets_curve}). By comparing our models with the original M3GNet model (without fine-tuning), we note that fine-tuning on defect data reduces the errors on the test set (Table XX). However, training on larger defect subsets does not improve the test errors. 
% %
% \begin{figure}[ht]
% \centering\includegraphics[width=0.45\textwidth]{Figures/maml_m3gnet_splits.pdf}
% \caption{Subsets of the host compositions used for training models on increasingly large fractions of the dataset. The legend indicates the number of compositions included in each training subset. Different tones of green-blue illustrate the compositions that are \emph{added} in each subset.    
% }
% \label{fig:subsets_curve}
% \end{figure}
% %

\FloatBarrier
\section{Extension to Alloys}\label{ssec:alloy}

\begin{figure}[ht]
    %\centering
    \includegraphics[width=\linewidth]{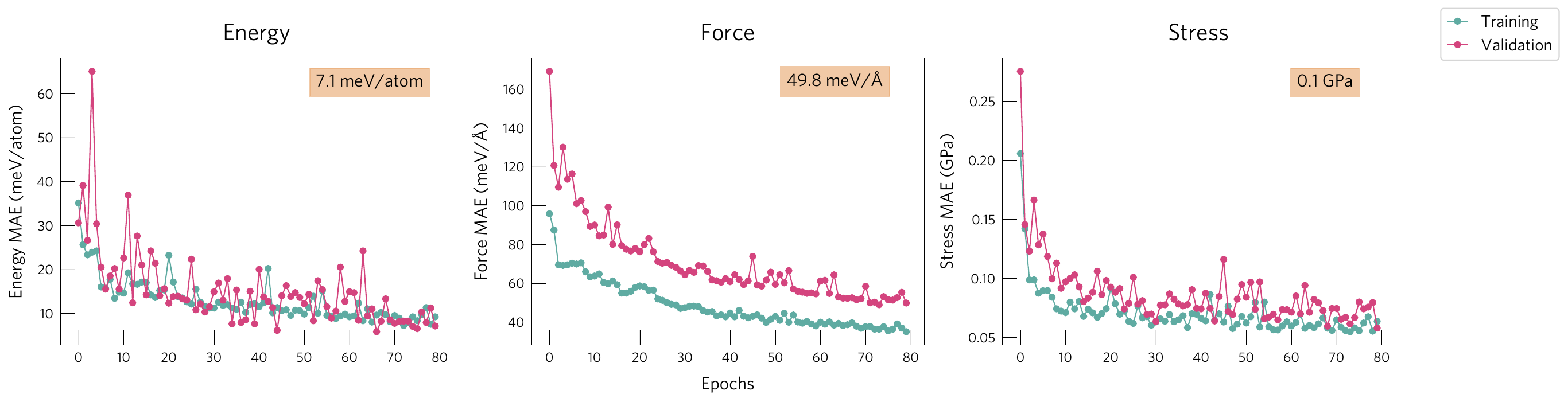}
    \caption{Evolution of training and validation errors with epoch number. The validation errors obtained for the epoch with the lowest validation loss are shown at the top right of each subplot.
    }\label{sfig:training_evolution_alloy}
\end{figure}

\begin{table}[ht!]
\fontsize{10.75pt}{10.75pt}\selectfont
\caption{Performance of the MLFF model on the CdSe$_x$Te$_{(1-x)}$ systems. For each inequivalent defect in each alloy, we show the number of local minima identified by the DFT-based (coarser) search and by the MLFF finer search. When a more favourable ground state (GS) configuration is identified by the MLFF search, the energy lowering and the reconstruction driving it are shown. 
%$T_d$ denotes the tetrahedral (high-symmetry) coordination of the defect environment when no anion-anion bonds are formed.
}
\label{stab:performance_alloy}
\setlength{\tabcolsep}{6.5pt} % Default value: 6pt
\renewcommand{\arraystretch}{0.98} % Default value: 1
\begin{tabular}{cccccc}
\toprule[1pt]
x   & Defects  & \makecell{Num. local minima\\ in DFT PES} & \makecell{Num. local minima\\ in MLFF PES} & Novel GS? (eV) & Reconstruction                  \\
\midrule[0.2pt]
0.2 & \textit{V}$_{\rm Cd_{0}}$  & 2 & 5  & -0.3 & Se-Te $\rightarrow$ Te-Te       \\
0.2 & \textit{V}$_{\rm Cd_{10}}$ & 2 & 6  & -0.5 & Se-Se $\rightarrow$ Se-Te       \\
0.2 & \textit{V}$_{\rm Cd_{1}}$  & 2 & 4  & -0.6 & Se-Se $\rightarrow$ Te-Te       \\
0.2 & \textit{V}$_{\rm Cd_{4}}$  & 2 & 4  & 0.0  & Te-Te $\rightarrow$ Te-Te       \\
0.3 & \textit{V}$_{\rm Cd_{0}}$  & 2 & 5  & -0.4 & Se-Te $\rightarrow$ Te-Te       \\
0.3 & \textit{V}$_{\rm Cd_{21}}$ & 3 & 5  & -0.2 & Se-Se $\rightarrow$ Se-Te       \\
0.3 & \textit{V}$_{\rm Cd_{2}}$  & 2 & 6  & 0.0  & Se-Se $\rightarrow$ Se-Se       \\
0.3 & \textit{V}$_{\rm Cd_{4}}$  & 2 & 4  & -0.2 & Se-Se $\rightarrow$ Se-Te       \\
0.3 & \textit{V}$_{\rm Cd_{5}}$  & 2 & 4  & 0.0  & Te-Te $\rightarrow$ Te-Te       \\
0.5 & \textit{V}$_{\rm Cd_{0}}$  & 1 & 6  & -0.7 & T$_{\rm d}$\footnotemark $\rightarrow$ Te-Te \\
0.5 & \textit{V}$_{\rm Cd_{4}}$  & 1 & 5  & -0.5 & T$_{\rm d}$ $\rightarrow$ Te-Te \\
0.5 & \textit{V}$_{\rm Cd_{1}}$  & 2 & 3  & -0.5 & Se-Te $\rightarrow$ Te-Te       \\
0.5 & \textit{V}$_{\rm Cd_{3}}$  & 2 & 4  & -0.3 & Se-Se $\rightarrow$ Se-Te       \\
0.5 & \textit{V}$_{\rm Cd_{5}}$  & 2 & 4  & -0.4 & Se-Se $\rightarrow$ Se-Se       \\
0.6 & \textit{V}$_{\rm Cd_{0}}$  & 2 & 4  & -0.5 & Se-Se $\rightarrow$ Se-Te       \\
0.6 & \textit{V}$_{\rm Cd_{2}}$  & 3 & 10 & -0.1 & Se-Se $\rightarrow$ Se-Te       \\
0.6 & \textit{V}$_{\rm Cd_{3}}$  & 2 & 5  & 0.0  & Te-Te $\rightarrow$ Te-Te       \\
0.6 & \textit{V}$_{\rm Cd_{4}}$  & 3 & 3  & 0.0  & Se-Se $\rightarrow$ Se-Se       \\
0.8 & \textit{V}$_{\rm Cd_{0}}$  & 3 & 5  & -0.6 & Se-Se $\rightarrow$ Se-Te       \\
0.8 & \textit{V}$_{\rm Cd_{1}}$  & 3 & 7  & 0.0  & Se-Se $\rightarrow$ Se-Se       \\
0.8 & \textit{V}$_{\rm Cd_{25}}$ & 1                           & 3                             & -0.6      & T$_{\rm d}$ $\rightarrow$ Te-Te \\
0.8 & \textit{V}$_{\rm Cd_{3}}$  & 3 & 4  & -0.3 & Se-Te $\rightarrow$ Te-Te       \\
0.9 & \textit{V}$_{\rm Cd_{0}}$  & 1 & 4  & -0.7 & T$_{\rm d}$ $\rightarrow$ Se-Te \\
0.9 & \textit{V}$_{\rm Cd_{0}}$  & 3 & 6  & 0.0  & Se-Se $\rightarrow$ Se-Se \\
\bottomrule[1pt]
\end{tabular}
\vskip 1em
\footnotetext[1]{$T_d$ denotes the tetrahedral (high-symmetry) coordination of the defect environment when no anion-anion bonds are formed.}
\end{table}

\end{document}